\documentclass[fleqn,usenatbib]{mnras}
\usepackage{graphicx} 
\usepackage{txfonts} 
\usepackage{natbib}               
\usepackage{rotating}
\usepackage{multicol}
\usepackage{longtable}
\usepackage{multirow}

\title[] {Multiple Stellar Populations in Globular Clusters with JWST:  a NIRCam view of 47\,Tucanae}
\author[ A.\,P.\,Milone et al.] 
       {A.\,P.\,Milone$^{1,2}$, A.\,F.\,Marino$^{2,3}$, A.\,Dotter$^{4}$, T.\,Ziliotto$^{1}$, E.\,Dondoglio$^{1}$, G.\,Cordoni$^{1}$, \newauthor S.\,Jang$^{5}$,  E.\,P.\,Lagioia$^{1}$, 
       M.\,V.\,Legnardi$^{1}$, A.\,Mohandasan$^{1}$,  M.\,Tailo$^{6}$, D.\,Yong$^{7}$, \newauthor S.\,Baimukhametova$^{1}$, M.\,Carlos$^{8}$  \\ 
$^{1}$Dipartimento di Fisica e Astronomia ``Galileo Galilei'', Univ. di Padova, Vicolo dell'Osservatorio 3, Padova, IT-35122\\
$^{2}$Istituto Nazionale di Astrofisica - Osservatorio Astronomico di Padova, Vicolo dell'Osservatorio 5, Padova, IT-35122\\
$^{3}$ Istituto Nazionale di Astrofisica - Osservatorio Astrofisico di Arcetri, Largo Enrico Fermi, 5, Firenze, IT-50125 \\
$^{4}$Department of Physics and Astronomy, Dartmouth College, 6127 Wilder Laboratory, Hanover, NH 03755, USA \\
$^{5}$ Center for Galaxy Evolution Research and Department of Astronomy, Yonsei University, Seoul 03722, Korea\\
$^{6}$ Dipartimento di Fisica e Astronomia Augusto Righi, Università degli Studi di Bologna, Via Gobetti 93/2, 40129, Bologna, Italy\\
$^{7}$Research School of Astronomy and Astrophysics, Australian National University, Canberra, ACT 2611, Australia\\
$^{8}$Theoretical Astrophysics, Department of Physics and Astronomy, Uppsala University, Box 516, SE-751 20 Uppsala, Sweden
}
\begin{document} 
%\date{Accepted 2016 March 10. Received 2016 March 8; in original form 2016 January 7} 
 
%\pagerange{\pageref{firstpage}--\pageref{lastpage}} \pubyear{2017} 
\maketitle 
\label{firstpage}

\begin{abstract}
We use images collected with the near-infrared camera (NIRCam) on board the James Webb Space Telescope and with the Hubble Space Telescope (HST) to investigate multiple populations at the bottom of the main sequence (MS) of 47 Tucanae. 
 The $m_{\rm F115W}$ vs.\,$m_{\rm F115W}-m_{\rm F322W2}$ CMD from NIRCam shows that, below the knee, the MS stars span a wide color range, where the majority of M-dwarfs exhibit blue colors, and a tail of stars are distributed toward the red. 
 A similar pattern is observed from the $m_{\rm F160W}$ vs.\,$m_{\rm F110W}-m_{\rm F160W}$ color-magnitude diagram (CMD) from HST, and multiple populations of M-dwarfs are also visible in the optical $m_{\rm F606W}$ vs.\,$m_{\rm F606W}-m_{\rm F814W}$ CMD. The NIRCam CMD shows a narrow sequence of faint MS stars  with masses smaller than $0.1\mathcal{M}_{\odot}$.
We introduce a chromosome map of M-dwarfs that reveals an extended first population and three main groups of second-population stars. By combining isochrones and synthetic spectra with appropriate chemical composition, we simulate colors and magnitudes of different stellar populations in the NIRCam filters (at metallicities [Fe/H]=-1.5 and  [Fe/H]=-0.75) and identify the photometric bands that provide the most efficient diagrams to investigate the multiple populations in globular clusters. 
 Models are compared with the observed CMDs of 47 Tucanae to constrain M-dwarfs' chemical composition. 
Our analysis suggests that the oxygen range needed to reproduce the colors of first- and second-population M-dwarfs is similar to that inferred from spectroscopy of red giants, constraining the proposal that the chemical variations are due to mass transfer phenomena in proto-clusters.
\end{abstract} 
 

\section{Introduction}\label{sec:intro}

Nearly all Globular Clusters (GCs) host two main  distinct stellar populations with different content of the elements involved in the hot H-burning (e.g.\,He, C, N, O, Na, Al, and in some cases Mg, Si, and K). The light-element abundance of the first stellar population (1P) resembles that of Galactic field stars with similar metallicities. On the contrary, second population (2P) stars are enhanced in He, N, Na, and Al and depleted in C and O. Both 1P and 2P stars can host subpopulations of stars \citep[see reviews by][]{kraft1994a, bastian2018a, gratton2019a, milone2022a}. 

Despite intensive investigation on GCs, the origin of their multiple populations is not understood. The phenomenon has been interpreted in terms of successive stellar generations, i.\,e.\, multiple bursts of star formation. This scenario implies that the GC progenitors were substantially more massive at the formation and lost most of their 1P stars into the halo before delivering the naked present-day GCs. As a consequence, GCs  could have provided a significant contribution to the assembly of the Milky-Way halo, and possibly, to the reionization of the Universe \citep[e.g.][]{cottrell1981a, dantona1983a, dantona2016a, decressin2007a, denissenkov2014a, renzini2015a, renzini2022a}.
An alternative scenario assumes that 1P and 2P stars are coeval and the chemical variations are due to mass accreted onto existing low-mass stars  \citep{bastian2013a, gieles2018a}. 

By far, photometry is one of the main techniques to study the multi-population phenomenon. Work based on multi-band images reveals that the multiple populations define distinct stellar sequences in various photometric diagrams constructed with magnitudes taken in appropriate filters \citep[][and  references therein]{milone2022a}.  The multiple sequences are better visible in photometric diagrams composed of ultraviolet filters from the {\it Hubble Space Telescope} ({\it HST}\,) and from ground-based facilities. 
 In these diagrams, the 1P and 2P sequences can be followed continuously along various evolutionary phases, from the upper main sequence (MS) to the sub-giant branch (SGB), the red giant branch (RGB), the horizontal branch (HB) and the asymptotic giant branch \citep[AGB, e.g.\,][]{marino2008a, yong2008a, milone2012a, milone2017a, piotto2015a, lee2017a, dondoglio2021a, lagioia2021a, jang2022a}. 

 The observational results are supported by 
 %works based on 
 simulated photometry derived from isochrones and synthetic spectra that account for the chemical composition of the multiple populations of GCs. The reason why UV filters are efficient tools to identify multiple populations is that they include various spectral features. As an example, the $U$ band, and the equivalent {\it HST} filter F336W, includes NH and CN molecular bands, whereas the F275W filter encompasses OH bands. The $B$ filter includes CH molecular bands, whereas the narrow-band F280N filter encloses Mg lines \citep[e.g.\,][]{marino2008a, sbordone2011a, milone2012a, milone2018a, milone2020a, dotter2015a, li2022a, jang2022a, vandenberg2022a}.
The multiple stellar populations are also visible among M-dwarfs by using photometry obtained with the near-infrared channel of the Wide Field Camera 3 (NIR/WFC3) on board {\it HST}. Below the knee, the MS of various GCs is either split or exhibits a broad F110W$-$F160W color distribution. This phenomenon is mostly associated with the effect of various molecules composed of oxygen, which strongly affect the spectral region covered by the F160W filter. The 2P M-dwarfs, which are depleted in oxygen, have weaker molecular bands than the 1P stars. Hence, they exhibit brighter F160W magnitudes and redder F110W$-$F160W colors \citep{milone2012b, milone2019a, dotter2015a, dondoglio2022a}.

  Recently, \citet{salaris2019a} started investigating the effect of multiple populations in the filters of the near-infrared camera (NIRCam) of the James Webb Space Telescope ({\it JWST}) by using theoretical stellar spectra and stellar evolution models.
  In their analysis, which is limited to the RGB, they identified various colors to distinguish multiple populations near the RGB tip.
  
 47\,Tucanae is one of the most-studied clusters in the context of multiple populations by means of spectroscopy \citep[e.g.][]{carretta2013a, cordero2014a}, photometry \citep[e.g.][]{anderson2009a,  dicriscienzo2010a, milone2022a}, and kinematics \citep[e.g.][]{richer2013a, milone2018b, cordoni2020a}.
 The 1P and 2P stars appear as discrete stellar populations in the ChMs derived from the upper MS and the RGB and define distinct sequences in the photometric diagrams that are commonly used to identify multiple populations \cite[e.g.\,][]{milone2012a, milone2022a, dondoglio2021a, jang2022a, lee2022a}. A visual inspection at the ChM reveals that the sequence of 2P stars exhibits at least four stellar clumps that correspond to stellar populations with different content of helium, carbon, nitrogen, and oxygen \citep{milone2017a, marino2019a, milone2022a}. 

 The maximum star-to-star variation of [C/Fe], [N/Fe], and [O/Fe] are $\sim$0.5, $\sim$1.0, and $\sim$0.5 dex, respectively, \citep{carretta2009a, marino2016a, dobrovolskas2014a}, whereas helium spans an interval of $\delta Y=$0.05 in mass fraction \citep{milone2018a}. 
  The 1P sequence of the ChM exhibits a significant  color broadening, which is consistent with an iron  variation by [Fe/H]$\sim$0.09 dex \citep{milone2018a, legnardi2022a, jang2022a}.

 In this work, we investigate the behavior of multiple populations in photometric diagrams of GCs constructed with photometry from NIRCam/JWST and from {\it HST}.   
  The paper is organized as follows. Section\,\ref{sec:data} describes the data of 47\,Tucanae, and the data reduction. The photometric diagrams of 47\,Tucanae are presented in Section\,\ref{sec:47tuc}. 
  Section\,\ref{sec:iso} presents synthetic spectra and isochrones in the NIRCam and {\it HST} filters that account for the chemical composition of 1P and 2P stars in GCs, while Section\,\ref{sec:iso47t} compares the isochrones and the observations of 47\,Tucanae. The summary and conclusions are provided in Section\,\ref{sec:conclusion}. 
  
\section{Data and data analysis}\label{sec:data}
\begin{centering} 
\begin{figure} 
  \includegraphics[width=8.75cm]{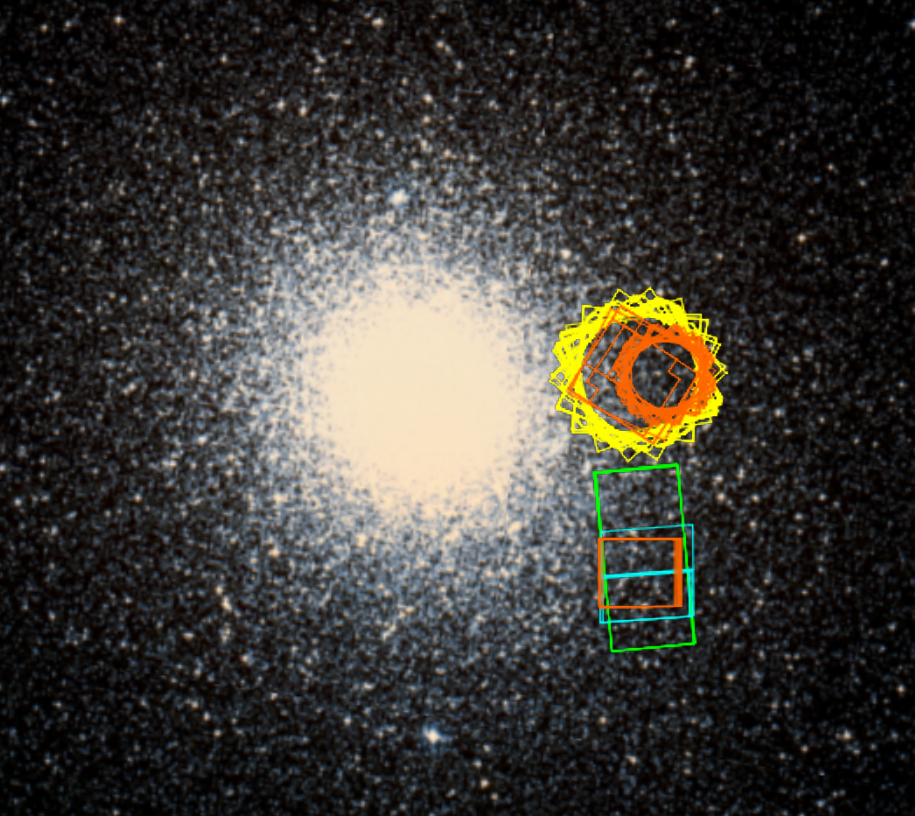}
 % Emanuele D.
 \caption{ Footprints of the images used in this paper. North is up and east is left. The green rectangle marks the NIRCam FoV, whereas {\it HST} data are colored orange (NIR/WFC3), yellow (WFC/ACS), and cyan (UVIS/WFC3). 
 }
 \label{fig:footprints} 
\end{figure} 
\end{centering}

To investigate multiple stellar populations at the bottom of the MS of 47\,Tucanae, we used deep images of two distinct fields, namely A and B, collected with {\it HST} and {\it JWST}.
 As illustrated in Figure\,\ref{fig:footprints}, where we show the footprints of these images, field A is located $\sim$7 arcmin west with respect to the cluster center, whereas field B is $\sim 8.5$ arcmin to the south-west.
 The main observations of the field A (RA$\sim$00$^{\rm h}$22$^{\rm m}$37$^{\rm s}$, DEC$\sim -72^{\rm d}$04$^{\rm m}$06$^{\rm s}$) are obtained with the Wide Field Channel of the Advanced Camera for Survey (WFC/ACS) through the F606W and F814W filters and the near-infrared channel of WFC3 (IR/WFC3) in the F110W and F160W bands. 
  Moreover, we analyzed IR/WFC3 images collected through the F105W and F140W filters.
  Details on the dataset are provided in Table\,\ref{tab:data}, whereas in the following we describe the methods to measure stellar fluxes and positions.
 Field B (RA$\sim$00$^{\rm h}$22$^{\rm m}$36$^{\rm s}$, DEC$\sim -72^{\rm d}$09$^{\rm m}$27$^{\rm s}$) has been observed with NIRCam/JWST as part of GO-2560 (PI A.\,F.\,Marino). 
% {\color{red} NON VEDO IN FIG 1 INDICATI I FIELDS CON A E B.}
 Moreover, we used images collected through the F606W filter of the Ultraviolet and Visual Channel of the Wide Field Camera 3 (UVIS/WFC3)  on board {\it HST} and F110W, and F160W IR/WFC3 data. 

%%%%%%%%%%%%%%%%%%%     

 \onecolumn
\begin{center}
\begin{longtable}{cccllll}
\hline
\caption{Description of the images used in the paper. For each dataset, we indicate the mission ({\it JWST} or {\it HST}), the camera, the filter, the date, the exposure times, the GO program, and the principal investigator. }\label{tab:data}\\
\multicolumn{7}{c}%
%\endhead

%\hline \multicolumn{7}{|r|}{{Continued on next page}} \\ \hline
\endfoot
%{{\bfseries \tablename\ \thetable{} -- continued from previous page}} \\

\hline \hline
\endlastfoot
\hline \hline
 MISSION & CAMERA & FILTER  & DATE & N$\times$EXPTIME & GO & PI \\
\hline
         &          &          &     Field A                &                   &        & \\
  HST    &  WFC/ACS &  F606W   &  February, 13, 2010 &  1s$+$1261s$+$1303s$+$1442s$+$1456s$+$1457s &  11677      &  H.\,B.\,Richer   \\
  HST    &  WFC/ACS &  F606W   &  March, 04, 2010    &  1442s$+$1456s$+$1457s$+$1470s$+$1498s      &  11677      &  H.\,B.\,Richer   \\
  HST    &  WFC/ACS &  F606W   &  March, 16, 2010    & 10s$+$1206s$+$1298s$+$1396s$+$1442s$+$1457s &  11677      &  H.\,B.\,Richer   \\
  HST    &  WFC/ACS &  F606W   &  April, 10, 2010    &  1442s$+$1456s$+$1457s$+$1470s$+$1490s      &  11677      &  H.\,B.\,Richer   \\
  HST    &  WFC/ACS &  F606W   &  June, 12, 2010     &  1306s$+$1320s$+$1442s$+$1457s$+$1498s      &  11677      &  H.\,B.\,Richer   \\
  HST    &  WFC/ACS &  F606W   &  June, 18, 2010     &  1s$+$1261s$+$1303s$+$1442s$+$1456s$+$1457s &  11677      &  H.\,B.\,Richer   \\
  HST    &  WFC/ACS &  F606W   &  July, 29, 2010     &  1226s$+$1442s$+$1457s                      &  11677      &  H.\,B.\,Richer   \\
  HST    &  WFC/ACS &  F606W   &  August, 05, 2010   & 10s$+$1266s$+$1298s$+$1442s$+$1456s$+$1457s &  11677      &  H.\,B.\,Richer   \\
  HST    &  WFC/ACS &  F606W   &  August, 14-15, 2010   &  1442s$+$1456s$+$1457s$+$1470s$+$1490s   &  11677      &  H.\,B.\,Richer   \\
  HST    &  WFC/ACS &  F606W   &  September, 19, 2010& 100s$+$1252s$+$1253s$+$1442s$+$1456s$+$1457s&  11677      &  H.\,B.\,Richer   \\
  HST    &  WFC/ACS &  F606W   &  October, 01, 2010  & 1442s$+$1456s$+$1457s$+$1470s$+$1498s       &  11677      &  H.\,B.\,Richer   \\
  HST    &  WFC/ACS &  F606W   &  January, 16, 2010  &  1s$+$1217s$+$1303s$+$1371s$+$1442s$+$1457s &  11677      &  H.\,B.\,Richer   \\ 
  HST    &  WFC/ACS &  F606W   &  January, 17, 2010  &  1371s$+$1385s$+$1442s$+$1457s$+$1485s      &  11677      &  H.\,B.\,Richer   \\
  HST    &  WFC/ACS &  F606W   &  January, 18, 2010  & 10s$+$1208s$+$1303s$+$1371s$+$1442s$+$1457s &  11677      &  H.\,B.\,Richer   \\
  HST    &  WFC/ACS &  F606W   &  January, 19, 2010  &  1371s$+$1385s$+$1442s$+$1457s$+$1485s      &  11677      &  H.\,B.\,Richer   \\
  HST    &  WFC/ACS &  F606W   &  January, 20, 2010  &100s$+$1118s$+$1303s$+$1371s$+$1442s$+$1457s &  11677      &  H.\,B.\,Richer   \\
  HST    &  WFC/ACS &  F606W   &  January, 21, 2010  &  1371s$+$1385s$+$1428s$+$1457s$+$1498s      &  11677      &  H.\,B.\,Richer   \\
  HST    &  WFC/ACS &  F606W   &  January, 25, 2010  &1s$+$1113s$+$1371s$+$1407s$+$1442s$+$1457s   &  11677      &  H.\,B.\,Richer   \\
  HST    &  WFC/ACS &  F606W   &  January, 23, 2010  &  1371s$+$1442s$+$1457s                      &  11677      &  H.\,B.\,Richer   \\
  HST    &  WFC/ACS &  F606W   &  January, 26, 2010  & 10s$+$1208s$+$1303s$+$1371s$+$1442s$+$1457s &  11677      &  H.\,B.\,Richer   \\
  HST    &  WFC/ACS &  F606W   &  January, 27, 2010  & 1371s$+$1385s$+$1442s$+$1457s$+$1485s       &  11677      &  H.\,B.\,Richer   \\
  HST    &  WFC/ACS &  F606W   &  January, 28, 2010  &100s$+$1118s$+$1303s$+$1371s$+$1442s$+$1457s &  11677      &  H.\,B.\,Richer   \\
  HST    &  WFC/ACS &  F606W   &  January, 21, 2010  &1371s$+$1385s$+$1442s$+$1443s$+$1457s$+$1498s&  11677      &  H.\,B.\,Richer   \\
  HST    &  WFC/ACS &  F606W   &  May, 03, 2010      &100s$+2\times$1253s$+$1442s$+$1456s$+$1457s  &  11677      &  H.\,B.\,Richer   \\
  HST    &  WFC/ACS &  F814W   &  February, 13, 2010 & 1443$+2\times$1456s$+2\times$1457s          &  11677      &  H.\,B.\,Richer   \\
  HST    &  WFC/ACS &  F814W   &  March, 04, 2010    & 1s$+$1261s$+$1303s$+$1443s$+$1456s$+$1457s  &  11677      &  H.\,B.\,Richer   \\
  HST    &  WFC/ACS &  F814W   & March, 16, 2010     &   1383s$+ 2\times$1396s$+ 2\times$1457s     &  11677      &  H.\,B.\,Richer   \\
  HST    &  WFC/ACS &  F814W   & April, 04, 2010     & 10s$+$1266s$+$1298s$+$1443s$+$1456s$+$1457s &  11677      &  H.\,B.\,Richer   \\
  HST    &  WFC/ACS &  F814W   & June, 06, 2010      &100s$+$1103s$+$1253s$+$1293s$+$1306s$+$1457s &  11677      &  H.\,B.\,Richer   \\
  HST    &  WFC/ACS &  F814W   & June, 18, 2010      &   1443s$+ 2\times$1456s$+ 2\times$1457s     &  11677      &  H.\,B.\,Richer   \\
  HST    &  WFC/ACS &  F814W   & July, 29, 2010      &1s$+$1031s$+$1213s$+$1226s$+$1254s$+$1303s$+$1457s$+$1484s   &  11677      &  H.\,B.\,Richer   \\
  HST    &  WFC/ACS &  F814W   & August, 05, 2010    &   1443s$+ 2\times$1456s$+ 2\times$1457s     &  11677      &  H.\,B.\,Richer   \\
  HST    &  WFC/ACS &  F814W   & August, 14, 2010    & 10s$+$1266s$+$1298s$+$1443s$+$1456s$+$1457s &  11677      &  H.\,B.\,Richer   \\
  HST    &  WFC/ACS &  F814W   & September, 19, 2010 &   1443s$+ 2\times$1456s$+ 2\times$1457s     &  11677      &  H.\,B.\,Richer   \\
  HST    &  WFC/ACS &  F814W   & October, 01, 2010   &100s$+$1252s$+$1253s$+$1443s$+$1456s$+$1457s &  11677      &  H.\,B.\,Richer   \\
  HST    &  WFC/ACS &  F814W   & January, 16, 2010   &   1358s$+ 2\times$1371s$+ 2\times$1457s     &  11677      &  H.\,B.\,Richer   \\
  HST    &  WFC/ACS &  F814W   & January, 17, 2010   & 1s$+$1217s$+$1303s$+$1358s$+$1371s$+$1457s &  11677      &  H.\,B.\,Richer   \\
  HST    &  WFC/ACS &  F814W   & January, 18, 2010   &   1358s$+ 2\times$1371s$+ 2\times$1457s     &  11677      &  H.\,B.\,Richer   \\
  HST    &  WFC/ACS &  F814W   & January, 19, 2010   & 10s$+$1208s$+$1303s$+$1358s$+$1371s$+$1457s &  11677      &  H.\,B.\,Richer   \\
  HST    &  WFC/ACS &  F814W   & January, 20, 2010   &   1358s$+ 2\times$1371s$+ 2\times$1457s     &  11677      &  H.\,B.\,Richer   \\
  HST    &  WFC/ACS &  F814W   & January, 20-21, 2010&100s$+$1118s$+$1303s$+$1371s$+$1372s$+$1457s &  11677      &  H.\,B.\,Richer   \\
  HST    &  WFC/ACS &  F814W   & January, 25, 2010   &   1358s$+ 2\times$1371s$+ 2\times$1457s     &  11677      &  H.\,B.\,Richer   \\
  HST    &  WFC/ACS &  F814W   & January, 23, 2010   &1s$+$1217s$+$1303s$+$1358s$+$1371s$+$1399s$+$1457s$+$1484s   &  11677      &  H.\,B.\,Richer   \\
  HST    &  WFC/ACS &  F814W   & January, 26, 2010   &   1358s$+ 2\times$1371s$+ 2\times$1457s     &  11677      &  H.\,B.\,Richer   \\
  HST    &  WFC/ACS &  F814W   & January, 27, 2010   & 10s$+$1208s$+$1303s$+$1358s$+$1371s$+$1457s &  11677      &  H.\,B.\,Richer   \\
  HST    &  WFC/ACS &  F814W   & January, 28, 2010   &   1358s$+ 2\times$1371s$+ 2\times$1457s     &  11677      &  H.\,B.\,Richer   \\
  HST    &  WFC/ACS &  F814W   & January, 15, 2010   &100s$+$1118s$+$1303s$+$1357s$+$1371s$+$1457s &  11677      &  H.\,B.\,Richer   \\
  HST    &  WFC/ACS &  F814W   & May, 03, 2010       &   1443s$+ 2\times$1456s$+ 2\times$1457s     &  11677      &  H.\,B.\,Richer   \\
  HST    &   IR/WFC3 & F110W   & July, 16-17, 2009   &   18$\times$149s                            &  11453        &  B.\,Hilbert       \\
  HST    &   IR/WFC3 & F110W   & April, 06, 2010     &    499s                                     &  11962        &  A.\,Riess        \\
  HST    &   IR/WFC3 & F160W   & July, 16-17, 2009   &   18$\times$274s                            &  11453        &  B.\,Hilbert       \\
  HST    &   IR/WFC3 & F160W   & July, 23, 2009      &   24$\times$274s                            &  11445        &  L.\,Dressel       \\
  HST    &   IR/WFC3 & F160W   &  Mar, 13 - Nov, 20, 2010  &   24$\times$92s$+$24$\times$352s  &  11931        &  B.\,Hilbert        \\
  HST    &   IR/WFC3 & F160W   &  Feb, 14 - Aug, 22, 2012  &   24$\times$92s$+$24$\times$352s  &  12696        & B.\,Hilbert         \\
  HST    &   IR/WFC3 & F160W   &  April, 18-19,  2013  &   4$\times$92s$+$2$\times$352s  &  13079        &    B.\,Hilbert      \\
  HST    &   IR/WFC3 & F160W   &  December, 21-23,  2013  &   4$\times$92s$+$2$\times$352s  &  13563        & B.\,Hilbert         \\
%  HST    &   IR/WFC3 & F160W   &     &                  &  11677        &         \\
% HST    &   IR/WFC3 & F160W   &                     &                  &  11677        &         \\
  HST    &  WFC/ACS &  F814W   &   October, 09, 2002 &   2$\times$1390$+$2$\times$1460s              &  9444        &  I.\,R.\,King       \\
   HST   &    IR/WFC3 &  F105W  &    April, 06, 2010 & 499s &  11926 &  S.\,Deustua \\
   HST   &    IR/WFC3 &  F140W  &    July, 16-17, 2009 & 18$\times$224s &  11453 &  B.\,Hilbert  \\
%  HST    &  WFC/ACS &  F814W   &                     &                  &  11677        &  H.\,B.\,Richer       \\
         \hline
                  &          &          &     Field B                &                   &        & \\
%         \hline
  JWST   &  NIRCam  &  F115W   &     July, 13, 2022  & 40$\times$1041s   &  2560        & A.\,F.\,Marino   \\
  JWST   &  NIRCam  &  F322W2  &     July, 13, 2022  & 40$\times$1041s   &  2560        & A.\,F.\,Marino   \\
%\hline
   HST   &  UVIS/WFC3 &  F606W  &    February, 13, 2010 & 2$\times$50s$+$1347s$+$1402s   &  11677 &  H.\,B.\,Richer  \\
   HST   &    IR/WFC3 &  F110W  &    February, 13, 2010 & 102s$+$174s$+$2$\times$1399s   &  11677 &  H.\,B.\,Richer  \\
   HST   &    IR/WFC3 &  F160W  &    February, 13, 2010 & 4$\times$299s$+$4$\times$1199s &  11677 &  H.\,B.\,Richer  \\
%  F275W & Jul 15 2009  & 35$+$9$\times$350s       & 11452 & J.\,Kim\,Quijano\\
     \hline\hline
\end{longtable} 
\end{center}    
\twocolumn      

\subsection{{\it HST} data}                \label{sub:HST}                                         
To measure the stellar fluxes and the positions from the {\it HST} images, we used the FORTRAN program KS2, which is developed by Jay Anderson and is the evolution of the computer program kitchen\_sinc \citep{anderson2008a}. KS2 uses three distinct methods to measure stars. Method\,I, which is optimal for bright stars, derives the magnitudes and the positions of the stars by fitting the best available effective Point Spread Function (PSF) model \citep[e.g.][]{anderson2000a}. These quantities are derived in each image, separately, and then are averaged together to determine improved magnitudes and positions.
Method II, which provides the best astrometry and photometry of faint sources, combines information from all the exposures at the same time. In this case, KS2 measures the flux of each star by subtracting neighbor stars and performing the aperture photometry of the star in the 5$\times$5 pixel raster. Method III is similar to method II but it calculates the aperture photometry over a circle with a radius of 0.75 pixels. Hence, it is optimal for deriving photometry in crowded regions \cite[see][for details]{sabbi2016a, bellini2017a, milone2022b}. 
We calibrate the {\it HST} photometry to the VEGA mag system as in \citet{milone2022b} and by using the photometric zero points provided in the Space Telescope Science Institute web page for WFC/ACS, UVIS/WFC3, and NIR/WFC3
\footnote{https://www.stsci.edu/hst/instrumentation/acs/data-analysis/zeropoints; https://www.stsci.edu/hst/instrumentation/wfc3/data-analysis/photometric-calibration}.
The stellar positions are corrected for geometric distortion by using the solution by \citet{anderson2006a} and \citet{bellini2011a}, and \citet{anderson2022a} for WFC/ACS, UVIS/WFC3, and NIR/WFC3, respectively.
To select the stars with the best photometry and astrometry, we used the procedures and the computer programs by \citet[][see their Section 2.4]{milone2022b}. We exploited the diagnostics of the astrometric and photometric quality of each source provided by KS2 to identify the isolated stars that are well-fitted by the PSF model and have small values of the root mean scatters in position and magnitude.

\subsection{NIRCam data}
\begin{centering} 
\begin{figure} 
  \includegraphics[width=8.5cm,trim={0.0cm 5.5cm 0.0cm 4.5cm},clip]{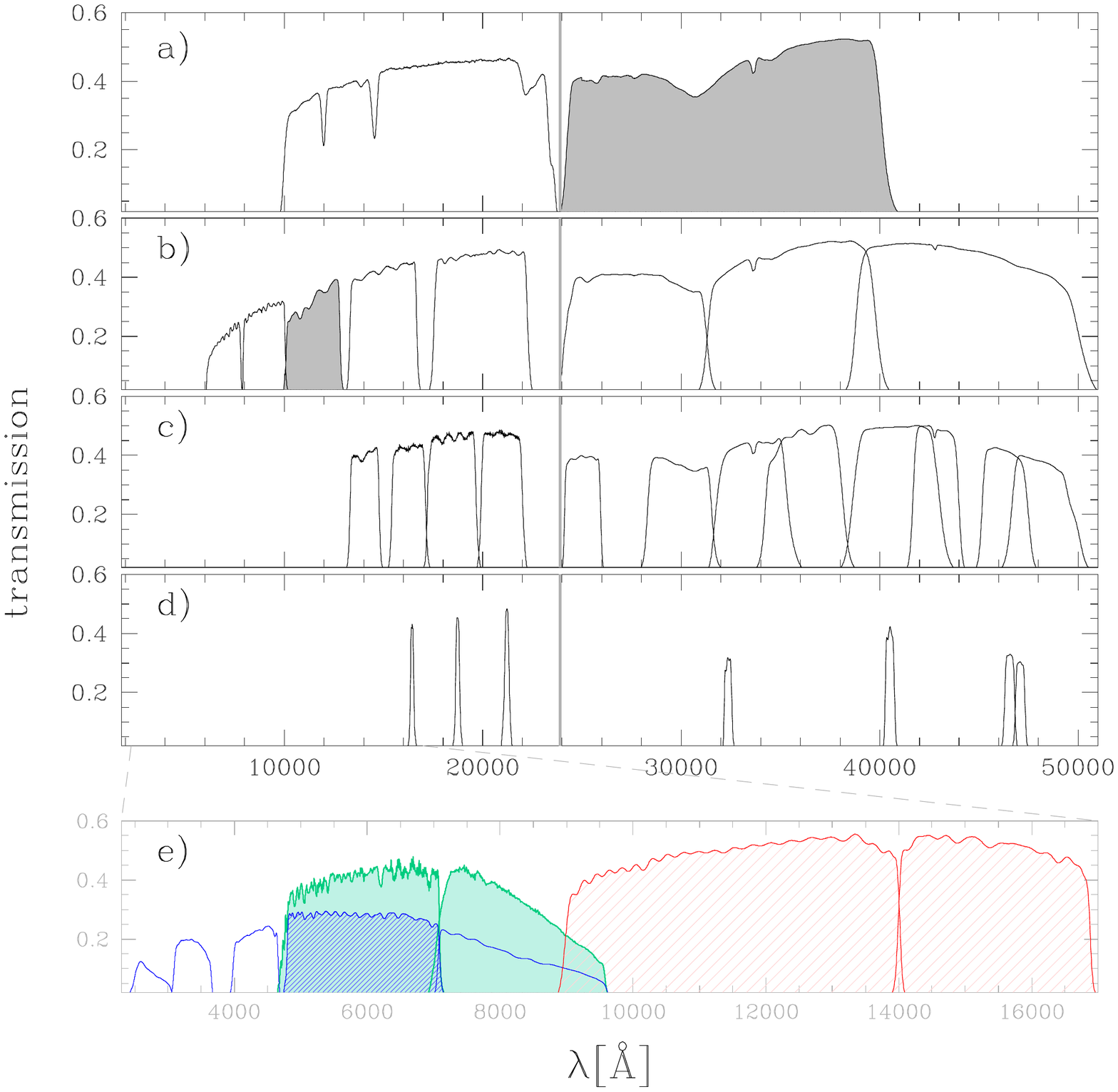}
 %/home/milone/WORKS/JWSTSPECTRA/JWST_25Gen/flussi1.macro fig2
 \caption{Panels a--d reproduce the transmission curves, throughput averaged over all detectors, of the extra-wide, wide, middle, and narrow passband  NIRCam filters, respectively. The vertical gray lines separate the short and long-wavelength channels. 
 Panel e provides the throughputs for the {\it HST} filters that are mostly used to study multiple stellar populations in GCs, including the F275W, F336W, F438W, F606W, and F814W UVIS/WFC3 filters (blue), the F606W and F814W filters of WFC/ACS (aqua), and the F110W and F160W NIR/WFC3 filters (red). The shaded areas indicate the filters for which photometry of 47\, Tucanae is available in this paper. }
 \label{fig:filters} 
\end{figure} 
\end{centering}

 NIRCam comprises short and long wavelength channels (SW and LW), with pixel scales of 0.031 and 0.063 arcsec, respectively, which cover the spectral regions $\lambda \sim$6,000-23,000 \AA\  and $\sim$24,000-50,000 \AA, respectively.
 NIRCam exploits a dichroic to allow the SW and LW channels to operate simultaneously. 
 Both channels are composed of two 2.2$\times$2.2 square-arcmin modules, A and B, which are separated by 44 arcsecs and operate in parallel. 
 NIRCam has ten detectors composed of 2040$\times$2040 pixels sensitive to light, including eight SW detectors (namely, A1, A2, A3, A4, and B1, B2, B3, B4) and 2 LW detectors. The detectors roughly cover the same field of view as modules A and B but the detectors A1--A4 and B1--B4 are separated from each other by 5 arcsec wide gaps. 
The short and long wavelength channels of JWST NIRCam are equipped with thirteen and sixteen bandpass filters, respectively. The total system throughputs for the NIRCam filters, including the contribution from the JWST  optical telescope element, are plotted in Figure\,\ref{fig:filters}. For completeness, we show the transmission curves of some UVIS/WFC3, WFC/ACS, and NIR/WFC3 filters that are commonly used to investigate multiple populations in GCs.

%%%%%%%%%%

%%%%%%%%%%

The dataset includes images taken with the NIRCam camera onboard JWST as part of the GO\,2560 program (P.I.\,A.\,F.\,Marino).
 We simultaneously collected images of field B, through the F115W filter of the SW channel  and the F322W2  filter of the LW channel. The images are taken with DEEP8 readout pattern and have been properly dithered to cover the gaps between the A and B detectors of the SW channel.    
  During the NIRCam observations, one of the two 3-mirror 'wings' on the primary mirror of JWST underwent a small but significant jump in position. A consequence of  the so-called wind-tilt event is that all sources in all F115W exposures exhibit a faint ghost shifted by $\sim$13 pixels on the top right. 

To reduce NIRCam data we used a modified version of the computer program described in \citet{anderson2006b}.
In a nutshell, we derived a grid of 3$\times$3 ePSFs for each exposure and detector by using isolated, bright, and not saturated stars. To measure each star, we derived the corresponding ePSF model by bi-linear interpolation of the closest four PSFs of the grid. Stellar fluxes and magnitudes are derived in each image separately, and the results are registered in a common master frame and averaged together. We computed the photometric zero points differences among the A1--A4  and the B1--B4 detectors of the SW channel by using the stars observed in different detectors and referred our instrumental F115W photometry to detectors A3 and B3, respectively. 
 To estimate the magnitude difference between the detectors A3 and B3, w and refer the F115W magnitudes to the A3 detector, we take advantage of the UVIS/WFC3 images that overlap NIRCam data. We derived the $m_{\rm F606W}$ vs.\,$m_{\rm F606W}-m_{\rm F115W}$ CMDs by using stars in the detectors A and B and derived the corresponding fiducial lines for MS stars as in \citep{milone2012a}. We assumed the color difference between the fiducial lines derived from stars in the detector B and A as the best estimate of the zero point difference in F115W.
 Similarly, we referred the F322W2 photometry to module A. 
 
 Photometry has been calibrated to the VEGA mag system by using the most-updated zero points provided by the STScI webpage\footnote{https:\,//jwst-docs.stsci.edu\,/jwst-near-infrared-camera\,/nircam-performance\,/nircam-absolute-flux-calibration-and-zeropoints} and the procedure by \citet{milone2022b}. Moreover, we corrected the stellar positions in the F115W images by using the distortion solution derived in the Appendex\,A for the NIRCam SW detectors. Stars with high-precision photometry and astrometry as selected as in Section\,\ref{sub:HST} \citep[see][for details]{milone2022b}

\subsection{Proper motions}
We derived stellar proper motions to separate the bulk of 47\,Tucanae members from the foreground and background stars in both fields A and B.
 In a nutshell, we averaged together the positions from all exposures of each epoch and compared the stellar positions in the different epochs to infer the displacements relative to the bulk of cluster stars. Specifically, we used all ACS/WFC images of field A listed in Table\,\ref{tab:data}, while for field B we compared the stellar position derived from the NIRCam images collected through the F115W filter and the UVIS/WFC3 images.
 To do that, we used the computer programs and the procedure described  by \citet{anderson2003a}, \citet{piotto2012a}, and \citet[][see their Section\,5]{milone2022b}. 
 The relative stellar displacements are converted into absolute proper motions as in \citet{milone2022b}, by using stars for which proper motions are available from both {\it HST}-{\it JWST} and from the Gaia data release 3 \citep[DR3,][]{gaia2021a}.

\begin{centering} 
\begin{figure*} 
  \includegraphics[height=7.75cm,trim={0.0cm 5.5cm 0.0cm 4.5cm},clip]{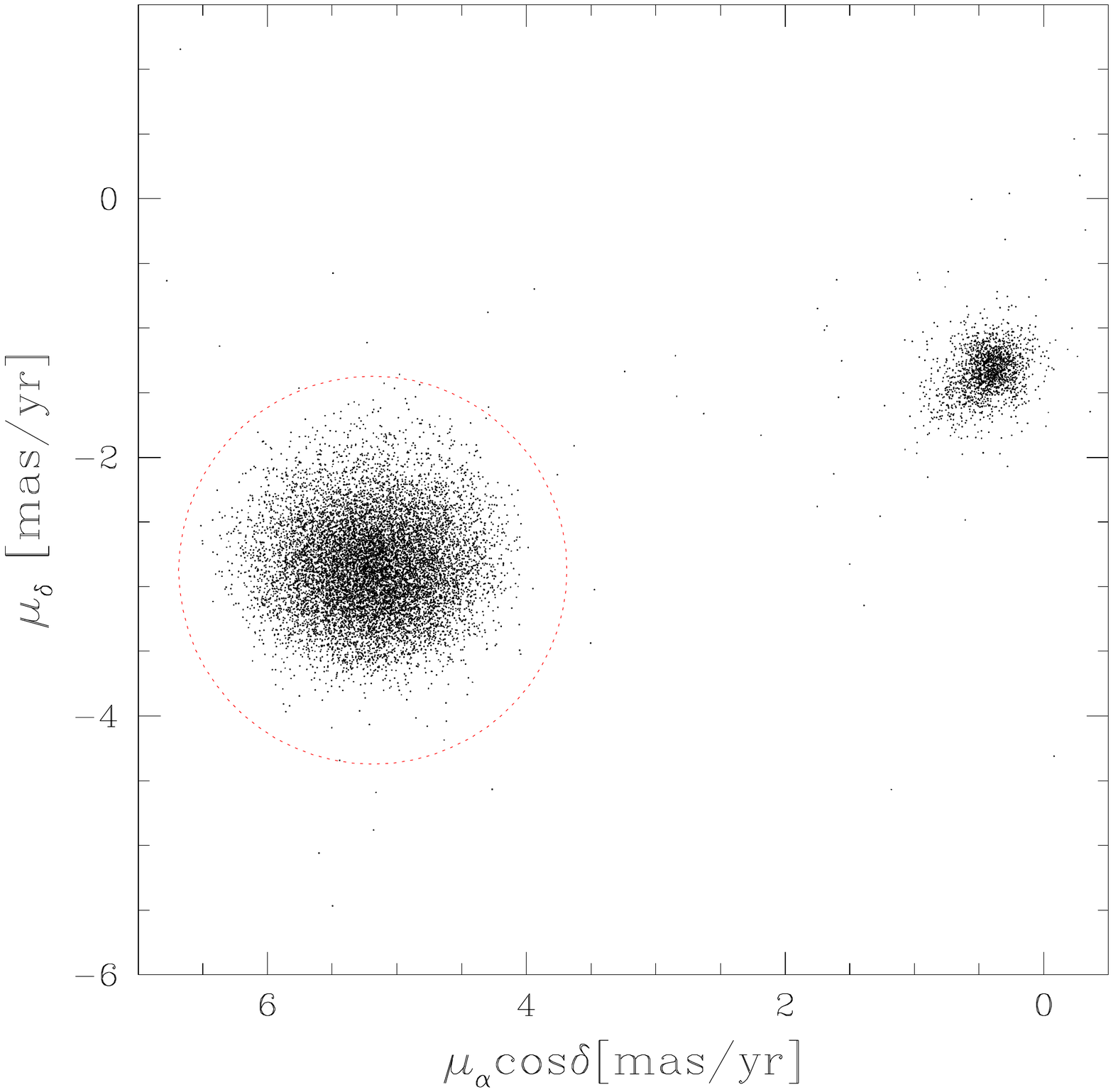}
%  /home/milone/GCs/NGC0104/WEST/test.macro pm
    \includegraphics[height=7.75cm,trim={3.0cm 5.5cm 0.0cm 4.5cm},clip]{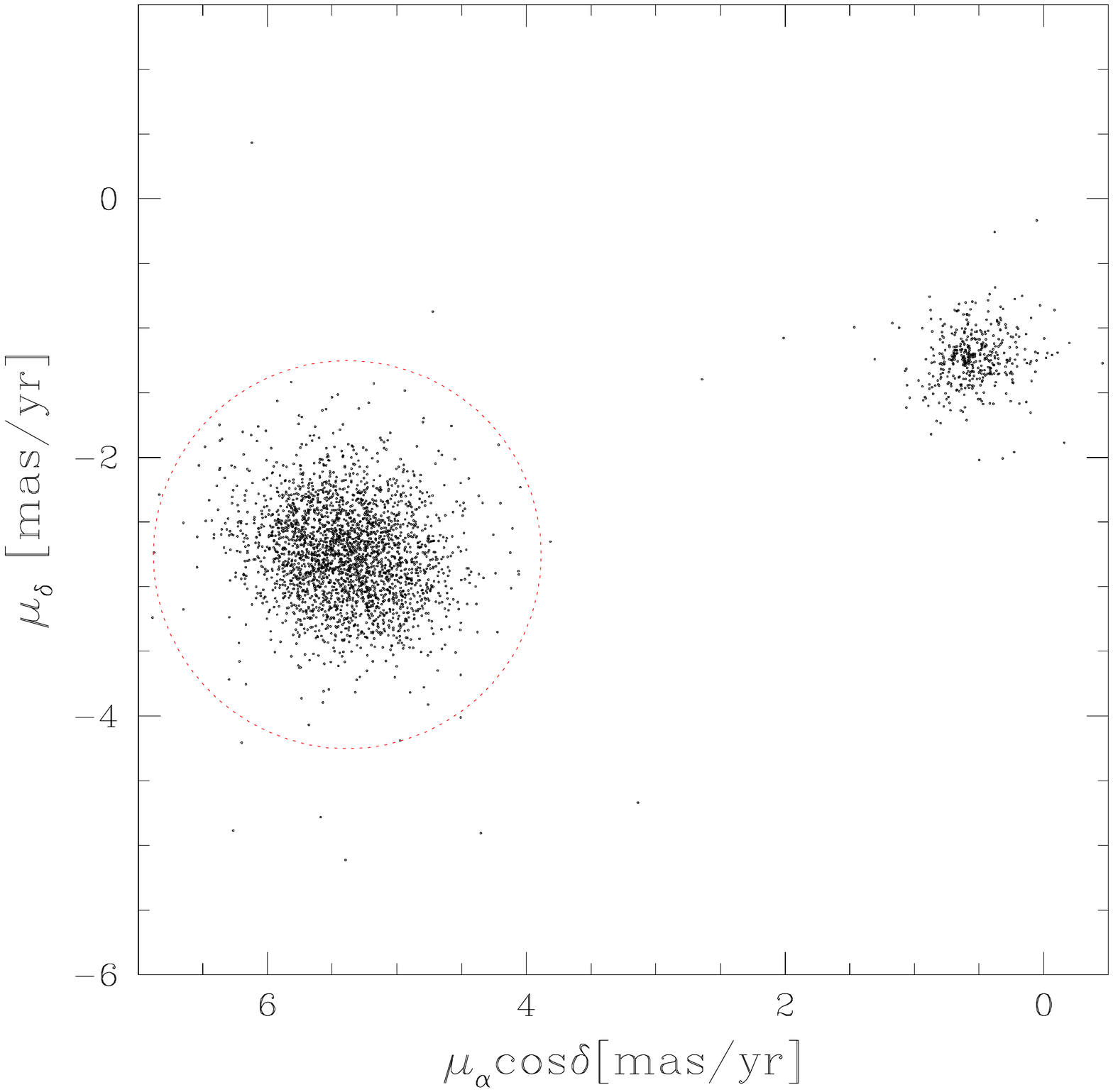}
% /home/milone/GCs/NGC0104/JWST/TOTORO/F115W/NRCB3/SAT/CLEAN/ALLdist/pm go    
 \caption{ Proper motions of stars in the fields A (left) and B (right). The red dotted circles separate the probable 47\,Tucanae from the bulk of field stars.
 }
 \label{fig:PMs} 
\end{figure*} 
\end{centering}

 The resulting proper-motion diagrams are plotted in Figure\,\ref{fig:PMs}, where the majority of cluster members and Small Magellanic Cloud (SMC) stars are clustered into two main stellar clumps. We used the red-dotted circles to separate the probable 47\,Tucanae members from the field stars.

%%%%%%%%%%

\section{Multiple populations among very-low mass stars of 47 Tucanae}\label{sec:47tuc}
This section presents the photometry derived in Section\,\ref{sec:data} and takes advantage of various photometric diagrams to investigate multiple populations among M-dwarfs of 47\,Tucanae. Specifically, Section\,\ref{sub:47TucHST} analyzes the photometric diagram derived from {\it HST} photometry of field-A stars, whereas Section\,\ref{sub:47TucJWST} is dedicated to the field B, observed with both {\it HST} and {\it JWST}.

\subsection{Results from {\it HST} observations of field-A stars}\label{sub:47TucHST}
\begin{centering} 
\begin{figure*} 
%/home/milone/GCs/NGC0104/WEST/test.macro figP1 figP2
  \includegraphics[height=8.9cm,trim={0.5cm 5.4cm 0.0cm 4.5cm},clip]{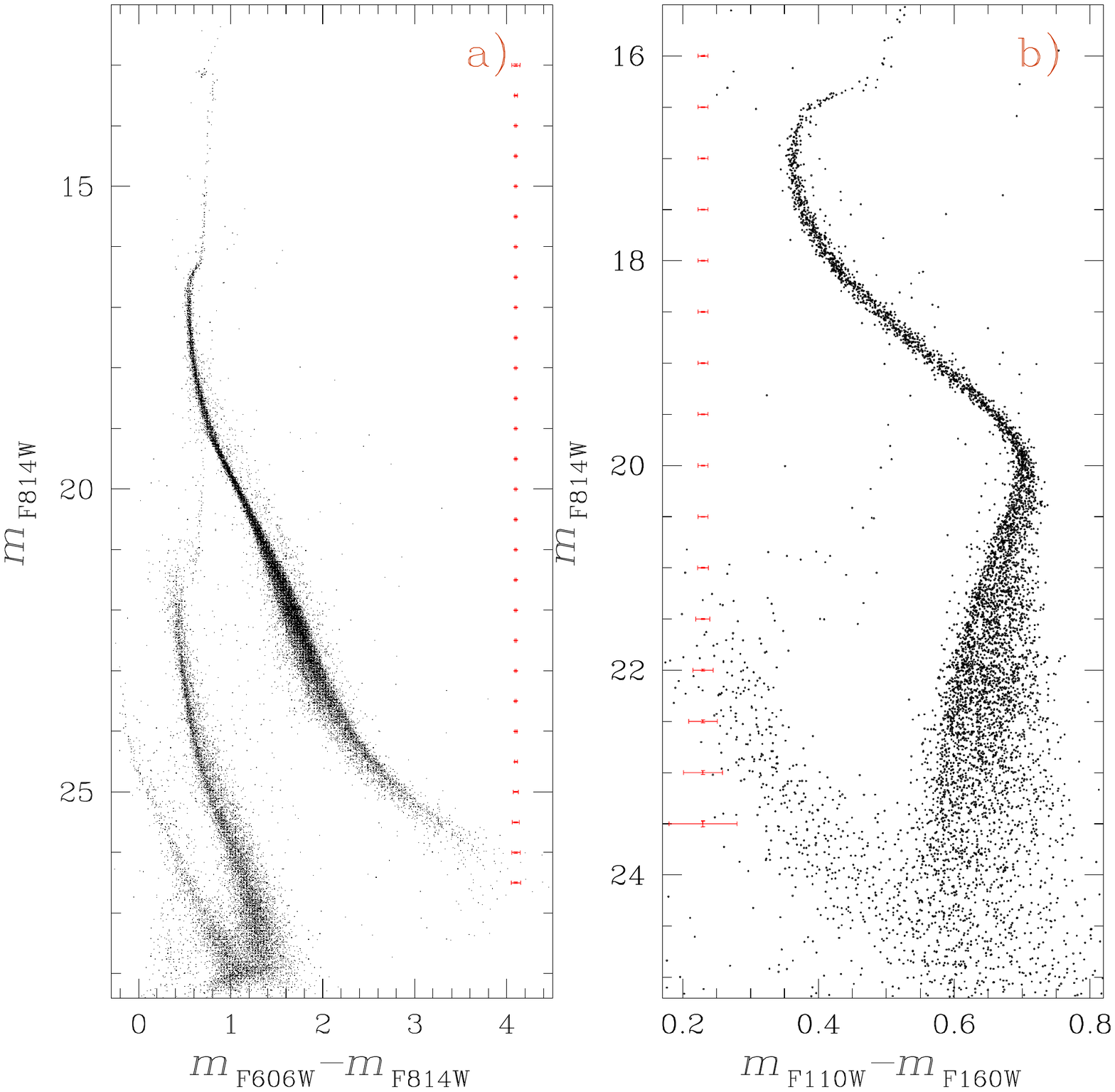}
  \includegraphics[height=8.9cm,trim={0.5cm 5.4cm 2.0cm 4.5cm},clip]{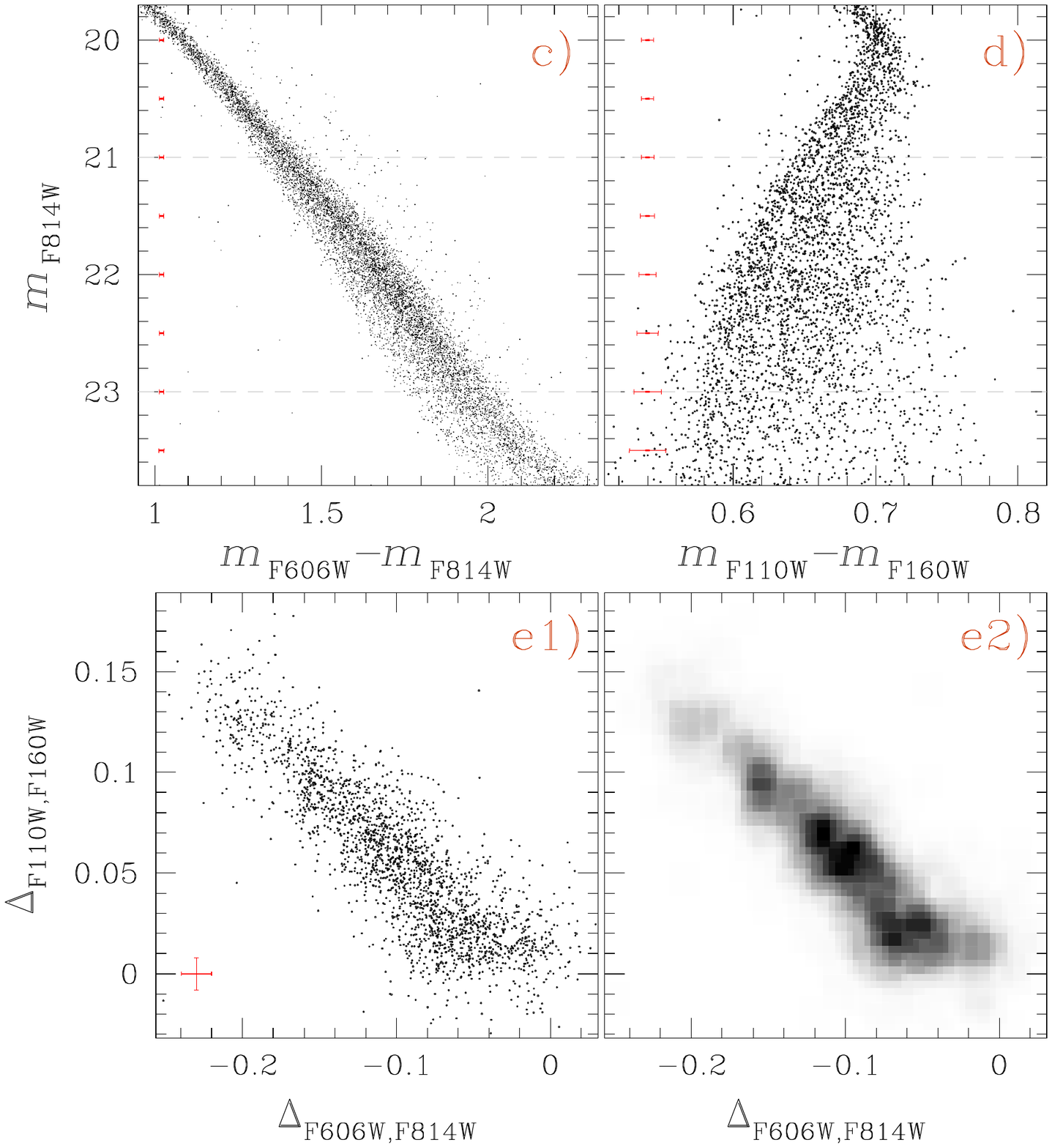}
 \caption{
 %{ \color{red}YLABEL SBAGLIATA IN PANEL b e d?}
 Optical (panel a) and NIR CMDs (panel b) of all stars in the field A derived from {\it HST} photometry. Panels c and d show the $m_{\rm F814W}$ vs.\,$m_{\rm F606W}-m_{\rm F814W}$ and  $m_{\rm F814W}$ vs.\,$m_{\rm F110W}-m_{\rm F160W}$ CMDs of proper-motion selected cluster members zoomed in the MS region below the knee. Panels e1 and e2 show the $\Delta_{\rm F110W,F160W}$ vs.\,$\Delta_{\rm F606W,F814W}$ ChM of the MS stars between the gray dashed lines of panels c and d.}
 \label{fig:CMDsHST} 
\end{figure*} 
\end{centering}

Panels a and b of Figure\,\ref{fig:CMDsHST} show the optical ($m_{\rm F814W}$ vs.\,$m_{\rm F606W}-m_{\rm F814W}$) and the NIR ($m_{\rm F814W}$ vs.\,$m_{\rm F110W}-m_{\rm F160W}$) color-magnitude diagrams (CMDs) for the stars in the field A. As discussed in previous work based on this dataset, the optical CMD clearly exhibits three main stellar sequences. The reddest sequence comprises the red horizontal branch (HB), red-giant branch (RGB), sub-giant branch (SGB), and main sequence (MS) of 47\,Tucanae, whereas the bluest sequence is the white-dwarf cooling sequence. The sequence in the middle is composed of Small Magellanic Cloud (SMC) stars \citep{richer2013a}. Similarly, the SMC MS is well separated by the MS of 47\,Tucanae in the NIR CMD but the two MSs merge together below $m_{\rm F160W} \sim 24$ mag.

In the panels c and d of Figure\,\ref{fig:CMDsHST} we show a zoom of the $m_{\rm F814W}$ vs.\,$m_{\rm F606W}-m_{\rm F814W}$ and  $m_{\rm F814W}$ vs.\,$m_{\rm F110W}-m_{\rm F160W}$ CMDs in the region below the MS knee. To minimize the contamination from field stars, we plot cluster members alone that we separated from field stars by using proper motions.
 A distinctive feature of both CMDs is that the MS broadening is much wider than the spread due to observational errors alone, thus revealing the multiple populations. 
Hints of parallel sequences are visible in the optical CMD. The most striking feature of the $m_{\rm F814W}$ vs.\,$m_{\rm F110W}-m_{\rm F160W}$ CMD is that the MS stars brighter than the knee at $m_{\rm F814W} \sim 20.0$ mag span a narrow color range, whereas the MS breadth suddenly increases from the MS knee towards fainter magnitudes.
A gradient in the $m_{\rm F110W}-m_{\rm F160W}$ color distribution is also evident, with the majority of stars having blue colors. 

The CMDs of panels c and d are used to construct the $\Delta_{\rm F110W,F160W}$ vs.\,$\Delta_{\rm F606W,F814W}$ ChM plotted in panel e1. To derive the ChM we only used the proper-motion selected M-dwarfs with $21.0<m_{\rm F814W}<23.0$ mag, and normalized the $\Delta_{\rm F110W,F160W}$ and $\Delta_{\rm F606W,F814W}$ quantities to the width of the MS at $m_{\rm F814W}=22.0$ mag, which corresponds to the luminosity of M-dwarfs that are two F814W mag fainter than the MS knee. As highlighted by the Hess diagram of panel e2, the ChM reveals an extended 1P sequence composed of stars with $\Delta_{\rm F110W,F160W} \lesssim 0.03$ mag and three main groups of 2P stars. 
 
\subsection{Results from NIRCam and UVIS observations of field-B stars}\label{sub:47TucJWST}
\begin{centering} 
\begin{figure*} 
  \includegraphics[height=8.3cm,trim={0.5cm 5.4cm 0.0cm 11.0cm},clip]{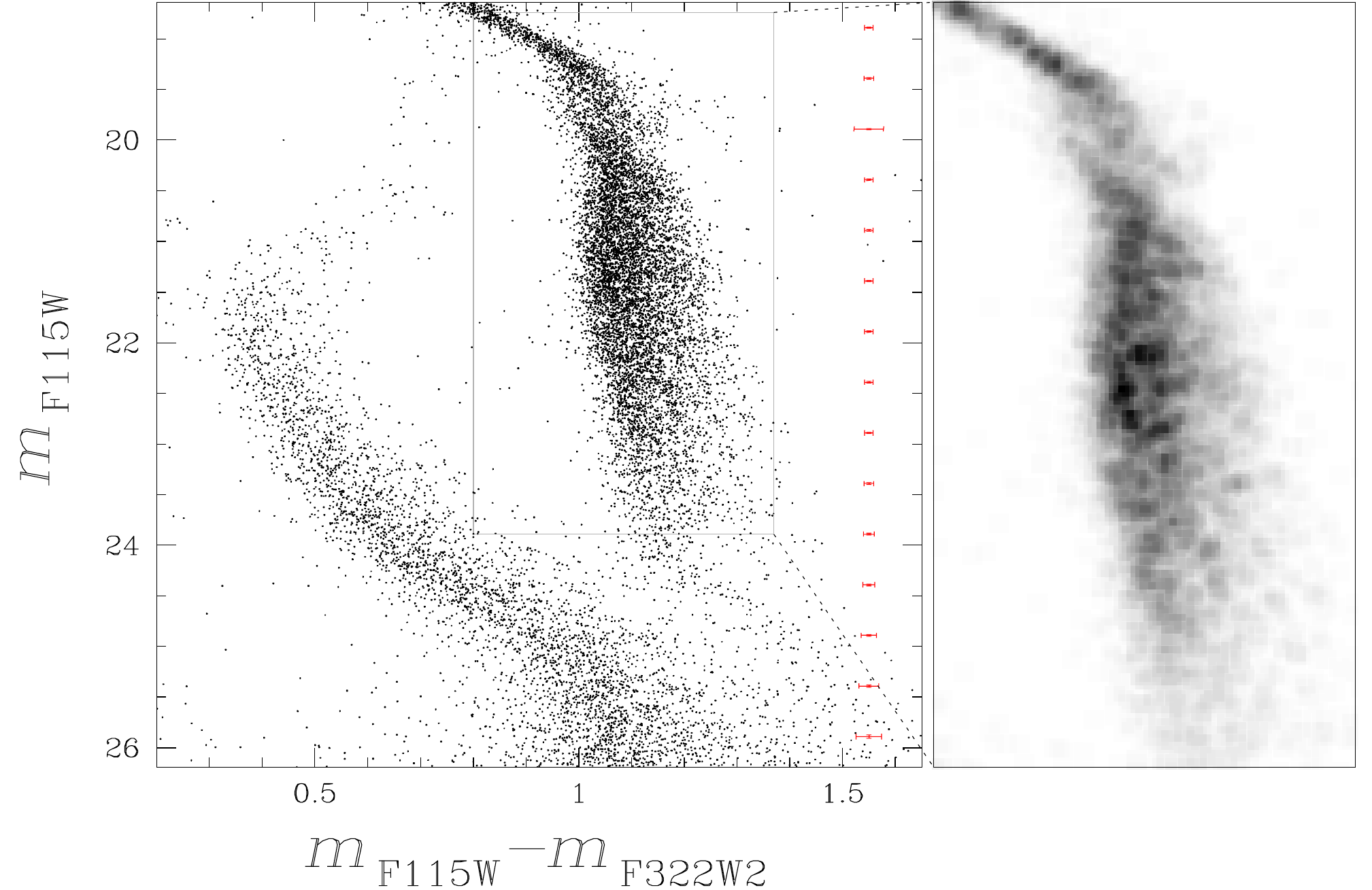}
 \caption{$m_{\rm F115W}$ vs.\,$m_{\rm F115W}-m_{\rm F322W2}$ CMD of all stars in the NIRCam FoV (left). The Hess diagram of the CMD region around the MS of 47\,Tucanae is shown in the right panel.}
 \label{fig:CMDjwst} 
\end{figure*} 
\end{centering}

The $m_{\rm F115W}$ vs.\,$m_{\rm F115W}-m_{\rm F322W2}$ CMD of all stars in the NIRCam FoV (field B) is plotted in Figure\,\ref{fig:CMDjwst}. Most 47\,Tucanae MS stars are enclosed by the gray rectangle, whereas the SMC stars distribute along the SGB and the MS that are visible on the blue side of the CMD. 

 The cluster CMD reveals that above the MS knee, the color broadening of MS stars with similar magnitudes is comparable with the color spread due to observational uncertainties, alone.
 Hence, the upper MS of 47\,Tucanae is narrow and well-defined and resembles what is expected from a single isochrone. 

The F115W$-$F322W2 color broadening dramatically increases below the MS knee, in the domain of M-dwarfs. As highlighted by the Hess diagram on the right, for a fixed F115W mag, the majority of MS M-dwarfs show blue colors, with a tail of stars distributed towards the red. 

Below the gray rectangle, we note a narrow tail of stars that seems mostly connected with the blue part of the M-dwarf MS. We associate this feature of the CMD with stars with masses smaller than $\sim$0.1 $\mathcal{M_{\odot}}$ \citep{dantona1994a, milone2012b, baraffe2015a}. In this case, it would be the first observational detection of such stars in a GC CMD.

\begin{centering} 
\begin{figure*} 
  \includegraphics[height=8.8cm,trim={0.5cm 5.0cm 9.75cm 4.25cm},clip]{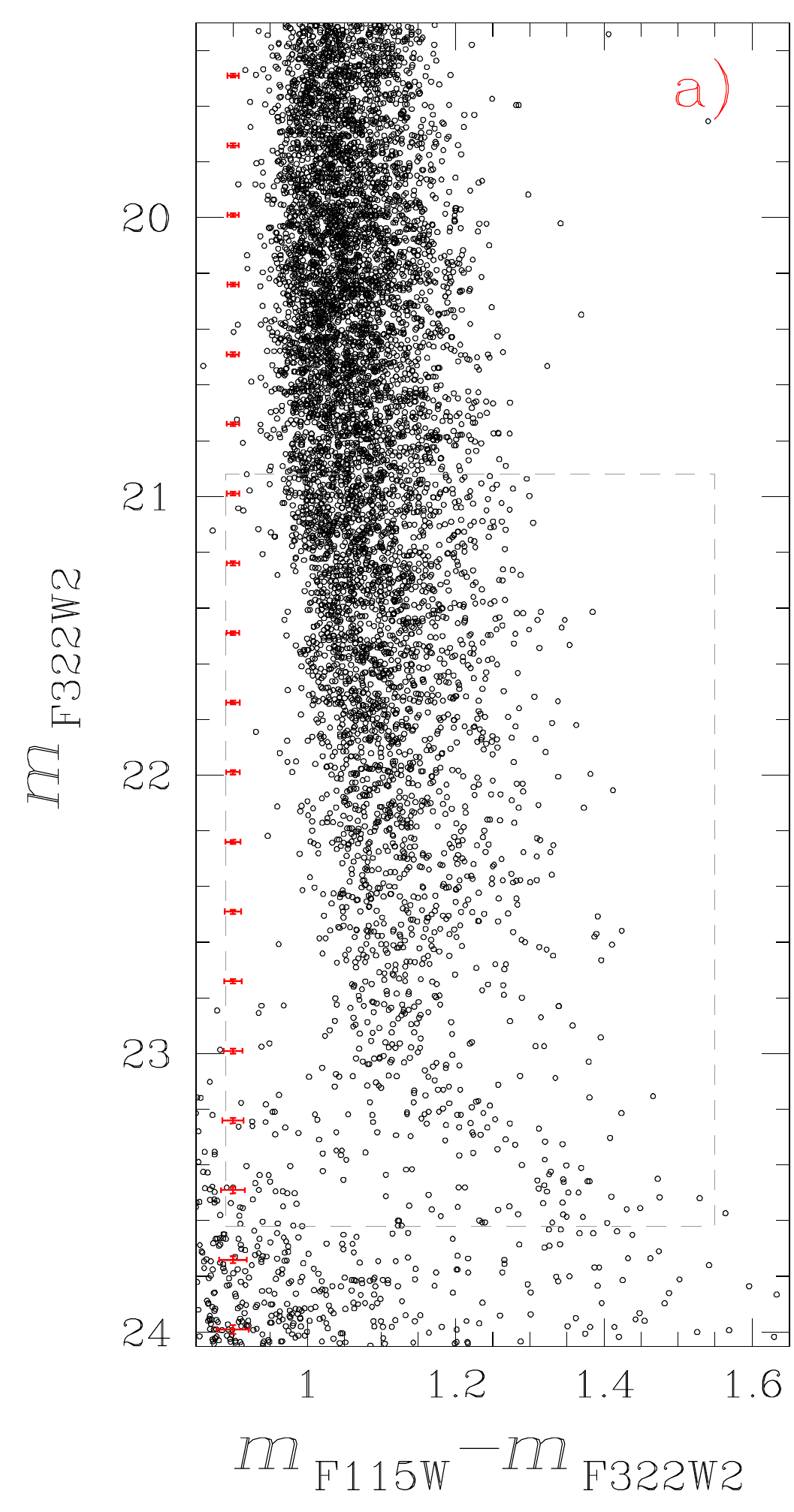}
  \includegraphics[height=8.8cm,trim={0.5cm 7.1cm 0.0cm 4.25cm},clip]{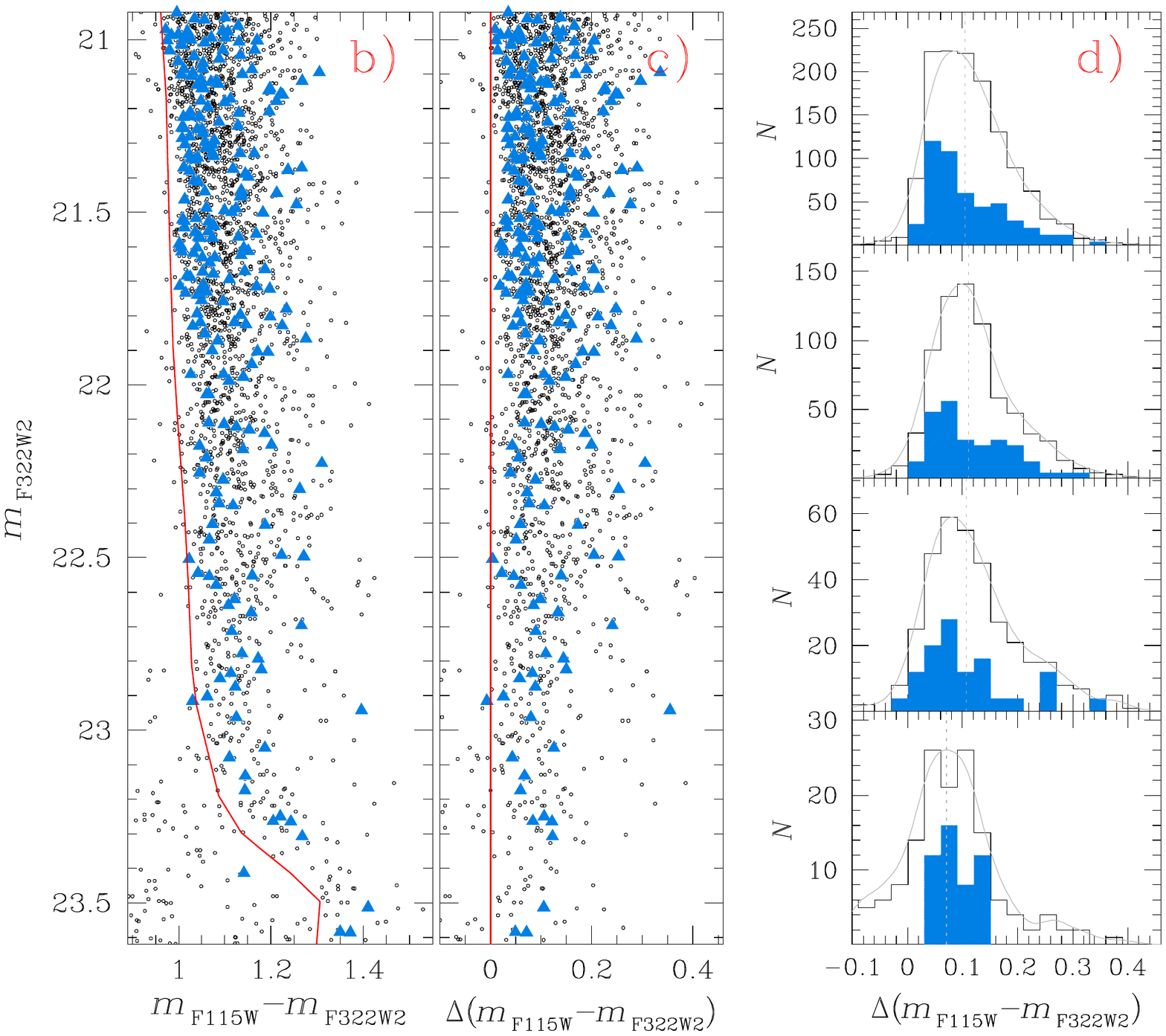}
 %/home/milone/GCs/NGC0104/JWST/TOTORO/fig.macro  pp pp2
 \caption{ $m_{\rm F322W2}$ vs.\,$m_{\rm F115W}-m_{\rm F322W2}$ CMD of all faint MS stars with available NIRCam photometry (panel a). Panel b is a zoom of the CMD region within the dashed box shown in panel a. The red line marks the blue boundary of the MS and is used to derive the verticalized $m_{\rm F322W2}$ vs.\,$\Delta$($m_{\rm F115W}-m_{\rm F322W2}$) diagram plotted in panel c. The azure triangles mark the proper-motion selected cluster members. Panels d show the $\Delta$($m_{\rm F115W}-m_{\rm F322W2}$) black histogram distributions for stars in four magnitude intervals and the corresponding kernel-density distributions. The median dotted lines mark the median values of $\Delta$($m_{\rm F115W}-m_{\rm F322W2}$). The distributions for cluster members alone are represented with azure-shaded histograms and, for clearness, the star counts are multiplied by a factor of four. See the text for details.}
 \label{fig:tail} 
\end{figure*} 
\end{centering}

 Figure\,\ref{fig:tail} further investigates the color distribution of stars at the bottom of the MS. The $m_{\rm F322W2}$ vs.\,$m_{\rm F115W}-m_{\rm F322W2}$ CMD of all stars with available  JWST photometry is plotted in the panel a, whereas panel b is a zoom in the F332W magnitude range between 20.9 and 23.6. We derived by hand the red fiducial line, which delimits the blue MS boundary, and used it to derive the verticalized $m_{\rm F322W2}$ vs.\,$\Delta$($m_{\rm F115W}-m_{\rm F322W2}$) diagram shown in Figure\,\ref{fig:tail}c. To derive the $\Delta$($m_{\rm F115W}-m_{\rm F322W2}$) quantity, we subtracted from the color of each star, the color of the red fiducial corresponding to the same F322W2 magnitude.
The azure triangles plotted in panels a and b of Figure\,\ref{fig:tail} mark the stars that, according to their proper motions, are probable cluster members. 

In the panels d, we analyzed the color distributions of M-dwarfs. We divided the magnitude interval shown in panels b and c into four equal-size bins, and for each bin, we derived the histogram distribution of $\Delta$($m_{\rm F115W}-m_{\rm F322W2}$) for all the stars in the corresponding luminosity interval. The gray lines superimposed on the histogram are the kernel-density distributions of the $\Delta$($m_{\rm F115W}-m_{\rm F322W2}$) quantities and are derived by assuming  Gaussian kernels with dispersion, $\sigma$=0.025 mag. The histogram and kernel-density distributions corresponding to the top three bins exhibit a peak around $\Delta$($m_{\rm F115W}-m_{\rm F322W2}$)$\sim$0.1 mag and a tail towards red colors. 
 The fractions of stars with $\Delta$($m_{\rm F115W}-m_{\rm F322W2}$)=0.15 mag in these three bins are very similar and correspond to 27$\pm$2 \%, 27$\pm$3 \%, and 29$\pm$4 \%.
 Intriguingly, the red tail seems poorly populated for stars with $m_{\rm F322W2} \gtrsim 23.0$, where the stars with $\Delta$($m_{\rm F115W}-m_{\rm F322W2}$)=0.15 mag include 14$\pm$3\% of the total number of stars. The conclusion is confirmed by the color distribution of the proper-motion selected cluster members (azure histograms in Figure\,\ref{fig:tail}d).

\begin{centering} 
\begin{figure*} 
  \includegraphics[height=8.8cm,trim={0.5cm 5.0cm 0.5cm 4.75cm},clip]{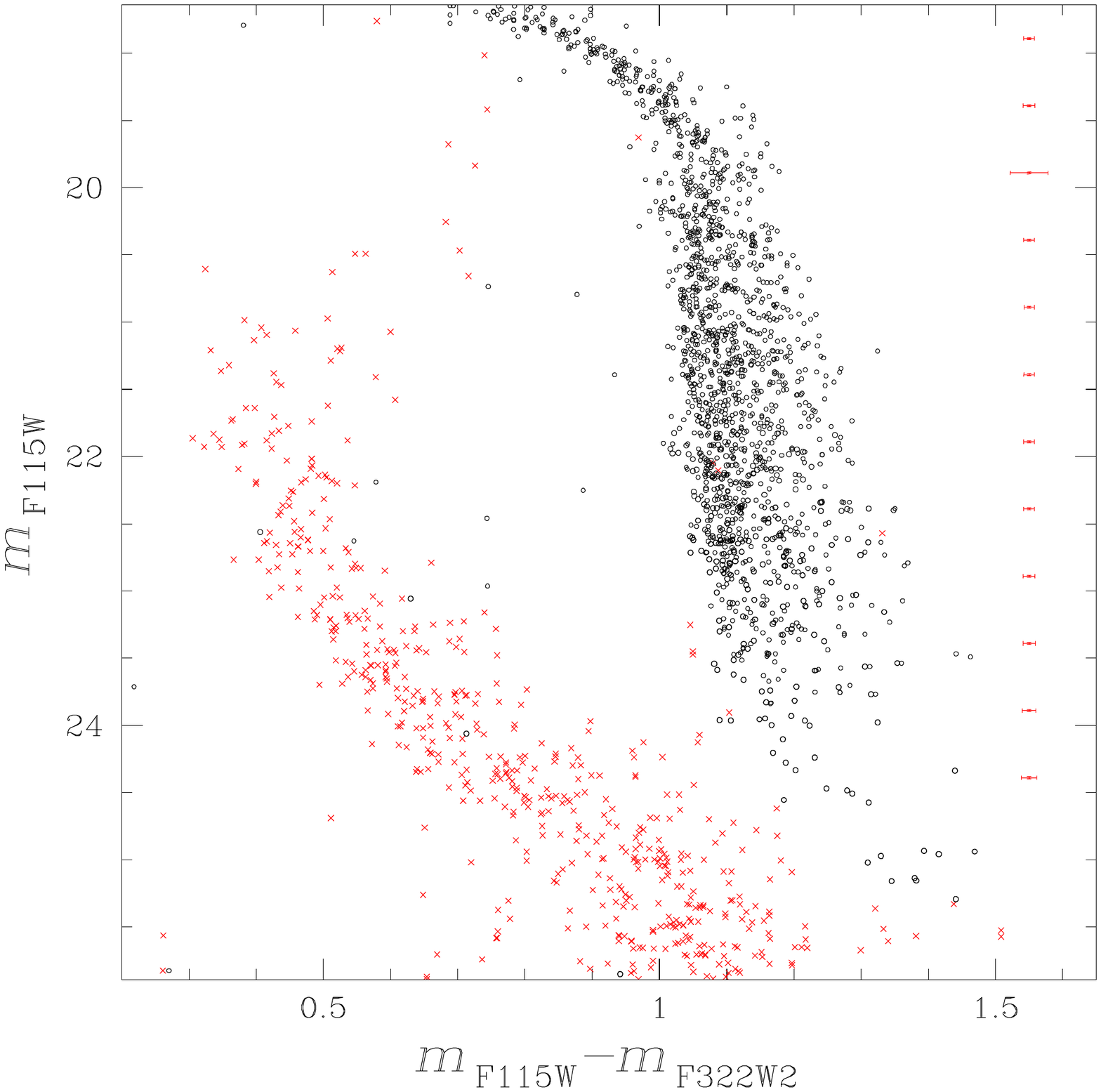}
  %/home/milone/GCs/NGC0104/JWST/TOTORO/fig.macro fig1n (o fig1)  proc
  \includegraphics[height=8.8cm,trim={0.5cm 5cm 0.5cm 4.75cm},clip]{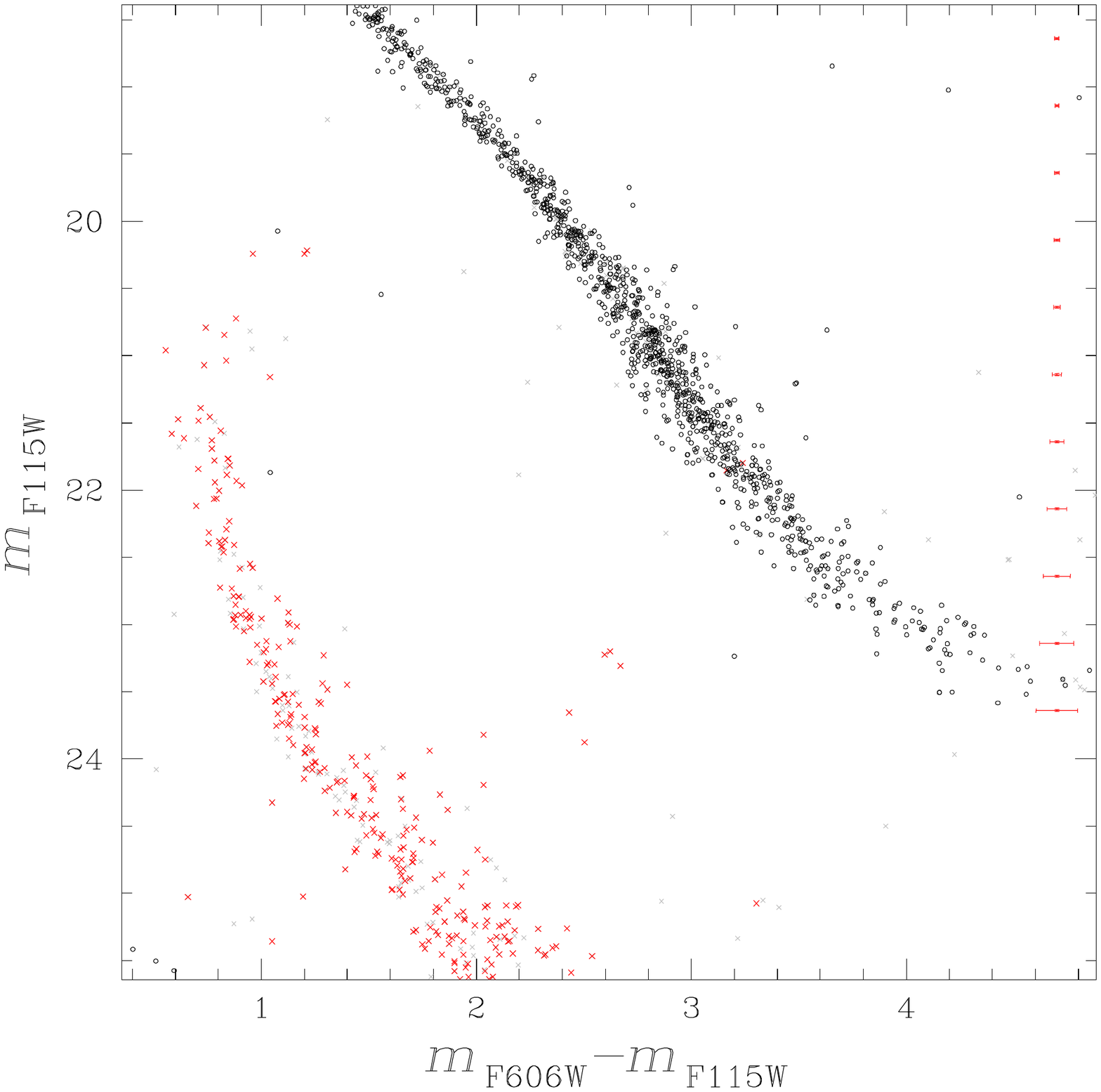}
 %/home/milone/GCs/NGC0104/JWST/TOTORO/fig.macro fig2
 \caption{$m_{\rm F115W}$ vs.\,$m_{\rm F115W}-m_{\rm F322W2}$ (left) and $m_{\rm F115W}$ vs.\,$m_{\rm F606W}-m_{\rm F115W}$ CMD (right) for the stars in the UVIS/WFC3 FoV. Stars that, based on proper motions, are probable cluster members, and field stars are colored black and red, respectively. %The gray points represent stars with no proper motion determination. 
 }
 \label{fig:CMDs} 
\end{figure*} 
\end{centering}

\begin{centering} 
\begin{figure*} 
    \includegraphics[height=8.0cm,trim={0.6cm 4.8cm 6.0cm 10.5cm},clip]{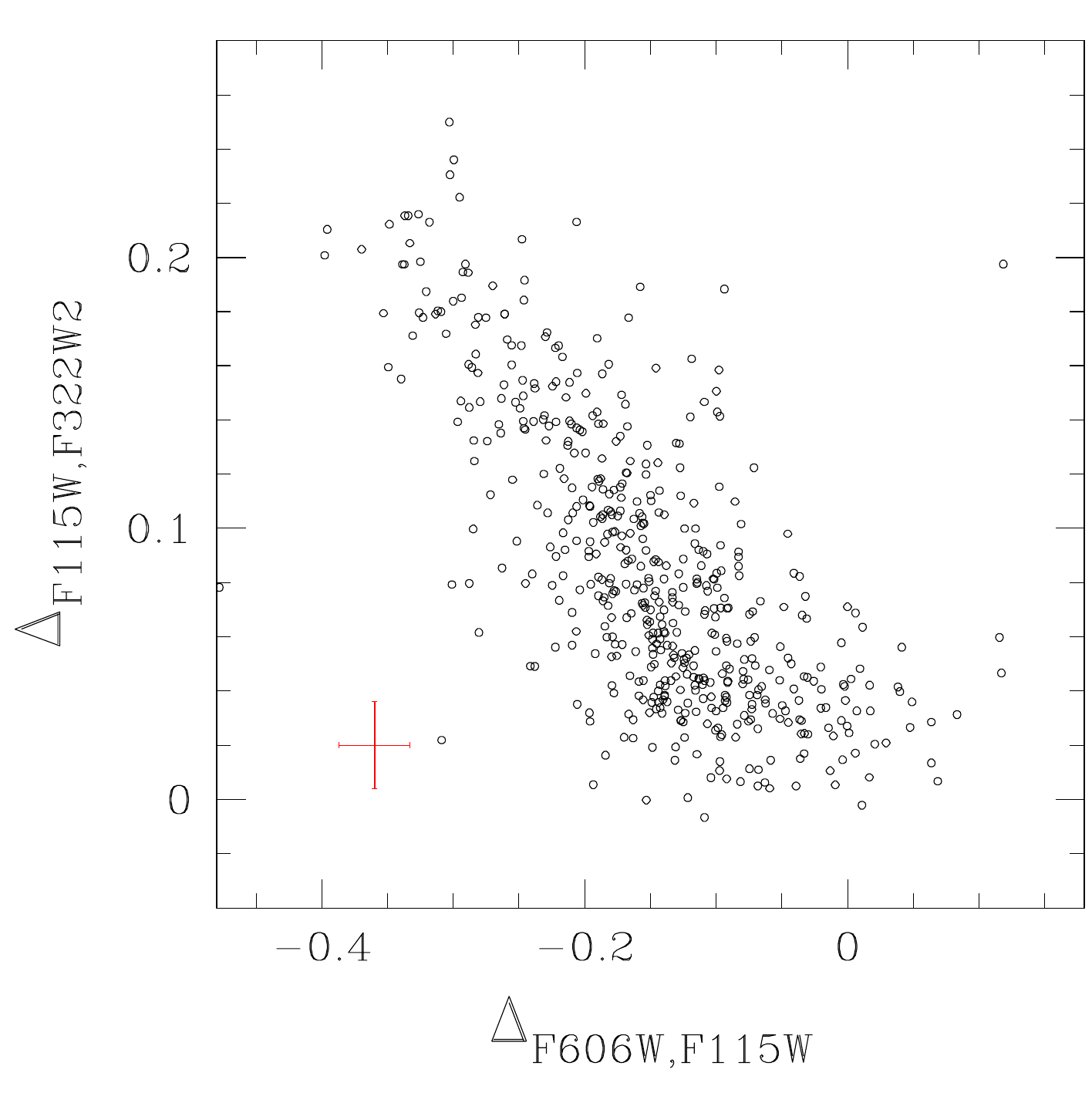}
        \includegraphics[height=8.0cm,trim={3.2cm 4.8cm 6.0cm 10.5cm},clip]{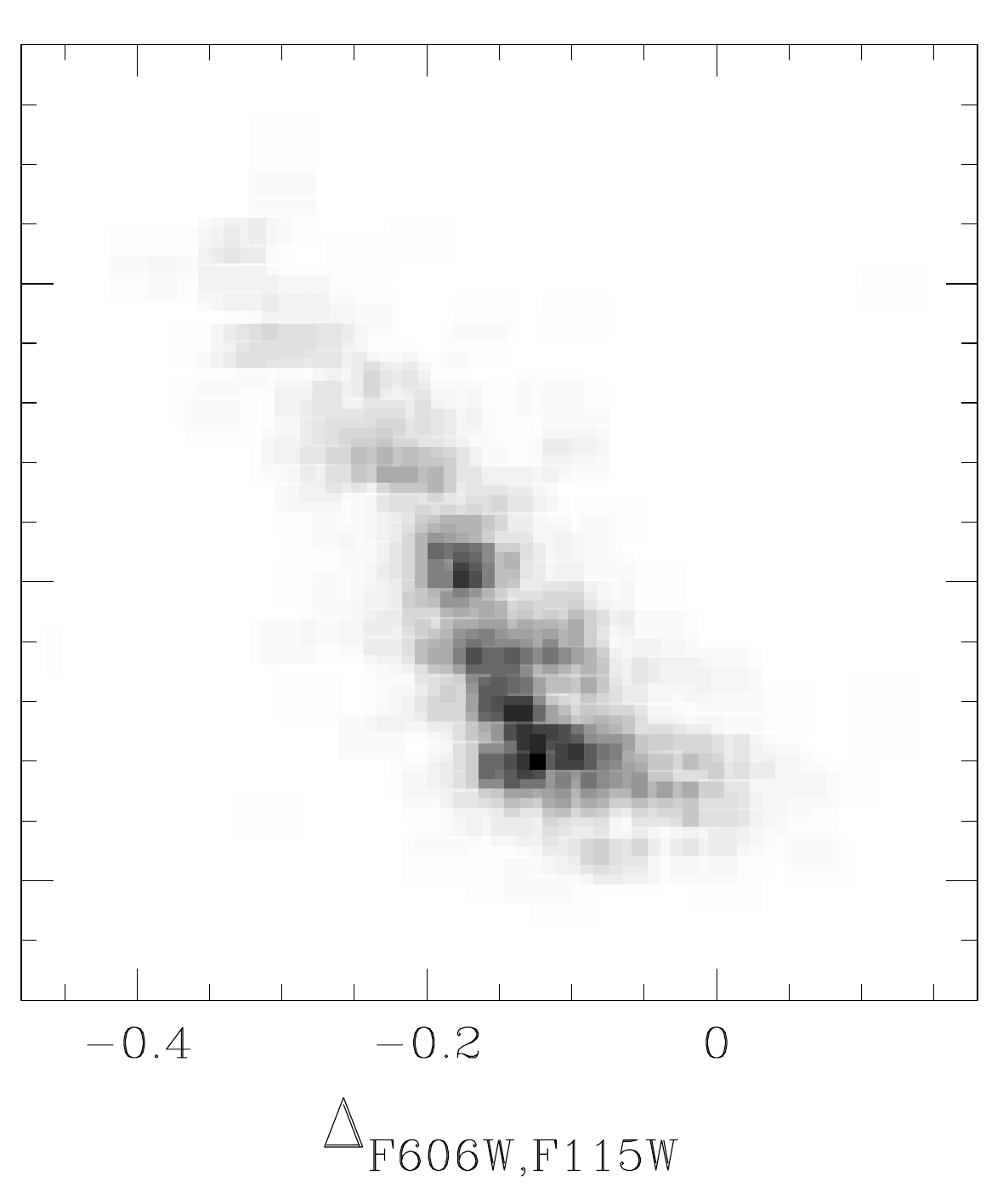}
 %/home/milone/GCs/NGC0104/JWST/TOTORO
 \caption{$\Delta_{\rm F115W,F322W2}$ vs.\,$\Delta_{\rm F606W,F115W}$ ChM (left) of 47\,Tucanae M-dwarfs in the field B. The corresponding Hess diagram is plotted in the right panel.   }
 \label{fig:chm} 
\end{figure*} 
\end{centering}

To further analyze stellar populations in 47\,Tucanae, we combine information from {\it JWST} and {\it HST} data. The left panel of Figure\,\ref{fig:CMDs} shows the same CMD of Figure\,\ref{fig:CMDjwst} but for the stars in the region that overlaps the UVIS/WFC3 FoV, only.   
Moreover, the $m_{\rm F115W}$ vs.\,$m_{\rm F606W}-m_{\rm F115W}$ CMD of stars in the UVIS/WFC3 FoV is plotted right panel of Figure\,\ref{fig:CMDs}. 
In both panels, we used stellar proper motions to separate the probable cluster members, which are colored black, from field stars (red crosses). 
The M-dwarfs define a wide MS in both CMDs of Figure\,\ref{fig:CMDs}. However, in the ${\rm F115W}$ vs.\,$m_{\rm F606W}-m_{\rm F115W}$ the bulk of MS populates the middle of the MS, and the color distribution exhibits tails of stars with blue and red $m_{\rm F606W}-m_{\rm F115W}$ colors.
To combine the information on multiple populations from the two CMDs of Figure\,\ref{fig:CMDs}, we construct the $\Delta_{\rm F115W,F322W2}$ vs.\,$\Delta_{\rm F606W,F115W}$ ChM of Figure\,\ref{fig:chm}. We restrict the analysis to M-dwarf stars with $20.64<m_{\rm F115W}<22.4$ mag, which is the magnitude interval where multiple populations are more clearly visible in the CMDs. 
%%%%%%%%%%%%%%%%%%%%%%%%%%%%%%%%%%
%%%%%%%%%%%%%%%%%%%%%%%%%%%%%%%%%%

The 1P stars are located around the origin of the ChM, whereas 2P stars define the sequence of stars that ranges from ($\Delta_{\rm F115W,F322W2}$,$\Delta_{\rm F606W,F115W}$)$\sim$($-$0.1,0.1) towards large values of $\Delta_{\rm F115W,F322W2}$.
In this ChM there is no evidence for a sharp separation between 1P and 2P stars, in contrast with  the traditional ChM for RGB stars that exhibits discrete sequences of 1P and 2P stars \citep[e.g.\,][]{milone2017a}.  
Since the position of a star in the $\Delta_{\rm F115W,F322W2}$vs.\,$\Delta_{\rm F606W,F115W}$ ChM of M-dwarfs is mostly due to its oxygen abundance, the partial overlap of 1P stars and 2P stars is possibly due to 2P stars that are slightly oxygen depleted with respect to the 1P. Conversely, large nitrogen differences are the main reasons for the sharp separation of 1P and 2P stars in the ChM of RGB stars.

%{\color{red}{Ti riferisci alla mappa delle RGB qui? Se si' bisogna specificarlo perche' si potrebbe pensare che parli della mappa delle M dwarfs precedente fatta solo con HST. Un'altra cosa e' se questa differenza e' fisica o e' dovuta agli errori, si puo' dire qualcosa riguardo agli errori qui?}}
Similarly to what is observed along the RGB and the upper MS, the 1P stars define an extended sequence in the ChM. 
To quantify the color extension of the 1P we followed the recipe by \citet{milone2017a} and computed the difference between the 90$^{\rm th}$ and the 10$^{\rm th}$ percentile of the $\Delta_{\rm F606W,F115W}$ distribution of 1P stars. The intrinsic width has been estimated by subtracting the color errors in quadrature and corresponds to 0.10$\pm$0.01 mag.
Similarly, the sequence of 2P stars is not consistent with a simple population but shows hints of stellar overdensities around $\Delta_{\rm F115W,F322W2}=$0.15, 0.25, and 0.3 mag.
%{\color{red} FORSE FACENDO UN HESS DIAGRAM ANCHE QUI E' POSSIBILE VEDERE GRUPPI PIU' DISCRETI??}
%APM. L'ho aggiunto... ma non migliora...

%%%%%%%%%%%%%%%%%%%%%%%%%%%%%%%%%%

%%%%%%%%%%%%%%%%%%%%%%%%%%%%%%%%%%

%%%%%%%%%%%%%%%%%%%%%%%%%%%%%%%%%%
%%%%%%%%%%%%%%%%%%%%%%%%%%%%%%%%%%

%%%%%%%%%%%%%%%%%%%%%%%%%%%%%%%%%%
%%%%%%%%%%%%%%%%%%%%%%%%%%%%%%%%%%

\section{Comparison with theory}\label{sec:iso}
The chemical species that most shape the photometric patterns typical of different stellar populations in GCs are He, C, N, and O \citep[e.g.][]{marino2008a, milone2017a, marino2019a}. 
To explore the impact of variations in these elements %He, C, N, and O 
on the NIRCam stellar magnitudes, we followed the same recipes used in previous work of our team \citep[e.g.][]{milone2012a, milone2018a, dotter2015a}.
We first considered two isochrones from the Dartmouth database \citep{dotter2008a} with the same age of 13 Gyr 
%, iron abundance, [Fe/H]=$-$1.50, 
and $\alpha$ enhancement of [$\alpha$/Fe]=$+$0.4 dex, but with different helium contents (helium mass fractions of Y=0.246 and Y=0.33).
  These isochrones include stars with masses bigger than 0.1 solar masses.
We inferred the colors and magnitudes of stars with different C, N, and O abundances by combining information from isochrones and from model atmospheres and synthetic spectra of stars with appropriate chemical compositions. 
 For this procedure, we considered two different iron abundances, namely [Fe/H]$=-0.75$ and [Fe/H]$=-1.50$, with the most Fe-rich case resembling the metallicity of 47~Tucanae.

 To do this, we first selected fifteen points along each isochrone. The effective temperature ($T_{\rm eff}$) and surface gravity ($g$) of each selected point are then used to compute a reference stellar spectrum with similar chemical composition as 1P star (i.e.\,Y=0.246, solar-scaled abundances of C and N, and [O/Fe]=$+$0.40). Similarly, we simulated the  reference spectra with the abundances of C, N, and O which are indicative of 2P stars. %Specifically, 
 We adopted different chemical compositions for the simulated spectra with different metallicities.
 The chemical compositions of the spectra with [Fe/H]=$-$0.75 are provided in Table\,\ref{tab:iso} and resemble the chemical composition of 1P and 2P stars of 47\,Tucanae.
  For the spectra with [Fe/H]=$-$1.5 we assumed that 2P stars are depleted in both [C/Fe] and [O/Fe] by 0.5~dex and enhanced in [N/Fe]=$+$1.2~dex, with respect to the 1P. These elemental variations are comparable with those inferred for NGC\,6752 \citep{yong2005a, yong2008a, yong2015a}. 
 %{\color{red} PER 47TUC ABBIAMO USATO LE STESSE ABBONDANZE PER CNO NO?} 
  We verified that the adopted microturbolence value does not significantly change the spectra \citep[see also]{sbordone2011a}. For simplicity, 
 we adopted for all models a microturbulence velocity of 2~km\,s$^{-1}$. 

  We derived the model atmospheres with the ATLAS\,12 computer program, which is based on the opacity-sampling technique and assumes local thermodynamic equilibrium \citep{kurucz1970a, kurucz1993a, sbordone2004a}.
  We included molecular line lists for all the diatomic molecules listed in Kurucz's website\footnote{\url{http://kurucz.harvard.edu}} plus the 
  %CO, C$_{2}$, CN, OH, MgH, SiH, 
  H$_{2}$O molecules
  %and TiO
  %, VO, ZrO from
 \citep{partridge1997a,schwenke1998a}.
 %, and B.\,Pletz (private communication).
  The spectra are computed with the SYNTHE programme \citep{kurucz1981a, castelli2005a, kurucz2005a, sbordone2007a} in the region between 1,500~\AA\, and 51,000~\AA\, that is covered by the UVIS/WFC3, NIR/WFC3, WFC/ACS, and NIRCam filters.
   The stellar magnitudes are then derived by integrating the spectra over the bandpasses of the filters. To derive the magnitudes of the 2P isochrone, we calculated the magnitude differences between 2P and 1P stars ($\delta m$) and added these quantities to the magnitudes of the 1P isochrone.
  
As an illustrative case, we show here the results for the generic case of [Fe/H]$=-1.50$. In the next section, we show instead results for the more metal-rich isochrones in comparison with the observations of 47~Tucanae.  %example, the black lines shown 

\begin{centering} 
\begin{figure*} 
  \includegraphics[height=9.5cm,trim={0.0cm 5cm 6.2cm 3.0cm},clip]{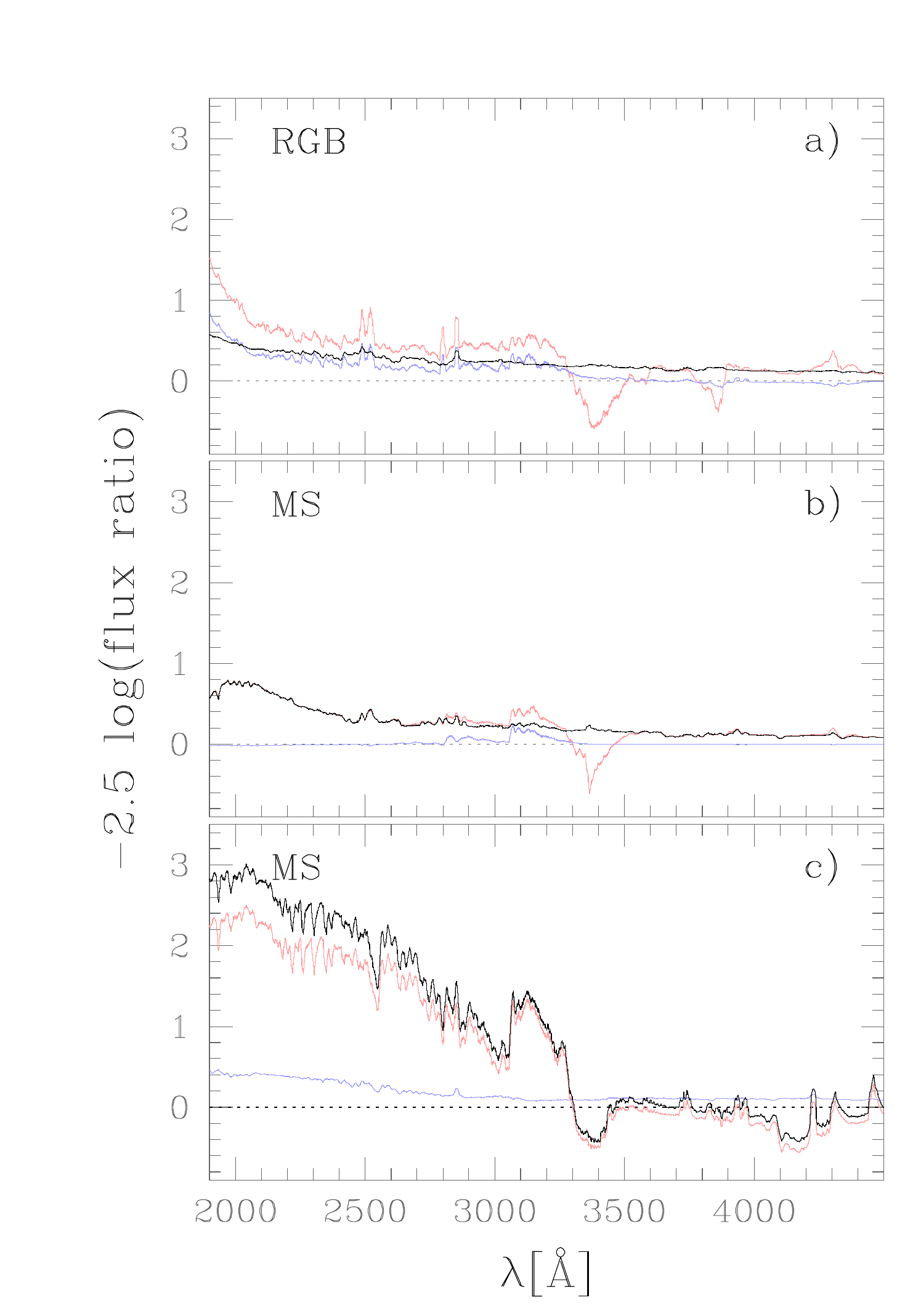}s
  \includegraphics[height=9.5cm,trim={2.0cm 5cm 0cm 3.0cm},clip]{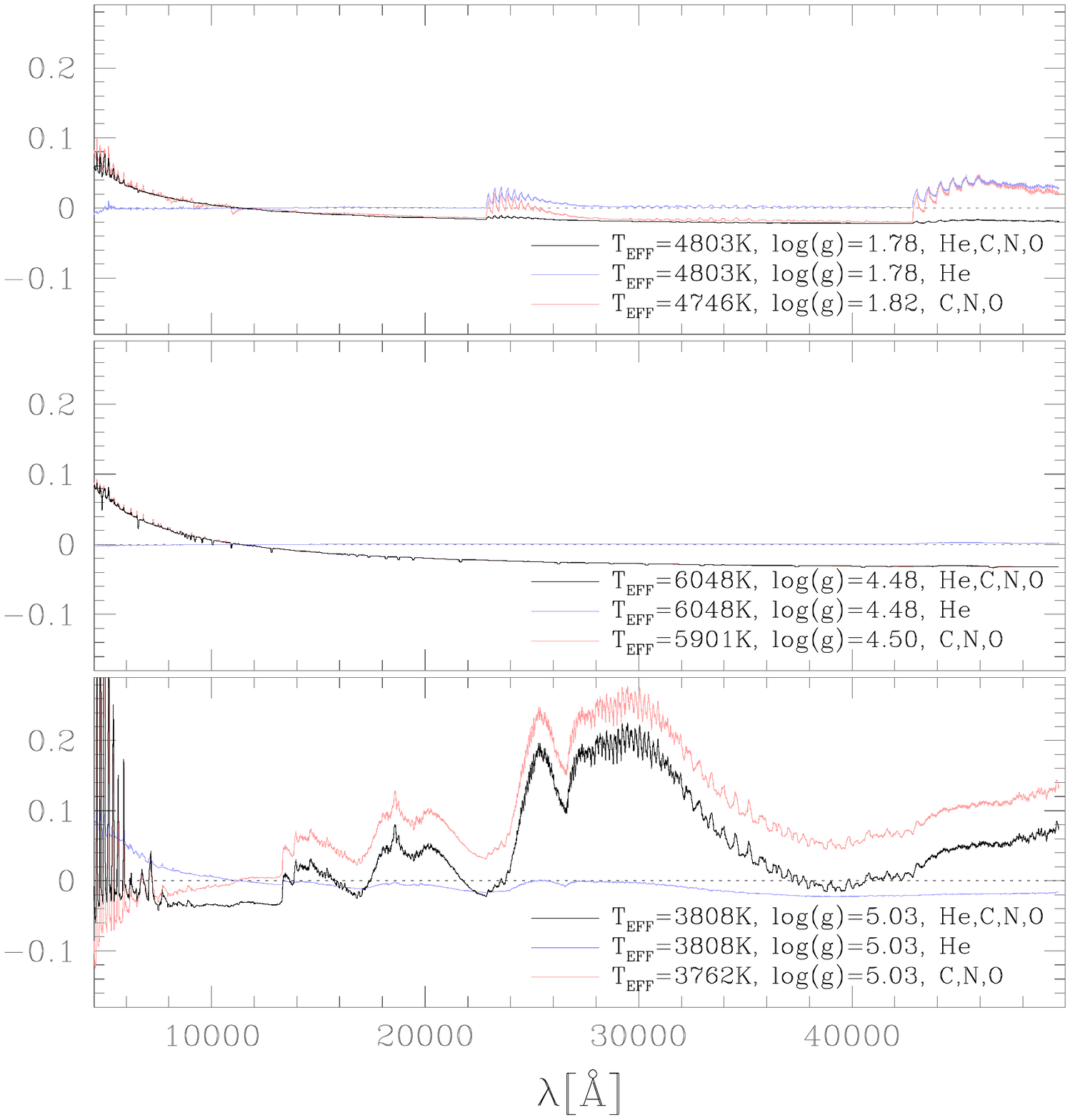}
  %\includegraphics[width=9.cm,trim={0.0cm 5cm 0.0cm 9.5cm},clip]{FigFITERS.pdf}
 %/home/milone/WORKS/JWSTSPECTRA/JWST_25Gen/flussi1.macro fig figB
 \caption{The black lines compare the fluxes of simulated spectra for stars with  the same F115W magnitude but 1P-like and 2P-like stellar compositions, with 2P stars being enhanced in helium and nitrogen, and depleted in carbon and oxygen with respect to the 1P. Panel a refers to RGB stars, whereas panels b and c are focused on a bright K-dwarf and M-dwarf, respectively. The pink %{\color{red} PINK?} 
  lines refer to stars with the same He content abundances as 1P stars but different abundances of C, N, and O, while the azure lines are derived from spectra with the same C, N, and O content as 1P stars but enhanced helium content. The left and right panels refer to the wavelength intervals with $\lambda < 4,500$~\AA\ and $\lambda > 4,500$~\AA, respectively. See the text for details.   }
 \label{fig:FigDF} 
\end{figure*} 
\end{centering}

\begin{centering} 
\begin{figure*} 
    \includegraphics[height=5.5cm,trim={0.6cm 6.8cm 0.0cm 10.5cm},clip]{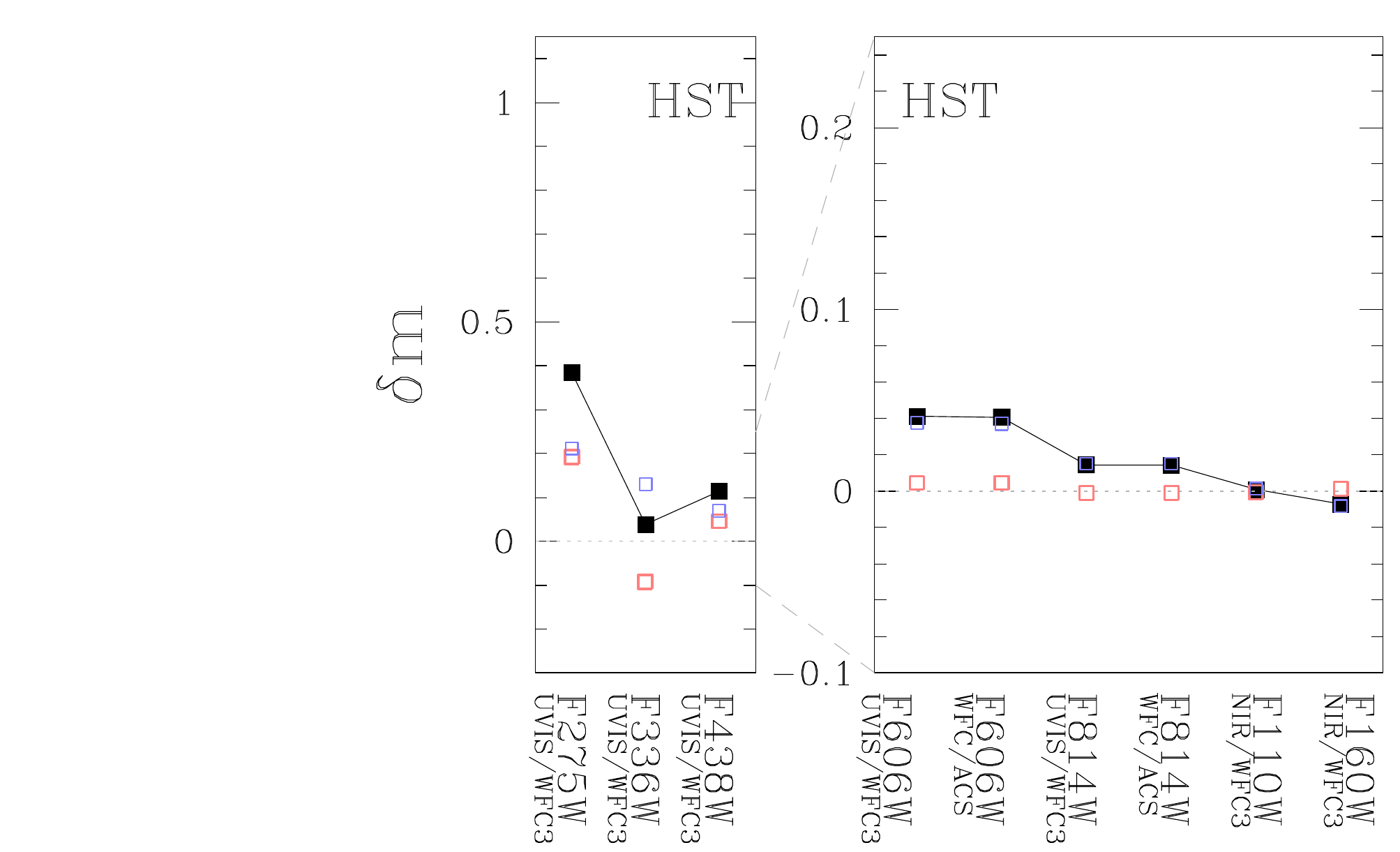}
  \includegraphics[height=5.5cm,trim={2.85cm 6.8cm 0.0cm 10.5cm},clip]{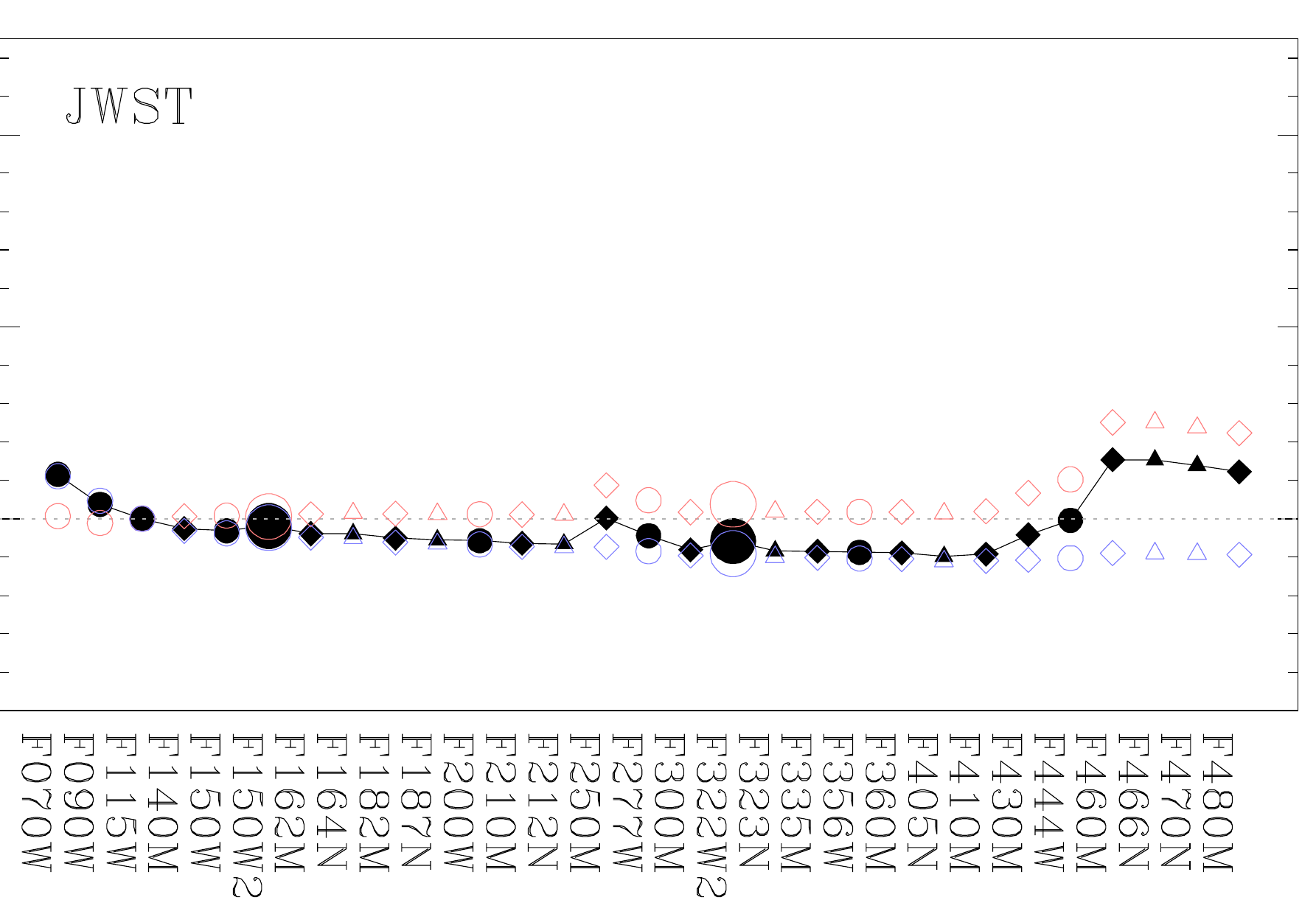}
      \includegraphics[height=5.5cm,trim={0.6cm 6.8cm 0.0cm 10.5cm},clip]{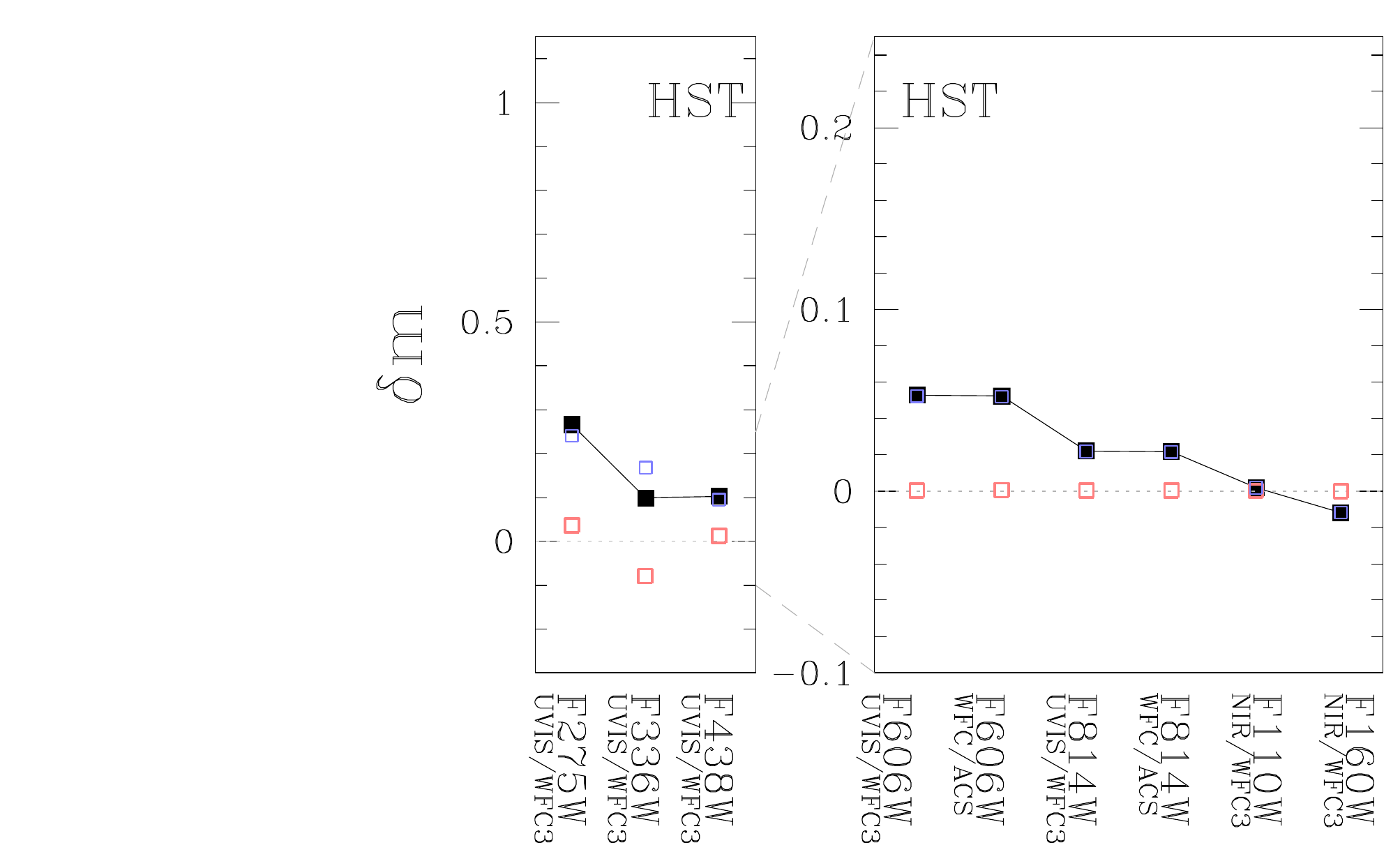}
  \includegraphics[height=5.5cm,trim={2.85cm 6.8cm 0.0cm 10.5cm},clip]{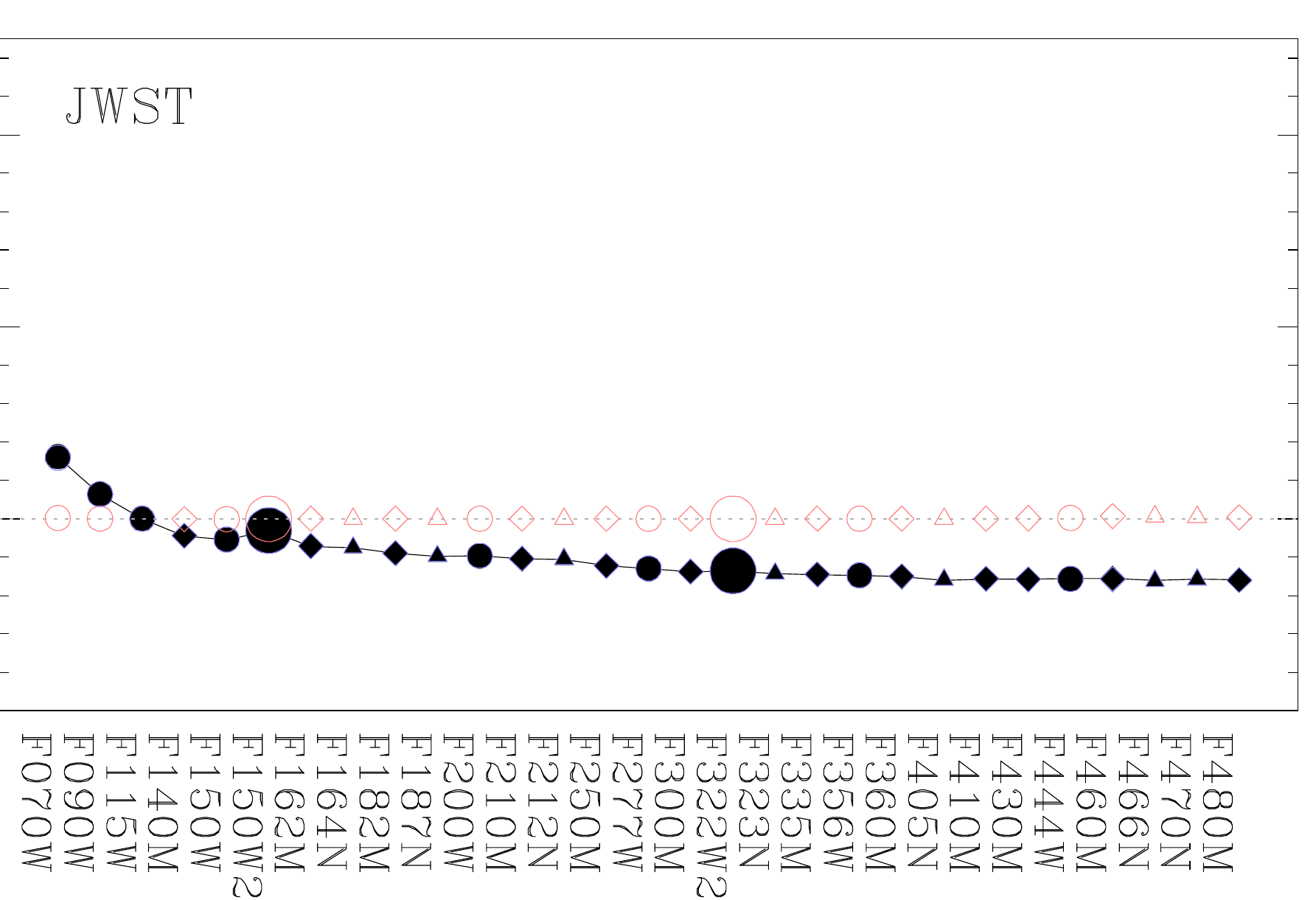}
       \includegraphics[height=5.5cm,trim={0.6cm 6.8cm 0.0cm 10.5cm},clip]{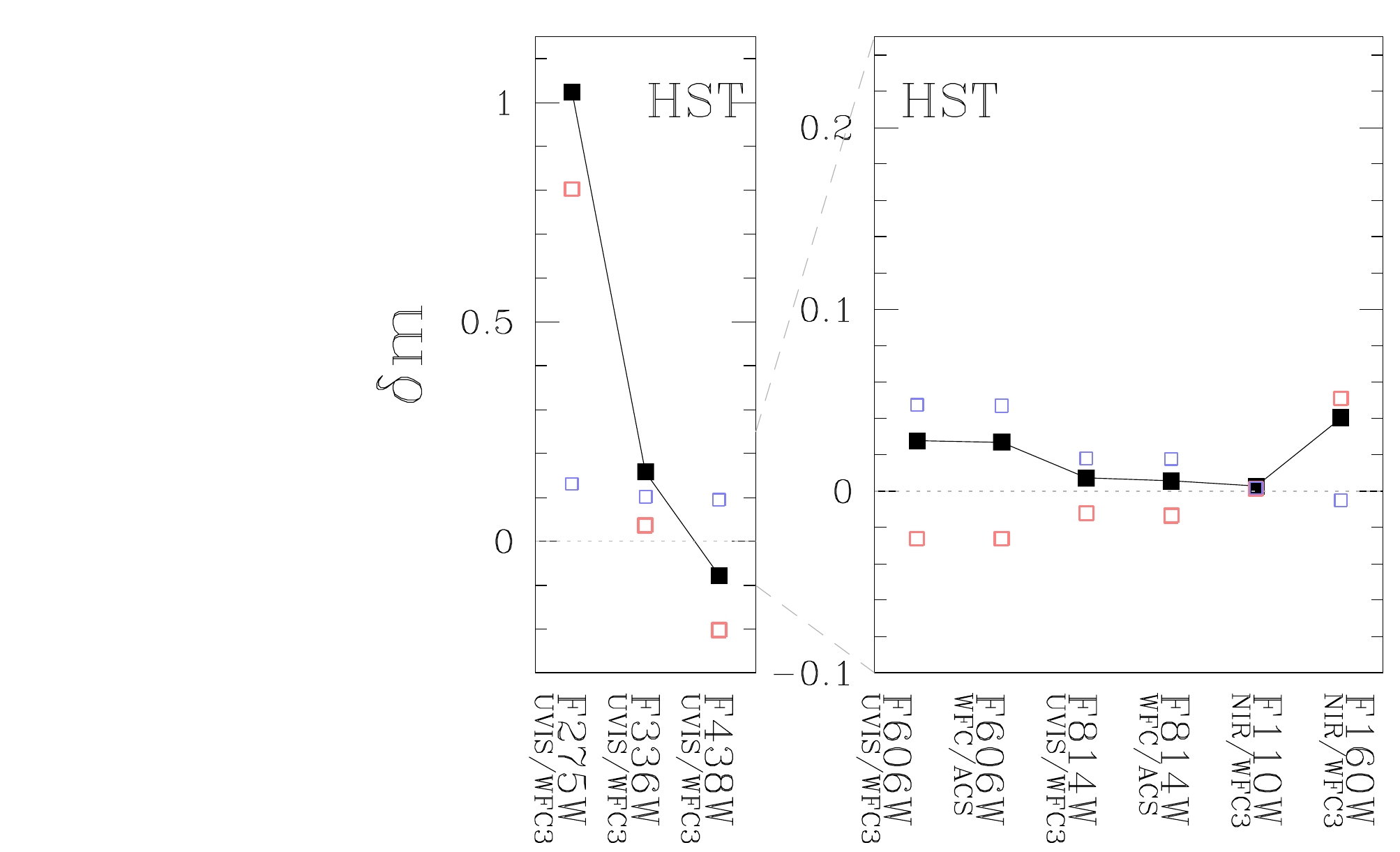} 
  \includegraphics[height=5.5cm,trim={2.85cm 6.8cm 0.0cm 10.5cm},clip]{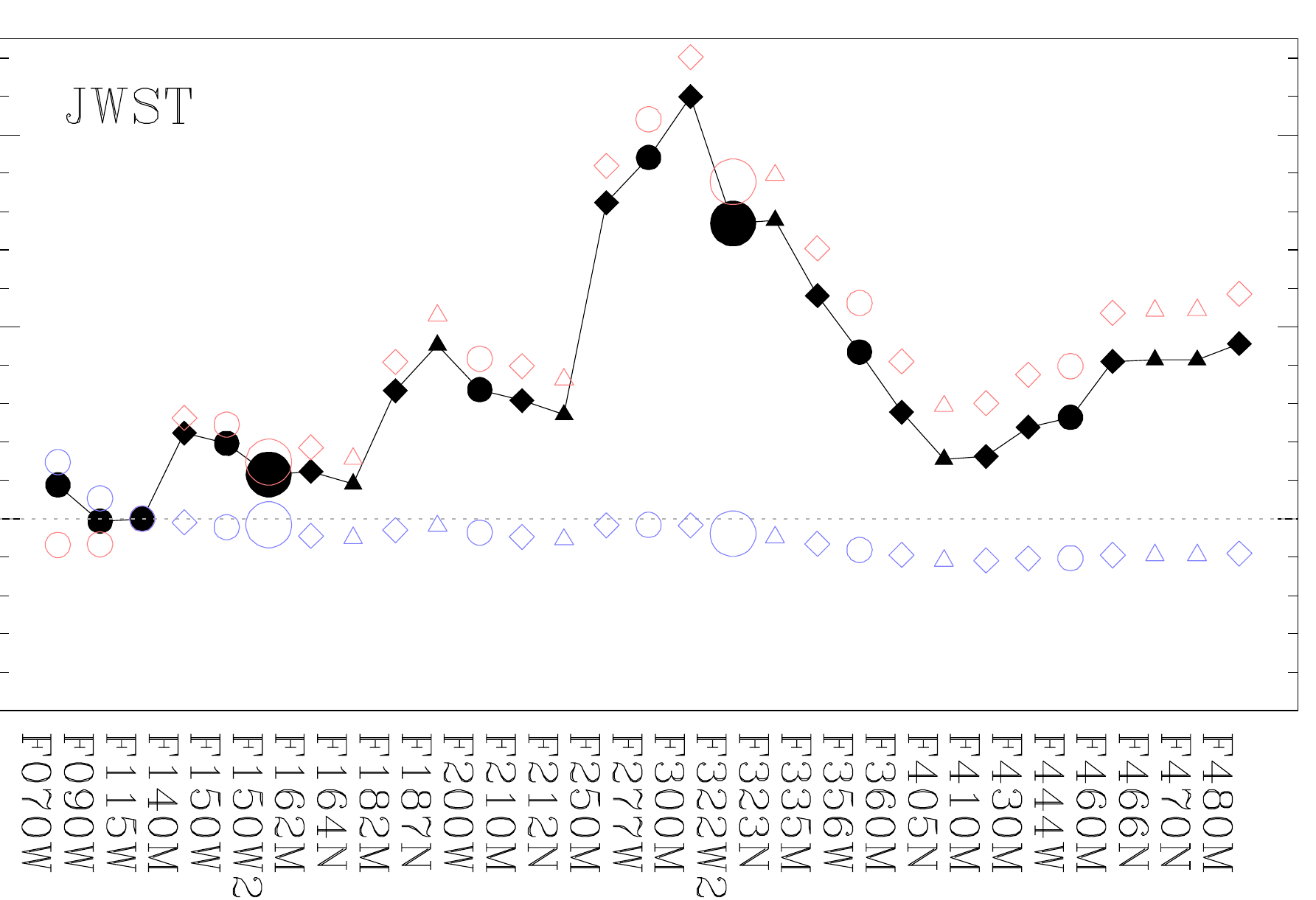}
 %/home/milone/WORKS/JWSTSPECTRA/47TSPEC/figDC0 figMMS e  figDC 
 \caption{Black symbols show the magnitude difference for simulated spectra of stars with the same F115W magnitude and [Fe/H]=$-$1.5, but 1P-like and 2P-like abundances of He, C, N, and O. 
 The pink points refer to stars with the same helium content but different abundances of C, N, and O, while the blue ones are derived from spectra with the same C, N, and O content as 1P stars but enhanced helium. The top, middle, and bottom panels refer to RGB stars, K-dwarfs, and M-dwarfs, respectively. The triangles, diamonds, circles, and large circles indicate the narrow, middle, wide, and extra-wide passbands  NIRCam filters, respectively.   }
 \label{fig:DMAG} 
\end{figure*} 
\end{centering}

In Figure\,\ref{fig:FigDF} the black lines compare the logarithm of the flux ratios of He-enhanced 2P-like stars with different light-element abundances with respect to 1P stars with Y=0.246. We also show the flux ratios, relative to the 1P spectrum, of stars with Y=0.246 but with the C, N, and O abundances that we assumed for 2P stars (pink lines), and the flux ratios derived from stars with Y=0.33 but the same C, N, O content as 1P stars (blue lines).
 The simulated spectra correspond to stars with the same F115W magnitude. % but different abundances of He, C, N, and O. 
  Panels a compare the spectra of RGB stars with absolute magnitude $M_{\rm F115W}$=$-$1.9 mag, while panels b refer to a bright MS star, with $M_{\rm F115W}$=4.3 mag, and panel c is focused on M-dwarfs with $M_{\rm F115W}$=8.1 mag. %For completeness, we plot in panels d1--d4 of Figure\,\ref{fig:FigDF} the transmission curves of the NIRCam filters.
  The resulting magnitude differences are plotted in Figure\,\ref{fig:DMAG} for all NIRCam filters and for the {\it HST} filters shown in Figure\,\ref{fig:filters}. 
  
From Figure\,\ref{fig:FigDF} it is immediately clear that the spectra of the 2P and 1P M-dwarfs strongly differ from each other along most of the analyzed spectral regions covered by NIRCam. The largest flux differences involve the long-wavelength channel at $\lambda \gtrsim$10,000~\AA. As illustrated by the black line in panel c, for an M-dwarf with $M_{\rm F115W}$=8.0 mag, the logarithm of the flux ratio, which is close to zero around 23,000~\AA, approaches its maximum between $\sim$25,000 and 32,000 \AA, with 2P stars having fainter fluxes than 1P stars with the same F115W magnitude. The logarithm of the flux ratio nearly drops to zero around $\lambda$=40,000~\AA\ and arises towards positive values at longer wavelengths. 
 In the spectral range of the short-wavelength channel, the logarithm of the flux ratio is nearly flat for $\lambda \lesssim 13,000$~\AA, while 1P stars are typically fainter than the 2P at longer wavelengths. 
 The fact that the spectra of 1P stars are more absorbed than those of the 2P for $\lambda \gtrsim 13,000$~\AA\ is mostly due to various molecules composed of oxygen, including H$_{2}$O. At variance with the aforementioned light elements, helium variation has a moderate effect on the luminosity difference of M-dwarfs in the NIRCam filters, as indicated by the azure curve of Figure\,\ref{fig:FigDF}c.
 
In contrast with the M-dwarfs, the flux difference between 2P and 1P stars with the same F115W magnitude is small for MS stars brighter than the MS knee and for giant stars. As shown in panels a and b of Figure\,\ref{fig:FigDF}, most of the flux variation is due to the different helium content, whereas C, N, and O variations affect the relative fluxes at long wavelengths alone, with $\lambda \gtrsim 43,000$\AA.  

\begin{centering} 
\begin{figure*} 
%/home/milone/WORKS/JWSTSPECTRA/JWST_25Gen/ISO/isocrone.macro
        \includegraphics[height=7.6cm,trim={0.6cm 5cm 4.7cm 4.5cm},clip]{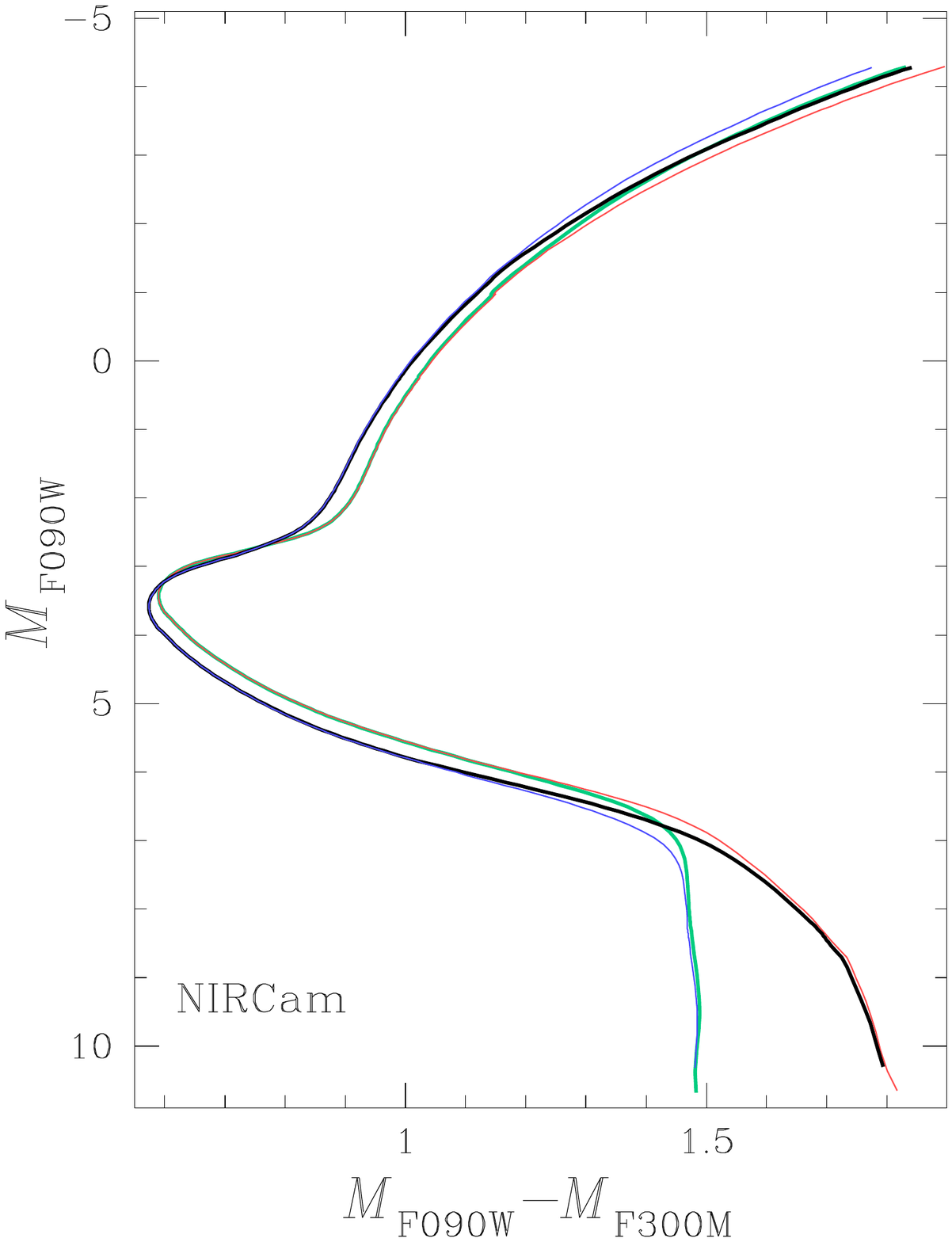} % go7
            \includegraphics[height=7.6cm,trim={0.6cm 5cm 4.7cm 4.5cm},clip]{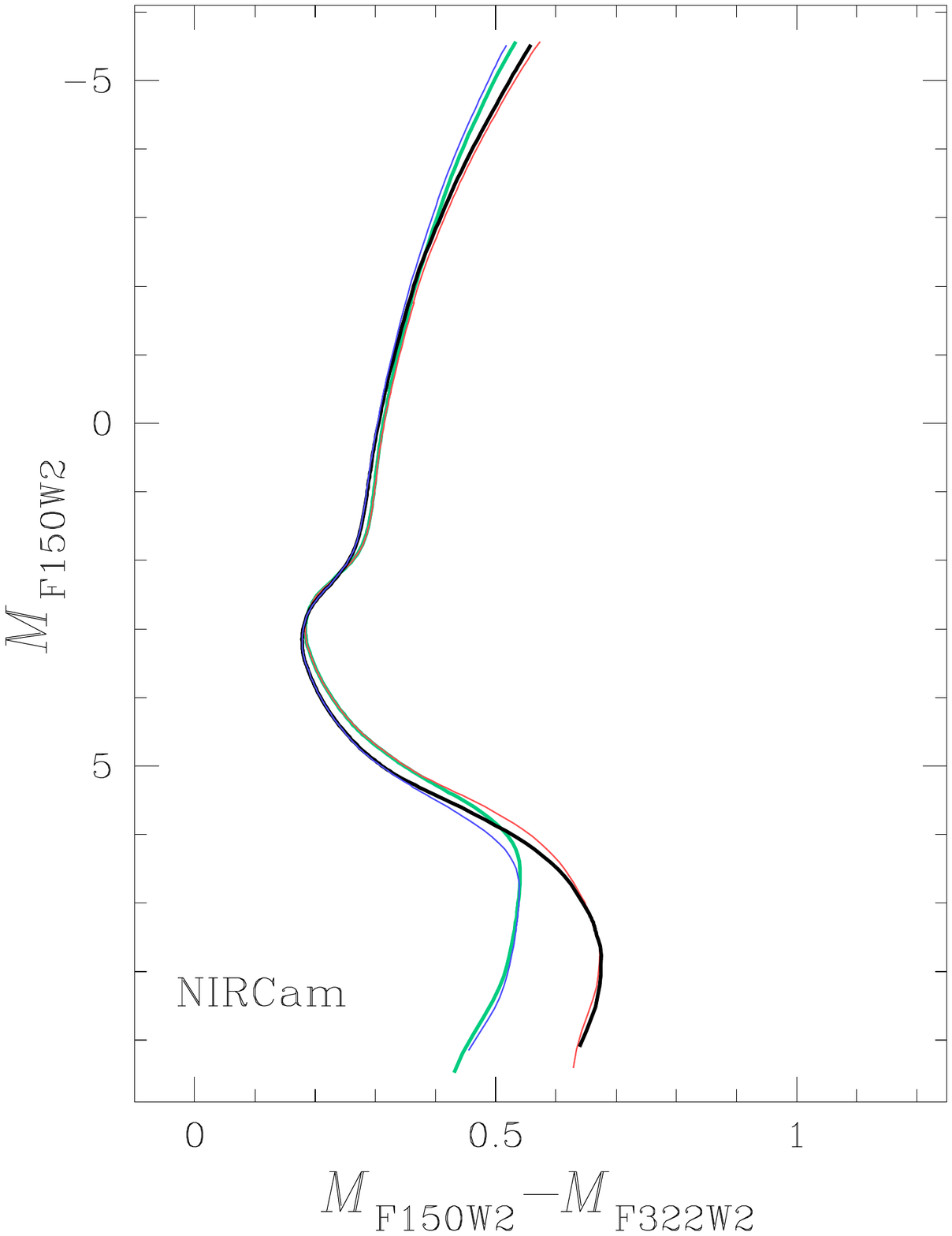} %go6
            \includegraphics[height=7.6cm,trim={0.6cm 5cm 4.7cm 4.5cm},clip]{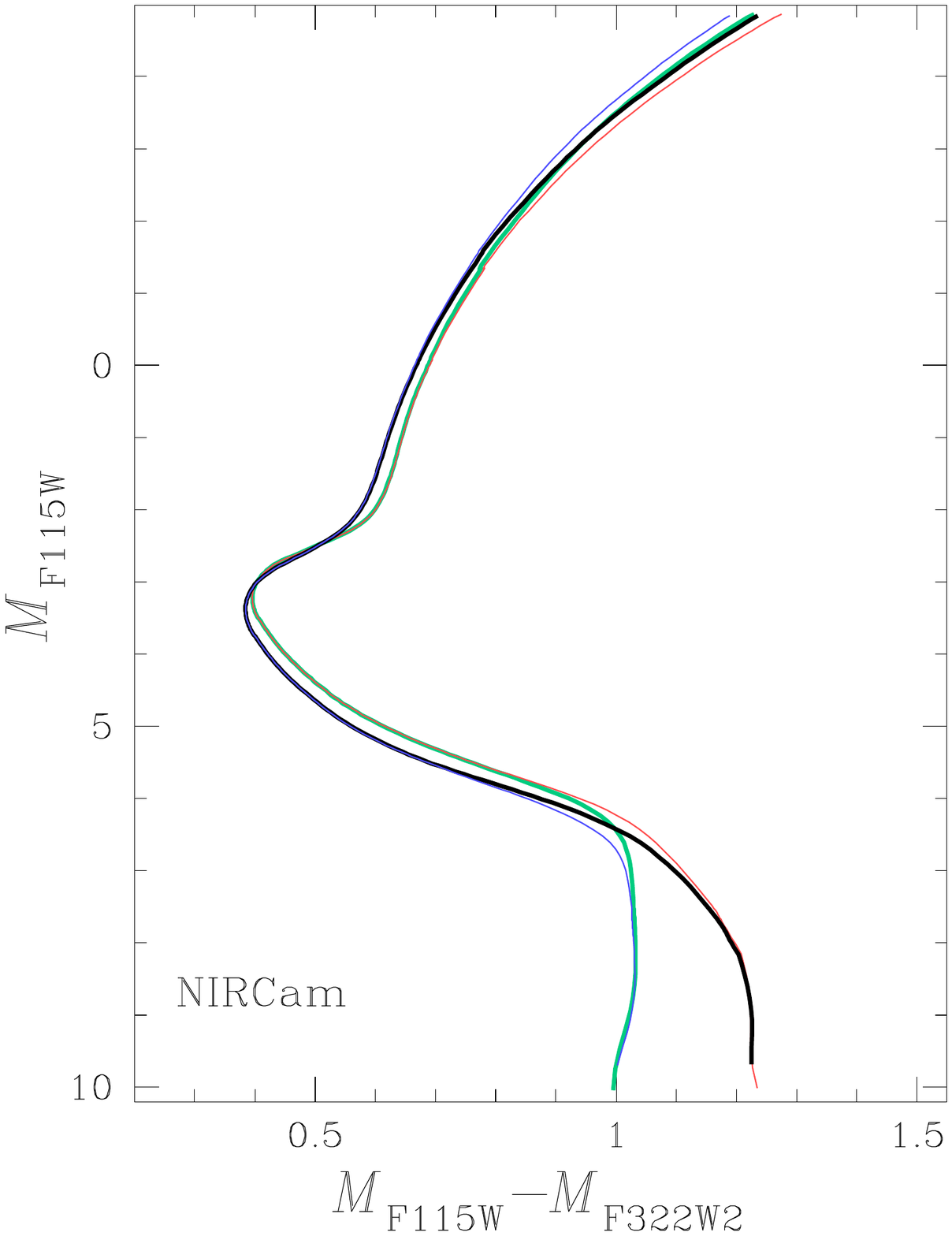} % go    
    \includegraphics[height=7.6cm,trim={0.6cm 5cm 4.7cm 4.5cm},clip]{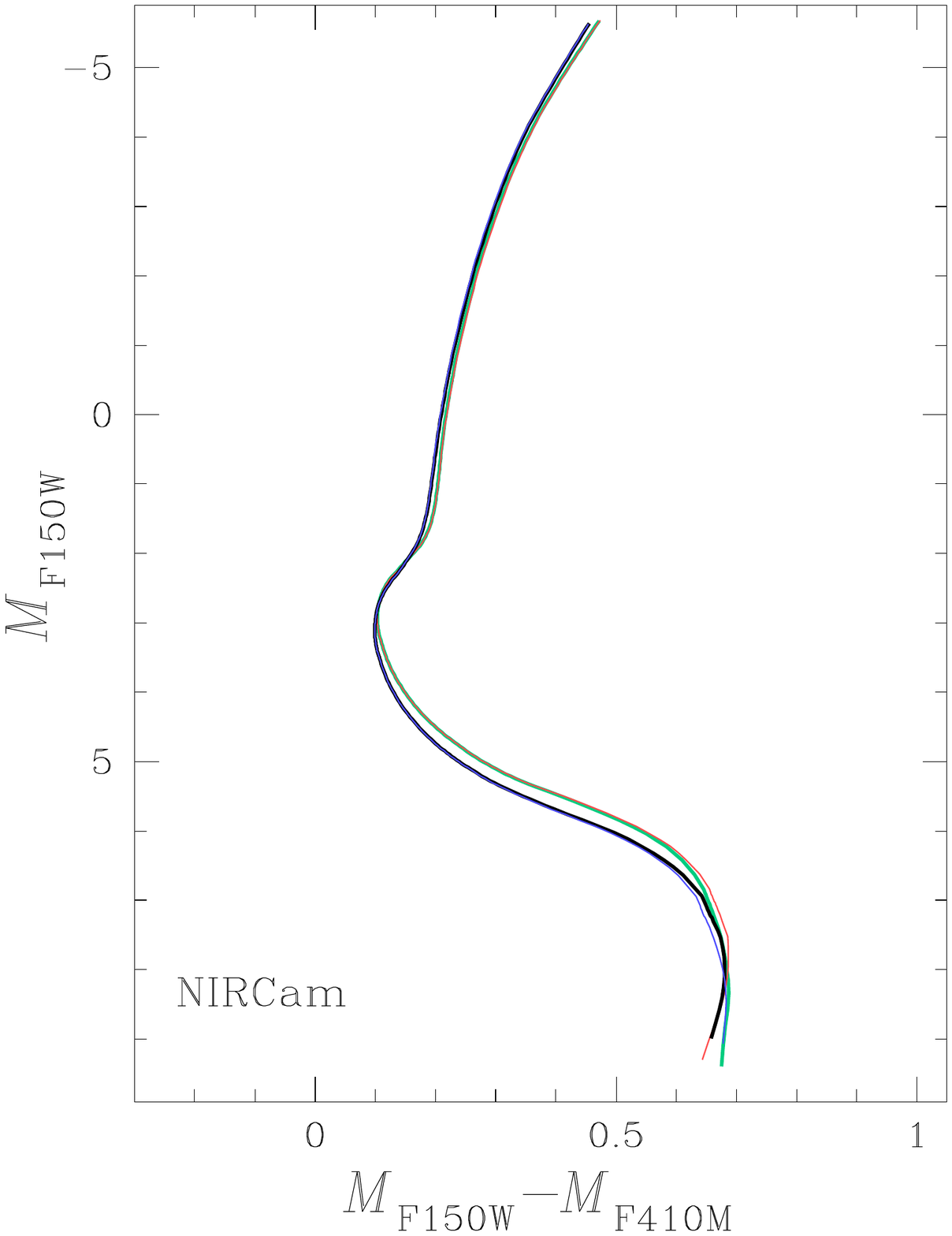} %go2
    \includegraphics[height=7.6cm,trim={0.6cm 5cm 4.7cm 4.5cm},clip]{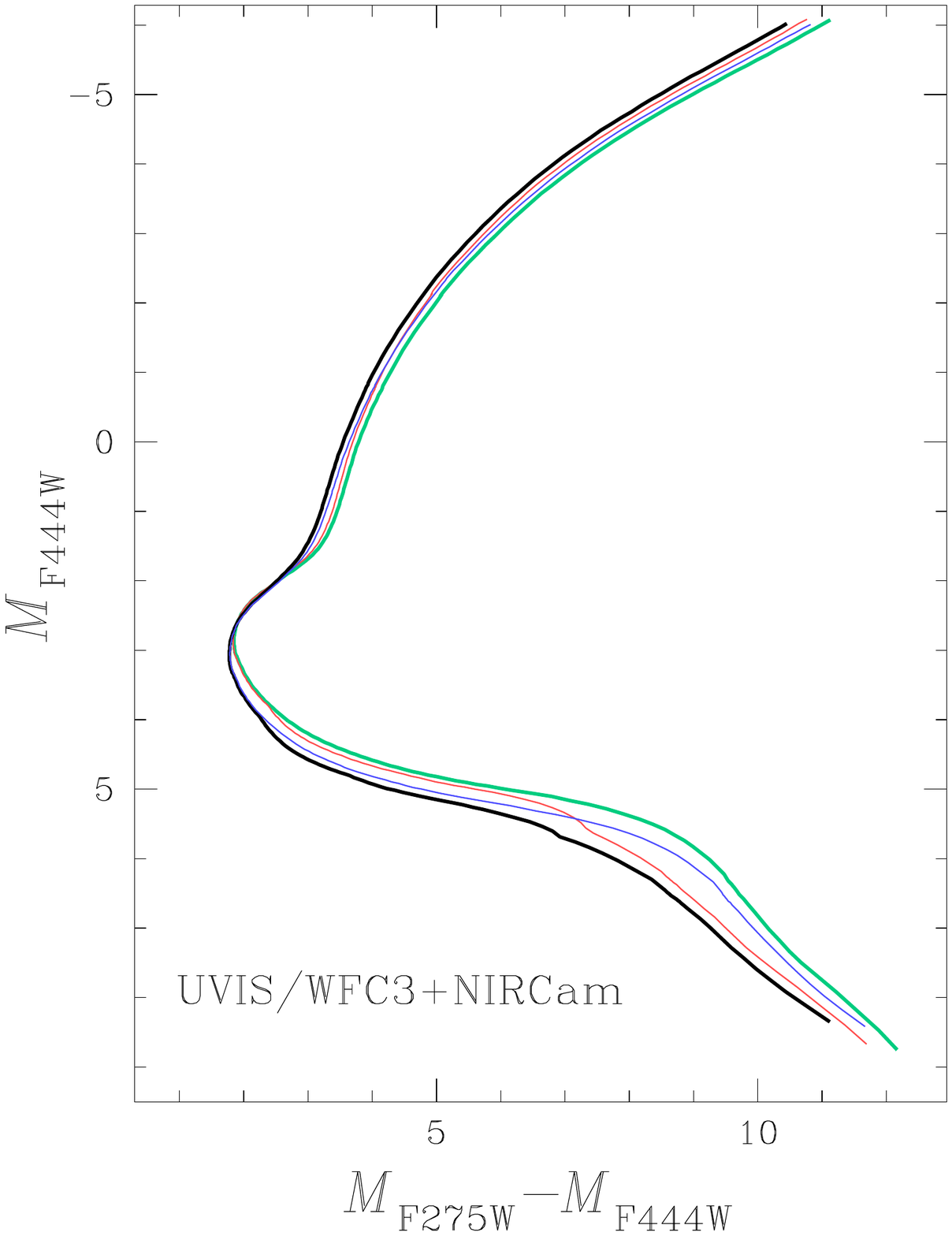} %go5
    \includegraphics[height=7.6cm,trim={0.6cm 5cm 4.7cm 4.5cm},clip]{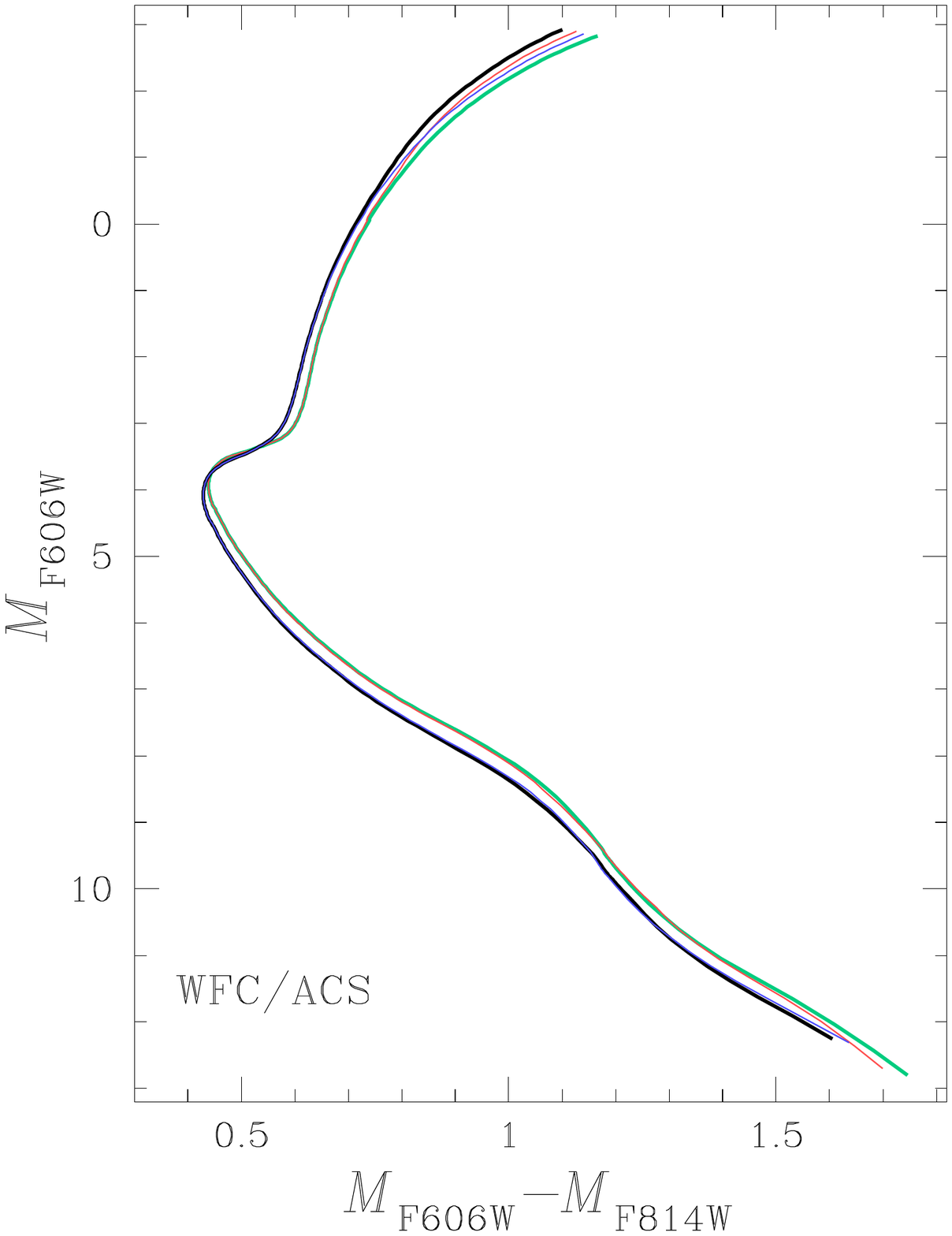}  %go4 
  \caption{ Isochrones from the Darthmouth database \citep{dotter2008a} with ages of 13 Gyr, [Fe/H]=$-$1.5, and [$\alpha$/Fe]=0.4.  The isochrones comprise stars with masses $\mathcal{M}>0.1 \mathcal{M}_{\odot}$. 
   The aqua and pink isochrones have pristine helium content Y=0.246, whereas blue and black isochrones are helium-enhanced (Y=0.33). The pink and black isochrones are depleted in carbon and oxygen by 0.5 and 0.9 dex, respectively, and enhanced by 1.2 dex in nitrogen compared to the other isochrones, which have [O/Fe]=0.4 and solar contents of carbon and nitrogen.}
 \label{fig:iso1} 
\end{figure*} 
\end{centering}

Figure\,\ref{fig:iso1} shows isochrones constructed with NIRCam and {\it HST} photometric bands in various CMDs. The colors indicate four stellar populations with different chemical compositions: the aqua isochrones share the same chemical composition as 1P stars, whereas the blue isochrones resemble the chemical abundances for C, N, and O of 2P stars and are He-enhanced (Y=0.33). Pink and black isochrones have the same C, N, and O contents that we adopted above for 2P stars but different helium abundances (Y=0.246 and 0.33, respectively). 

The upper panels of Figure\,\ref{fig:iso1} show three CMDs that are sensitive to M-dwarfs with different abundances of C, N, and O.
The $M_{\rm F090W}$ vs.\,$M_{\rm F090W}-M_{\rm F300M}$ CMD provides the widest color separation  between the M-dwarf sequences of isochrones with different C, N, and O abundances  (top-left panel of Figure\,\ref{fig:iso1}). Hence, it is the most-sensitive color in detecting multiple populations among low-mass stars. In contrast, these isochrones are almost superimposed on each other along the MS segment above the MS knee, the SGB, and most of the RGB. In the upper RGB segment, the isochrone with enhanced N and depleted C and O are slightly redder than the isochrones with the same He content but 1P-like C, N, and O abundances.

The patterns of the four isochrones in $M_{\rm F150W2}$ vs.\,$M_{\rm F150W2}-M_{\rm F322W2}$ CMDs resemble the $M_{\rm F090W}$ vs.\,$M_{\rm F090W}-M_{\rm F300M}$ CMD.
However, for a fixed $\log{g}$ value, the F150W2$-$F322W2 color separation between  isochrones with different C, N, and O abundances is significantly narrower than the F090W$-$F300M color distance. Conversely, the F150W2 and F322W2 bands are more efficient filters than the F090W/F300M filter pair. Hence, the first combination could be preferable due to the shorter exposure times needed to obtain a given signal-to-noise ratio.

A similar conclusion can be extended to the $M_{\rm F115W}$ vs.\,$M_{\rm F115W}-M_{\rm F322W2}$ CMD shown in  top-right panel of Figure\,\ref{fig:iso1}, which is based on the same filters that are available for 47\,Tucanae from GO-2560.
The F115W and F322W2 bands give intermediate color separations compared with the F150W2$-$F322W2 and F090W$-$F300M colors. Moreover, for a fixed exposure time, the F115W observations would provide intermediate signal-to-noise ratios, when compared to F090W and F150W2 images.

In contrast with what is shown in the upper panels of Figure\,\ref{fig:iso1}, some CMDs composed of NIRCam filters are poorly sensitive to multiple stellar populations.
As an example, the isochrones with the same helium content but different C, N, and O abundances are nearly superimposed with each other in the $M_{\rm F150W}$ vs.\,$M_{\rm F150W}-M_{\rm F410M}$ CMD (bottom-left panel of Figure\,\ref{fig:iso1}). 

We expect a very-wide color separation between multiple stellar populations with different chemical compositions by combining appropriate filters from {\it HST} and NIRCam. As an example, the CMD plotted in the bottom-middle panel of Figure\,\ref{fig:iso1}, which is obtained from the F275W filter of UVIS/WFC3 and the F444W NIRCam filter, would show extreme color separations of more than one magnitude between the MSs of M-dwarfs with different C, N, and O abundances. Moreover, large color separations are observed along the upper MS and the RGB for stellar populations with different helium abundances.

Finally, we show isochrones in the $M_{\rm F606W}$ vs.\,$M_{\rm F606W}-M_{\rm F814W}$ plane (bottom-right panel of Figure\,\ref{fig:iso1}). This CMD, which is constructed with ACS/WFC filters of {\it HST} that are used in various studies of faint GC stars, is a poor tool to identify stellar populations with different C, N, and O, abundances.

 %%%%%%%%%%%%%%%%%%%%%%%%%%%%%%%%%%%%%%%%%%%%%%%%%%%%%%%%%%%%%%%%%%%%%%%%%%%%

%%%%%%%%%%%%%%%%%%%%%%%%%%%%%%%%%%

%%%%%%%%%%%%%%%%%%%%%%%%%%%%%%%%%%

\section{Interpreting the observations of 47~Tucanae}\label{sec:iso47t}

\begin{centering} 
\begin{figure*} 
    \includegraphics[height=6.0cm,trim={0.7cm 5.4cm 0.0cm 15.0cm},clip]{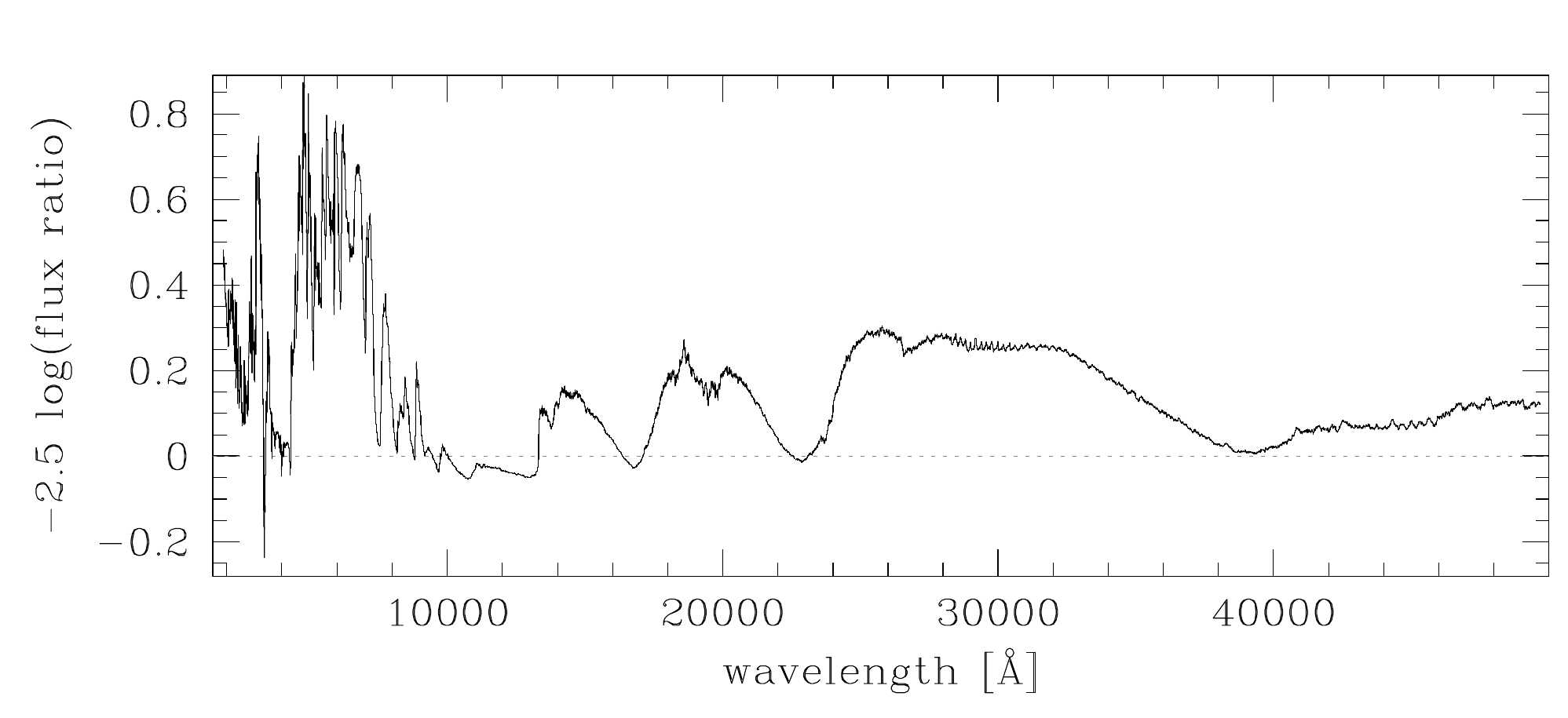}
    %/home/milone/WORKS/JWSTSPECTRA/47TSPEC/flussi1n.macro figT
    \includegraphics[height=5.2cm,trim={9.6cm 6.8cm 0.0cm 11.0cm},clip]{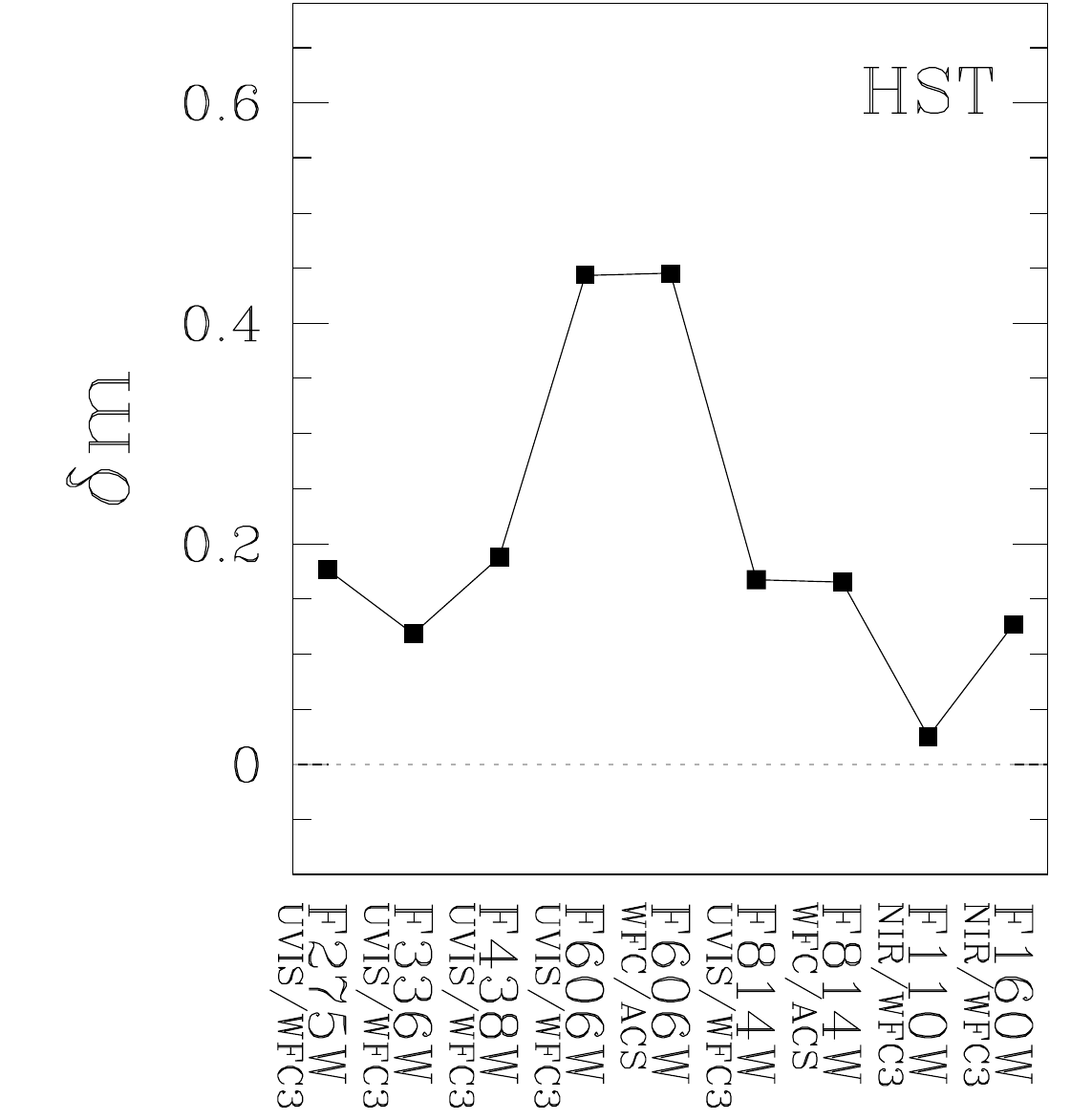}
\includegraphics[height=5.2cm,trim={2.85cm 6.8cm 0.0cm 11.0cm},clip]{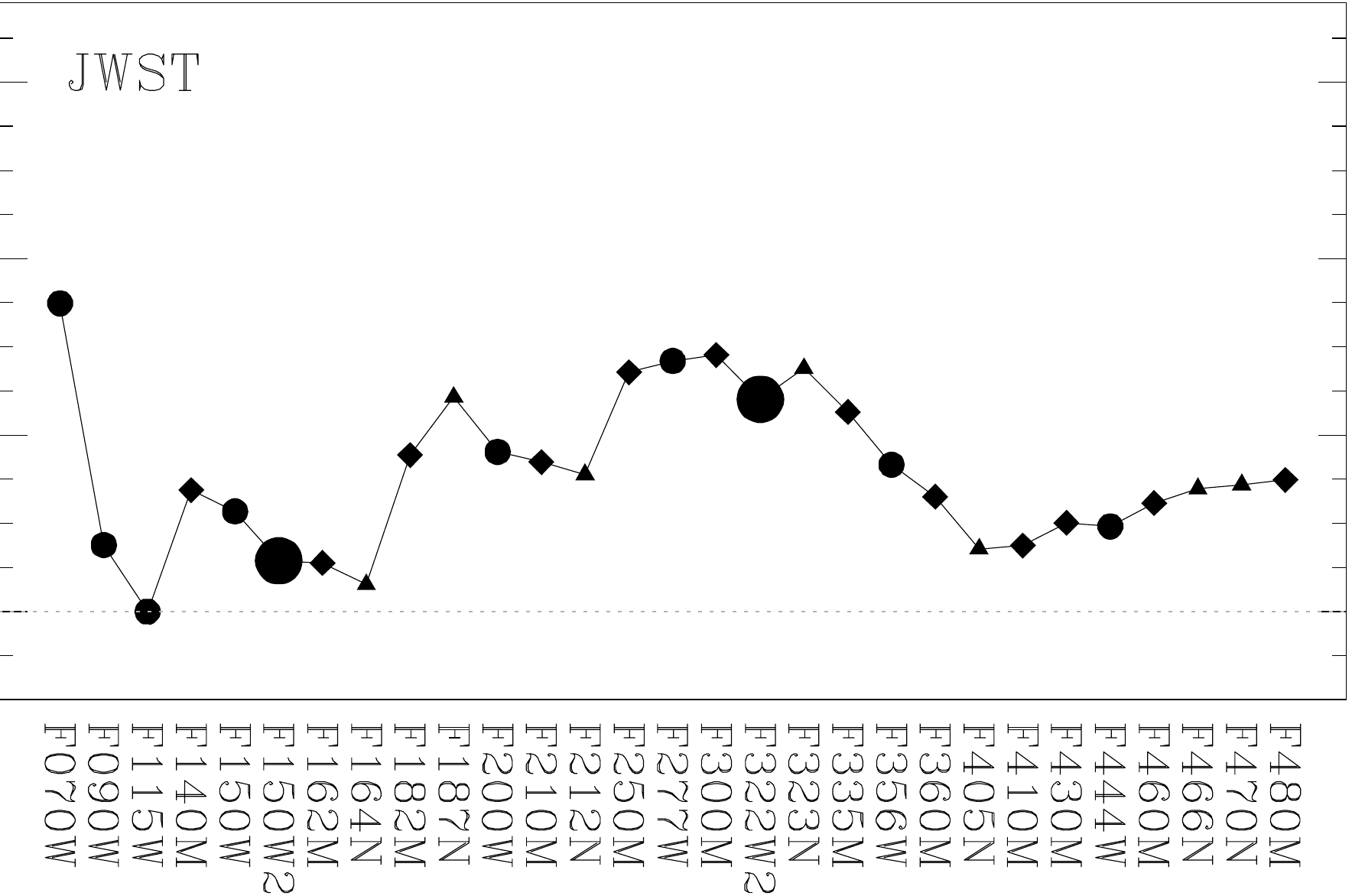}
 %/home/milone/WORKS/JWSTSPECTRA/47TSPEC/figDC047t figMMS
 \caption{Flux ratio between the simulated spectra of M-dwarf stars with $M_{\rm F115W}$=8.0 mag, [Fe/H]=$-$0.75 dex and the same chemical compositions of 1P and extreme 2P stars of 47\,Tucanae (top). The corresponding magnitude differences are plotted in the bottom panels for various {\it HST} filters and for all NIRCam filters. }
 \label{fig:spettri47t} 
\end{figure*} 
\end{centering}

%{\color{red} ChMs per RGB e Upper MS????}
To constrain the effect of variations in C, N, and O on the stellar magnitudes, we extend here to 47\,Tucanae the procedure discussed in Section\,\ref{sec:iso} for the metallicity case of [Fe/H]$=-1.50$. 
%APM. Tabella e descrizione delle composizioni chimiche.
The results on M-dwarfs are summarized in Figure\,\ref{fig:spettri47t}, where we compare the fluxes of a 2P star with extreme chemical composition  %{\color{red} with extreme chemical composition FORSE DARE DEI NUMERI PER ESTREMA} 
 and a 1P star (top panel) and show the magnitude differences in the NIRCam bands and in various {\it HST} filters (bottom panels, see Table\,\ref{tab:iso} for details on the chemical composition). 

Qualitatively, for wavelengths redder than $\sim$9,000 \AA, the flux-ratio behavior is comparable with that observed for M-dwarfs with [Fe/H]=$-$1.5. 
The main differences occur at optical and UV wavelengths. We observe flux differences of more than 0.5 and 0.6 mag in the F606W and F070W bands, respectively, which are up two times bigger than the F300M magnitude difference. 
%{\color{red} NON MI RICORDO SE SI POTEVA DIRE QUALCOSA A PROPOSITO DELLA CAUSA DI QUESTA DIFFEERENZA.}
Such large flux differences are not observed in the spectra with [Fe/H]=$-$1.5, where we find moderate magnitude differences between M-dwarfs with different abundances of He, C, N, and O. On the other side, the spectra with [Fe/H]=$-$1.5 provide larger UV flux differences than the 47\,Tucanae spectra.

\begin{centering} 
\begin{figure*} 
    \includegraphics[height=8.5cm,trim={0.6cm 4.8cm 0.0cm 12.8cm},clip]{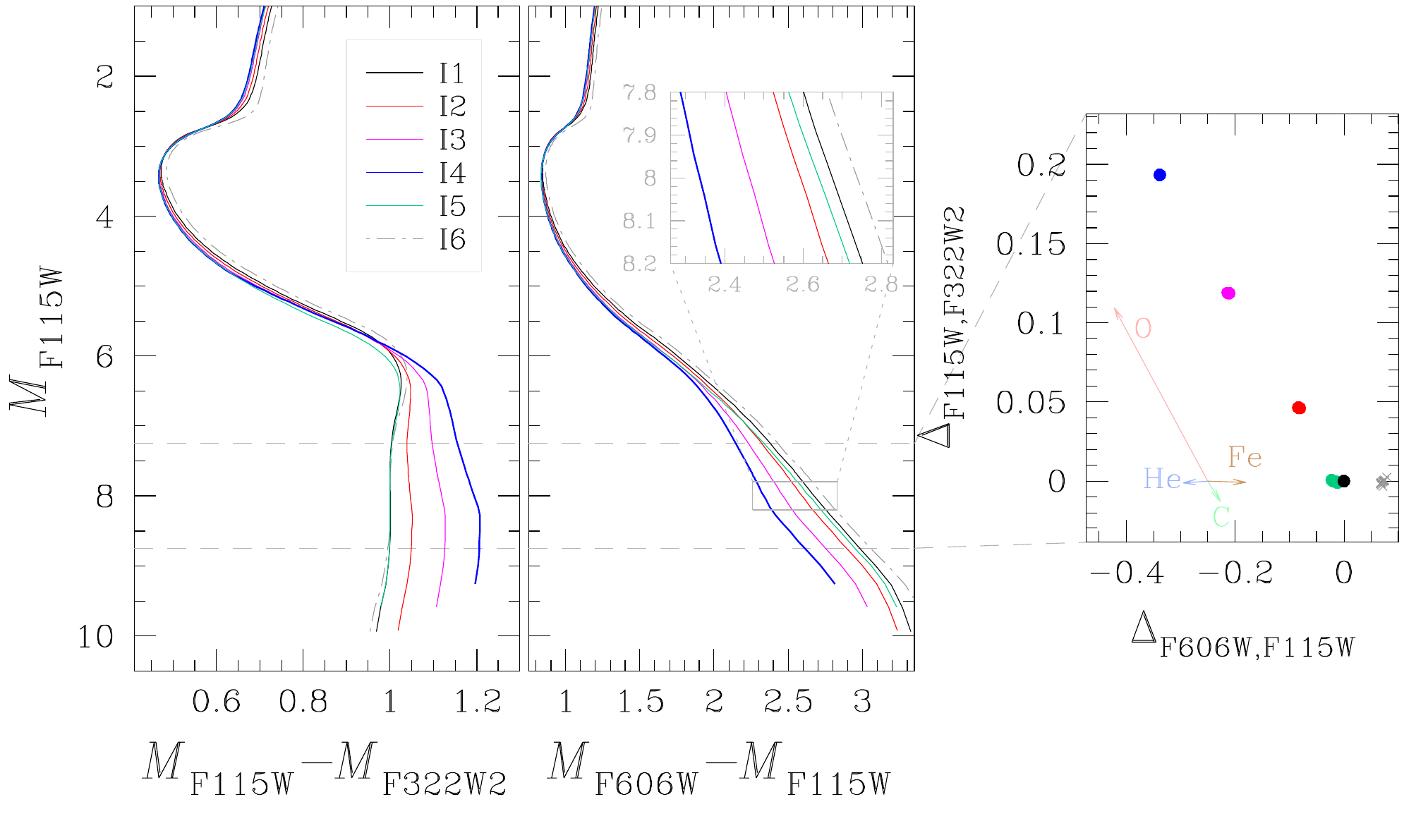}
 %/home/milone/WORKS/JWSTSPECTRA/JWST_25Gen/ISO/iso47t  figCOR (gof)
 \caption{Isochrones (I1--I5) with ages of 13 Gyr, [Fe/H]=$-$0.75 dex, [$\alpha$/Fe]=$+$0.40 dex, and different abundances  of He, C, N, and O in the $M_{\rm F115W}$ vs.\,$M_{\rm F115W}-M_{\rm F322W2}$ and $M_{\rm F115W}$ vs.\,$M_{\rm F606W}-M_{\rm F115W}$ CMDs. The dashed-dot isochrone, I6 has the same age and light-element abundance as I1 but is enhanced in [Fe/H] by 0.09 dex.
  The dashed horizontal lines delimit the MS region used to derive the ChM shown in the right panel. The colored arrows, which are shifted for clearness by $\Delta_{\rm F606W,F115W}=-0.25$ mag indicate the effect of changing the abundances of He, C, O, and Fe, one at a time. See the text for details.   }
 \label{fig:ChmTeo} 
\end{figure*} 
\end{centering}

Figure\,\ref{fig:ChmTeo} shows the $M_{\rm F115W}$ vs.\,$M_{\rm F115W}-M_{\rm F322W2}$ and $M_{\rm F115W}$ vs.\,$M_{\rm F606W}-M_{\rm F115W}$ CMDs for the six isochrones, I1--I6. The isochrones I1--I5 share the same iron content, [Fe/H]=$-0.75$, but have different abundances of He, C, N, and O, with I1 and I4 resembling the chemical composition of 1P and extreme 2P stars of 47\,Tucanae, respectively. The isochrone I5 is enhanced in helium mass fraction by $\Delta Y$=0.05 with respect to I1. Such helium difference corresponds to the variation that we obtain by assuming that the color extension of 1P stars in the RGB ChM is entirely due to star-to-star helium differences \citep{milone2018a}.
 The I6 isochrone has the same light-element abundances relative to iron as the I1 one but is slightly more metal-rich ([Fe/H]=$-0.66$). Such iron difference corresponds to the maximum [Fe/H] difference among 1P stars inferred by \citet{legnardi2022a} if metallicity variation is the only responsible for the extended 1P sequence of the RGB ChM.
  The elemental abundances of the I1-I6 isochrones are indicated in Table\,\ref{tab:iso}. 

\begin{table}
  \caption{Chemical compositions of the isochrones I1--I6. The elemental abundances of the isochrones I1 and I4 resemble 1P stars and 2P stars with extreme chemical compositions of 47\,Tucanae, respectively,  and are used to derive the spectra of  Figure\,\ref{fig:spettri47t}.}
\begin{tabular}{l c c c c l }
\hline \hline
 ID & Y & [C/Fe]  & [N/Fe] & [O/Fe] & [Fe/H]  \\
 I1 & 0.254 &    0.00    &  0.00  &     0.40  & $-$0.75  \\
 I2 & 0.261 & $-$0.05    &  0.70  &     0.30  & $-$0.75  \\
 I3 & 0.284 & $-$0.15    &  1.00  &     0.15  & $-$0.75  \\
 I4 & 0.300 & $-$0.35    &  1.20  &  $-$0.10  & $-$0.75  \\
 I5 & 0.300 &    0.00    &  0.00  &     0.40  & $-$0.75  \\
 I6 & 0.254 &    0.00    &  0.00  &     0.40  & $-$0.66  \\
%\hline
     \hline\hline
\end{tabular}
  \label{tab:iso}
 \end{table}

We used the isochrones in Figure\,\ref{fig:ChmTeo} to construct the $\Delta_{\rm F115W,F322W2}$ vs.\,$\Delta_{\rm F606W,F115W}$ ChM plotted in the right panel, which corresponds to the MS  segments between the dashed lines plotted in the left-hand and middle panels. 
The 1P stars  are clustered around the origin of the ChM, while most O-poor stars are distributed on the top-left extreme of the ChM. 2P stars with intermediate oxygen abundances exhibit less-extreme values of $\Delta_{\rm F606W,F115W}$ and $\Delta_{\rm F115W,F322W2}$. The colored arrows indicate the effect of changing the abundances of helium, carbon, oxygen, and iron, one at a time by $\Delta$Y=0.05, $\Delta$[C/Fe]=$-$0.25 dex, $\Delta$[O/Fe]=$-$0.25 dex, and $\Delta$[Fe/H]=0.1 dex. Nitrogen variations have a negligible effect on the location of the stars in this ChM.
 Noticeably, the isochrones I1 and I5 share nearly the same position in the ChM, thus suggesting that helium variation alone is not responsible for the extended 1P sequence of the ChM. Metallicity variations of [Fe/H]=0.09 dex %, as inferred from the RGB, 
 correspond to a difference of $\Delta_{\rm F606W,F115W} \sim 0.07$ mag. Hence, the observed F606W$-$F115W color extension of the ChM  $\Delta_{\rm F606W,F115W} = 0.10 \pm 0.01$ mag is consistent with a [Fe/H] variation of 0.12$\pm$0.01 dex, which is higher, at $2.5 \sigma$ level than the value inferred from the RGB width by \citet{legnardi2022a}.

%%%%%%%%%%%%%%%%%%%%%%%%%%%%%%%%%%

\begin{centering} 
\begin{figure*} 
    \includegraphics[height=8.2cm,trim={0.6cm 5.5cm 0.0cm 5.0cm},clip]{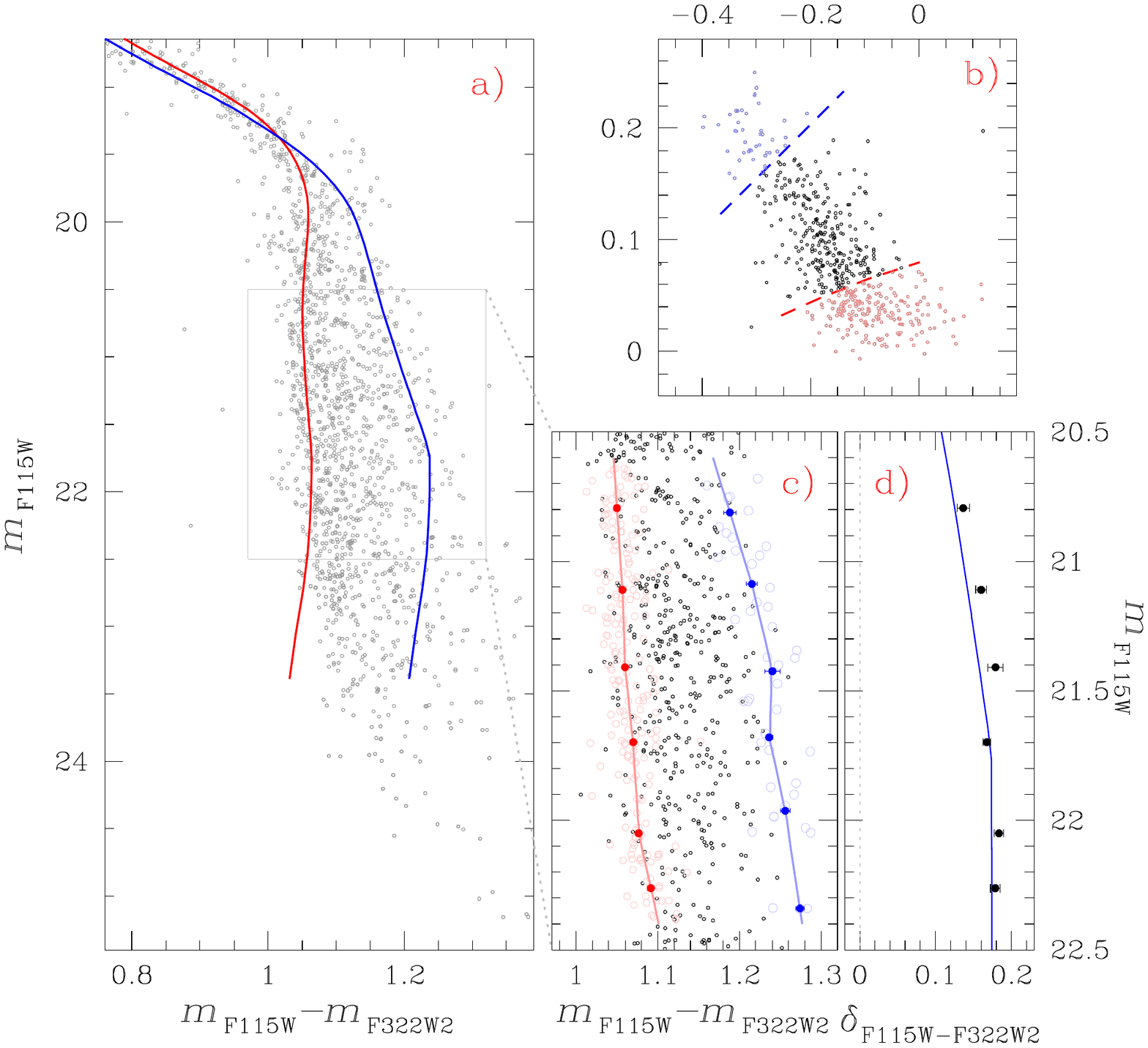}
        \includegraphics[height=8.2cm,trim={0.6cm 5.5cm 0.0cm 5.0cm},clip]{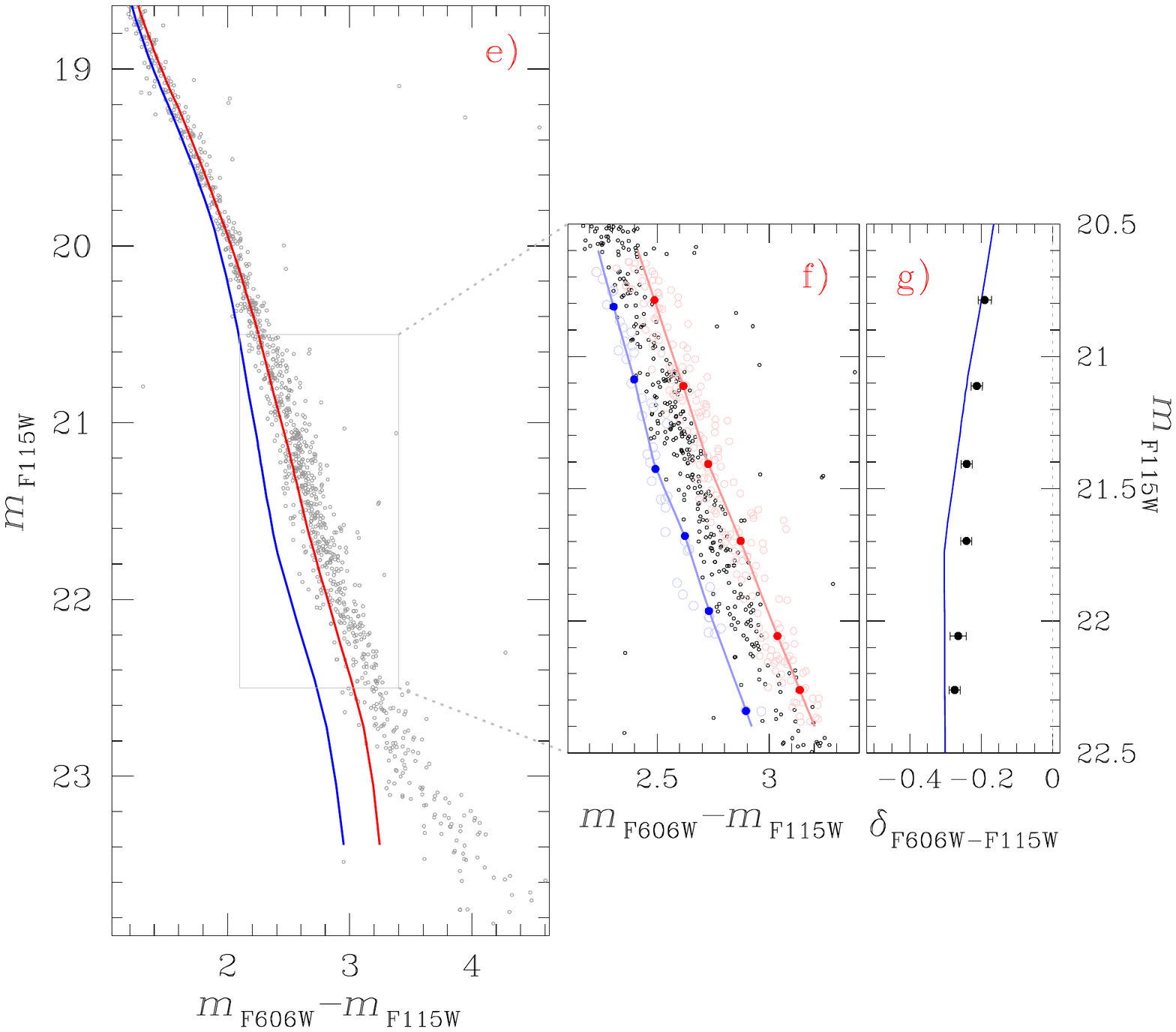}
 %/home/milone/WORKS/JWSTSPECTRA/JWST_25Gen/ISO/iso47t.macro con0 con
 \caption{Reproduction of the $m_{\rm F115W}$ vs.\,$m_{\rm F115W}-m_{\rm F322W2}$ CMD of Figure\,\ref{fig:CMDs} for probable cluster members alone (panel a). The red and blue lines superimposed on the CMD are the isochrones that provide the best fitting of 1P stars and 2P stars with extreme chemical composition, respectively and comprise stars more massive than 0.1 solar masses. The $\Delta_{\rm F115W,F322W2}$ vs.\,$\Delta_{\rm F606W,F115W}$ ChM of MS stars with $20.6<m_{\rm F115W}<22.4$ mag is plotted in the panel b, whereas panel c provides a zoom of the panel-a CMD around the low MS.  Panel d shows the $m_{\rm F115W}-m_{\rm F322W2}$ color distances between extreme 2P stars and 1P stars in five magnitude bins (black dots). 
  The comparison between the isochrones of 1P and extreme 2P stars and the $m_{\rm F115W}$ vs.\,$m_{\rm F606W}-m_{\rm F115W}$ CMD of 47\,Tucanae members is plotted in panel e, whereas panel f shows a zoom of the panel-e CMD below the MS knee. The $m_{\rm F606W}-m_{\rm F115W}$ color differences between extreme 2P stars and 1P stars are plotted in panel g as a function of the $m_{\rm F115W}$.
 The red and blue dots superimposed on the CMDs of panels c and f indicate the median colors of 1P and extreme 2P stars, respectively, in five F115W magnitude intervals, whereas the corresponding fiducial lines are indicated by continuous lines.
 The 1P and extreme 2P stars, selected in the ChM are colored pink and azure, respectively, in panels b, c, and f.
 The blue lines shown in panels d and g show the color distance between the blue and the red isochrones plotted in panel a.}
 \label{fig:comp} 
\end{figure*} 
\end{centering}

To further compare the isochrones and the NIRCam photometry of 47\,Tucanae, we superimposed the isochrones I1 and I4 to the $m_{\rm F115W}$ vs.\,$m_{\rm F115W}-m_{\rm F322W2}$ CMD of proper-motion selected cluster members (Figure\,\ref{fig:comp}a). To do that, we adopted a distance modulus ($m-M$)$_{0}$=13.38 mag and a foreground reddening E(B$-$V)=0.03 mag. 

To better compare the relative colors of the multiple stellar populations of 47\,Tucanae with the isochrones, we identified by eye the bulk of 1P stars and 2P stars with extreme chemical composition in the ChM shown in panel b of Figure\,\ref{fig:comp}. 
 We derived the fiducial lines of the selected stars in the $m_{\rm F115W}$ vs.\,$m_{\rm F115W}-m_{\rm F322W2}$ CMD, as shown in Figure\,\ref{fig:comp}c.
  To do this, we divided the F115W magnitude interval between 20.6 and 22.4 mag into six bins of the same size. We calculated the median color and magnitude of the stars in each bin and linearly interpolated these points. 
  The points that we used to derive the fiducial lines of 1P and extreme 2P stars are colored red and blue, respectively. The error bars are estimated as the dispersion of the colors of the stars in each magnitude bin divided by the square root of the number of stars minus one.
  The $m_{\rm F115W}-m_{\rm F322W2}$ color differences between the fiducials of 1P and extreme 2P stars, $\delta_{\rm F115W-F322W2}$, are plotted in panel d of Figure\,\ref{fig:comp} against the F115W magnitude. The blue line represents the corresponding color difference between the I5 and I1 isochrone and provides a good match of the observed points. 

  The comparison between the isochrones and the photometry of 47\,Tucanae in the $m_{\rm F115W}$ vs.\,$m_{\rm F606W}-m_{\rm F115W}$ plane is provided in  Figure\,\ref{fig:comp}e. The isochrones provide a reasonable fit for the observed MS above the knee but exhibit bluer colors than the bulk of MS data at fainter magnitudes. 
   Figure\,\ref{fig:comp}f shows a zoom of the panel-e CMD around the MS region below the knee. Here we plot the fiducial lines of 1P stars and extreme 2P stars that we derived with the same procedure described above for the $m_{\rm F115W}$ vs.\,$m_{\rm F115W}-m_{\rm F322W2}$ CMD, while the black dots shown in panel g indicate the $m_{\rm F606W}-m_{\rm F115W}$ color differences between extreme 2P and 1P stars, $\delta_{\rm F606W-F115W}$. In this case, the color difference between the I4 and I1 isochrones is slightly larger than the observed one.
    The best fit with the observations is provided by an isochrone with the same chemical composition as I5 but with $\sim$0.07 dex larger [O/Fe].

%%%%%%%%%%%%%%%%%%%%%%%%%%%%%%%%%%

\section{Summary and conclusions}\label{sec:conclusion}
Based on deep images collected with {\it HST} and {\it JWST}, we investigated the multiple populations at the bottom of the MS of 47\,Tucanae. We analyzed two distinct fields, namely A and B, with average radial distances of $\sim$7 and $\sim$8.5 arcmin, respectively, from the cluster center.
In addition, we computed synthetic spectra in the wavelength interval between $\sim$2,000 and 51,000 \AA\, for 13-Gyr old stars with [$\alpha$/Fe]=0.4 and with two metallicity values  corresponding to [Fe/H]=$-$1.50 and $-$0.75. We used these spectra, which have chemical compositions that are representative of  1P and 2P stars in GCs, to construct isochrones that account for multiple populations. We simulated photometry in all the filters of NIRCam and in various filters of UVIS/WFC3 (F275W, F336W, F438W, F606W, and F814W), NIR/WFC3 (F110W and F160W), and WFC/ACS (F606W and F814W), that are commonly involved in the investigation of multiple populations.

The main results on 47\,Tucanae can be summarized as follows:
\begin{itemize}
\item The CMD composed of NIRCam filters alone, $m_{\rm F115W}$ vs.\,$m_{\rm F115W }-m_{\rm F322W2}$ reveals that the MS of 47\,Tucanae exhibits a narrow color broadening for luminosities brighter than the MS knee. The MS broadening suddenly increases below the MS knee, thus revealing multiple populations among M dwarfs. Most M-dwarfs are distributed on the blue side of the MS, but a tail of stars is extended towards the red.
A similar pattern is observed in the $m_{\rm F160W}$ vs.\,$m_{\rm F110W}-m_{\rm F160W}$ CMD from {\it HST} photometry, although the maximum MS color broadening is smaller than that observed in the F115W and F322W2 filters of NIRCam.
 We detected a narrower sequence of faint stars with masses smaller than 0.1 $\mathcal{M}_{\odot}$. The F115W$-$F322W2 color distribution of these very-low-mass stars is mostly composed of blue MS stars, and the number of stars that populate the red tail seems two times smaller than that observed among stars with $\mathcal{M}>0.1 \mathcal{M}_{\odot}$.

We find that the $m_{\rm F115W}$ vs.\,$m_{\rm F606W }-m_{\rm F115W}$ CMD 
 is another efficient diagram to identify multiple populations among M dwarfs. In this CMD most M-dwarfs are located in the middle of the MS, but two less-numerous populations are distributed on the red and the blue side of the bulk of MS stars. Multiple sequences are also visible in the $m_{\rm F814W}$ vs.\,$m_{\rm F606W }-m_{\rm F814W}$ CMD from ACS/WFC photometry.

\item We introduced two ChMs that allowed us to better identify multiple populations among M-dwarfs. We defined $\Delta_{\rm F110W,F160W}$ vs.\,$\Delta_{\rm F606W,F814W}$ diagram, which is entirely constructed from {\it HST} photometry, and the $\Delta_{\rm F115W,F322W2}$ vs.\,$\Delta_{\rm F606W,F115W}$, where we combine UVIS/WFC3 and NIRCam photometry. The location of a star in both ChMs is mostly due to its oxygen abundance.
 Both ChMs reveal an extended 1P sequence and three main groups of 2P stars.

\item 
A similar conclusion of an extended 1P sequence in the ChM of 47\,Tucanae comes from RGB stars \citep{milone2017a, jang2022a} but this is the first evidence of chemical inhomogeneity among unevolved M-dwarfs. 
By comparing isochrones and observations of 47\,Tucanae, we find that the extended 1P sequence is consistent with an internal variation of $\Delta$[Fe/H]=0.12$\pm$0.01 dex. In this scenario, most 1P stars would share low iron content, and the remaining stars are distributed toward larger values of [Fe/H]. The comparison between isochrones with different helium abundances and the observed ChM rules out the possibility that helium variations are responsible for the color extension of the 1P, thus corroborating results based on spectroscopy and UV photometry of RGB and bright-MS stars \citep{marino2019a, marino2019b, tailo2019a, legnardi2022a}. However, the maximum iron variation derived in this paper is much larger than that inferred by Legnardi and collaborators from the ChM RGB stars ($\Delta$[Fe/H]=0.09$\pm$0.01 dex). 

\item
The width of the MS of M-dwarfs more massive than $\sim$0.1 $\mathcal{M}_{\odot}$ in the F115W$-$F322W2, F606W$-$F115W, and F606W$-$F115W colors is well fitted by isochrones of 1P and 2P stars with similar He, C, N, and O abundances as those inferred from RGB and bright MS stars. The evidence that the multiple populations of 47\,Tucanae share similar chemical compositions among stars with different masses is a strong constraint for those scenarios on the formation of multiple populations where all GC stars are coeval and the chemical composition of 2P stars is due to accretion of polluted material onto already existing pre-MS stars \citep{gieles2018a}.
 These results could imply that the amount of accreted material is proportional to the stellar mass. As an example, they would exclude a Bondi accretion, where the amount of accreted material is proportional to the square
of the stellar mass \citep{bondi1944a} and the very-low-mass stars accrete a smaller amount of polluted gas and exhibit smaller oxygen variations than RGB stars.

\end{itemize}

We used the isochrones to explore the impact of multiple populations with different abundances of He, C, N, and O on the photometric diagrams obtained with NIRCam bands. The CMDs constructed with NIRCam magnitudes are poorly sensitive to multiple stellar populations among the MS segment brighter than the knee, the SGB, and most RGB. The stellar populations with extreme helium variations, where we observed split MSs and RGBs are a remarkable exception. However, for a fixed luminosity, the color separation corresponding to a large helium difference of $\Delta Y \sim 0.08$ is typically smaller than 0.1 mag. 
Moreover, as pointed out by \citet{salaris2019a}, the spectra of 1P and 2P stars in the upper RGB exhibit significant flux  differences, which are most visible at wavelengths of $\lambda \sim$20,000-35,000 $\AA$ and for $\lambda \gtrsim 40,000 \AA$.\\
Below the knee, 1P and 2P stars define distinct sequences in various CMDs with the F090W$-$F300M color providing the maximum separation between the MSs corresponding to multiple populations of M-dwarfs. Other CMDs that are constructed with the F115W and F322W2 filters and with F150W2 and F322W2 filters are efficient diagrams to disentangle 1P and 2P M-dwarfs. Although these diagrams provide smaller color separations than that given by the F090W$-$F300M color, photometry in these filters requires shorter exposure times to get the same signal-to-noise ratio.

The isochrones with [Fe/H]=$-$0.75 show large F070W magnitude differences between 1P and 2P M-dwarfs, in contrast with what is observed for [Fe/H]=$-$1.5. A similar conclusion can be extended to the F606W bands of WFC/ACS and UVIS/WFC3. Hence, the F070W$-$F115W color and the C$_{\rm F070W,F115W,F322W2}$ pseudo-color would be powerful tools to identify multiple populations of M-dwarfs among metal-rich GCs.

\section*{acknowledgments} 
\small
%We are grateful to the anonymous referee for a constructive report that has improved the quality of the manuscript.
This work has received funding from the European Research Council (ERC) under the European Union's Horizon 2020 research innovation programme (Grant Agreement ERC-StG 2016, No 716082 'GALFOR', PI: Milone, http://progetti.dfa.unipd.it/GALFOR) and from the European Union’s Horizon 2020 research and innovation programme under the Marie Sklodowska-Curie Grant Agreement No. 101034319 and from the European Union – NextGenerationEU, beneficiary: Ziliotto.
APM, MT, and ED acknowledge support from MIUR through the FARE project R164RM93XW SEMPLICE (PI: Milone). APM and ED  have been supported by MIUR under PRIN program 2017Z2HSMF (PI: Bedin).

\section*{Data availability}
The data underlying this article will be shared upon reasonable request to the corresponding author.

\bibliography{ms}

\section{appendix A. Geometric distortion for the NIRCam SW detectors}
To derive stellar proper motions and separate field stars from cluster members in field B, we compare the distortion-free positions of stars observed by NIRCam as part of GO-2560 and by UVIS/WFC3 (GO-11677). While accurate solutions for geometric distortion of UVIS/WFC3 detectors are derived by \citet{bellini2009a} and \citet{bellini2011a}, high-precision maps for the geometric distortion of the NIRCam detectors are not available.

For this reason, we estimated the geometric distortion solution of the NIRCam SW detectors by using F090W images of a field in the Large-Magellanic Cloud located around (RA=05$^{\rm h}$:22$^{\rm m}$00$^{\rm s}$, DEC=$-69^{\rm d}$:$30^{\rm m}$:00$^{\rm s}$) that has been observed by JWST for calibration purposes.
This dataset consists of 9$\times$21s well-dithered images collected as part of GO\,1476 (PI\,M.\,Boyer) with RAPID readout pattern and two groups per integration. 
In addition, we used 15$\times$419s images from GO\,1144 (PI\,G.\,Hartig) that comprise three distinct regions in the calibration fields, with different orientations with respect to GO\,1476 images. Images of each field are collected by using five dithered points and  BRIGH\,1 readout pattern.

We derived photometry and astrometry of stars in each image by using the methods and the computer programs described in Section\,\ref{sec:data}.  
We build the astrometric master frame by cross-identifying the star catalogs from each individual exposure. Four-parameter linear transformations are used to transform the coordinate of the stars from the reference frame of each image into a common reference frame. To define the first master frame, we used the stars of the Gaia DR3 catalog, for which NIRCam measurements are available after projecting their coordinates into the tangential plane. Only used unsaturated stars that are relatively bright and well-fitted by the PSF model.
 We estimated the geometric-distortion solution by following the method introduced by \citet{anderson2003b} for the wide-field planetary camera 2 onboard {\it HST} and extended to various detectors from space and ground-based facilities \citep[e.g.\,][]{anderson2006a, bellini2011a, libralato2014a}.  
The distortion correction is provided by three terms: i) a linear transformation to put the eight detectors into a common reference frame, ii) a fifth-order polynomial correction that models the general optical distortion, and iii) a table of residuals that accounts for fine structure effects. 
In the following, we indicate the transformation from the detector $k$ of the coordinate system of image $j$ to the master reference frame as $T_{\rm j,k}$ \citep{libralato2014a}. 

The polynomial solution was computed for each detector, separately, by using an iterative approach. For convenience, we normalized the coordinates to the central pixel (1024, 1024), which allows us to better recognize the size of the
contribution of each term to the solution \citep{anderson2003b}.
\begin{itemize}
\item We first derived the linear transformation between the stars and the best-available master frame.
\item Hence, we transformed the position of each star in the master frame into the coordinate system by using the inverse transformation, $T_{\rm j,k}^{-1}$. The difference between the observed position of each star and the corresponding transformed reference-frame position is used to derive a pair of residuals ($\delta x, \delta y$).
\item 
We defined a look-up table made up of 14$\times$14 elements and computed the 3-$\sigma$-clipped median values of the residuals. The median residuals are fitted with two fifth-order polynomials by means of least squares, to derive the coefficients that best reproduce the observations.
\item We used the best-fit polynomials to correct the coordinates of the stars in each detector. We used a null correction at the first iteration and used half of the adjustment at the subsequent iterations. The new coordinates are used to derive an improved master frame. 
\end{itemize}
This procedure was iterated until the two subsequent determinations of the positions differ by less than 1\%.
To model the residual distortion we derived a look-up table of residuals by following the iterative procedure from previous work \citep{anderson2006b, bellini2011a, libralato2014a}.  The absolute values of the median residuals never exceed 0.008 pixels. The three steps of the iterative procedure can be summarized as follows.

i) We first derived a master frame by applying the polynomial solution and calculating the position residuals between each exposure and the master frame.
ii) Then, we divided each detector into a grid composed of 14$\times$14 cells and computed 3-$\sigma$-clipped median values of the positional residuals of stars in each cell. 
To associate a look-up table correction with each star in the detector, we used a bi-linear interpolation among the surrounding four grid points \citep[see][and references therein for details]{anderson2006b, libralato2014a}. iii) Finally, we corrected the stellar positions by using 75\% of the grid-point values and used the corrected positions to derive an improved master frame.
This procedure was iterated until the corrections are smaller than 0.003 pixels.

\section{appendix B. The four stellar populations of 47\,Tucanae}
\begin{centering} 
\begin{figure} 
    \includegraphics[height=5.0cm,trim={0.6cm 5.2cm 0.0cm 14.1cm},clip]{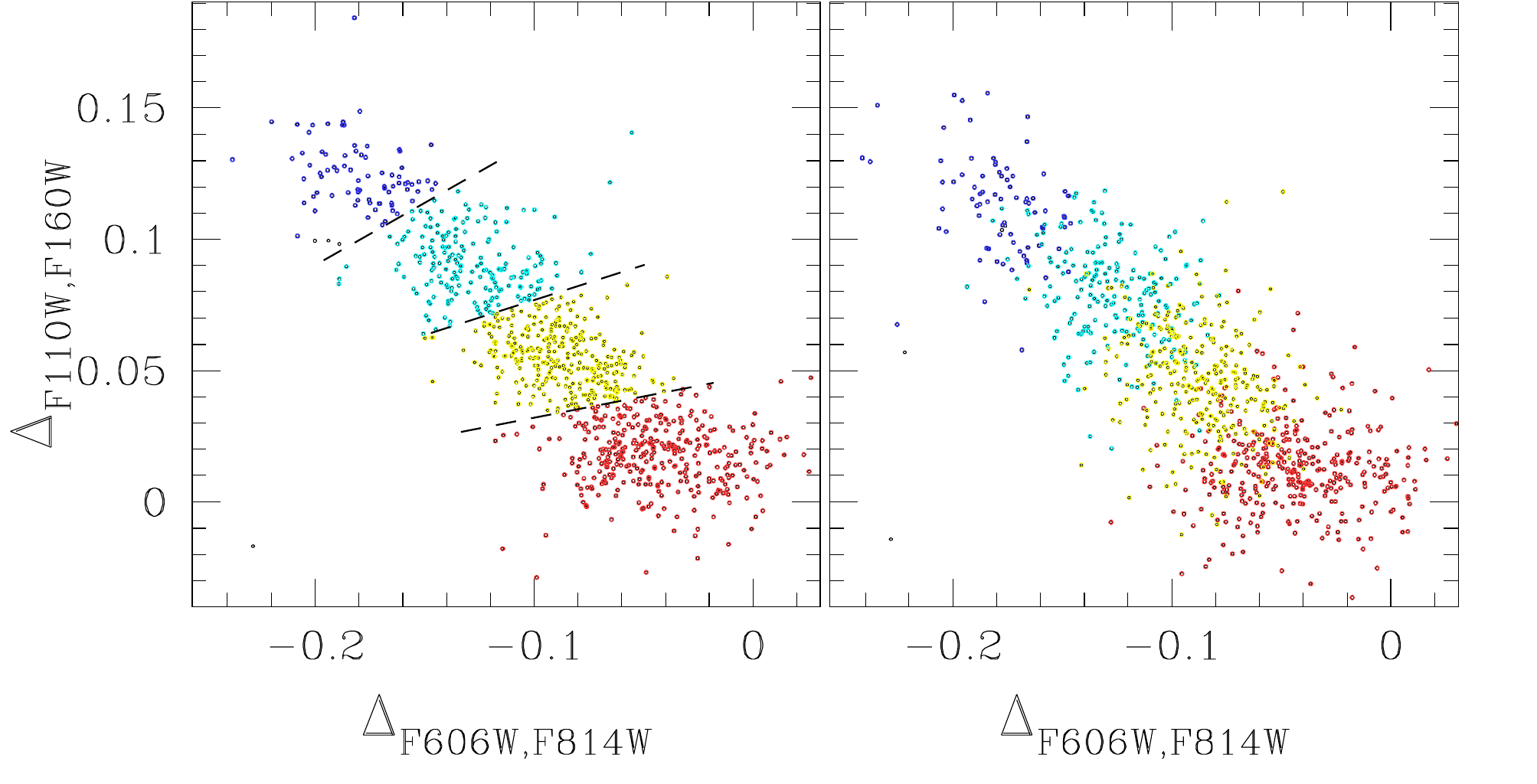}
    \includegraphics[height=5.0cm,trim={0.6cm 5.2cm 0.0cm 14.1cm},clip]{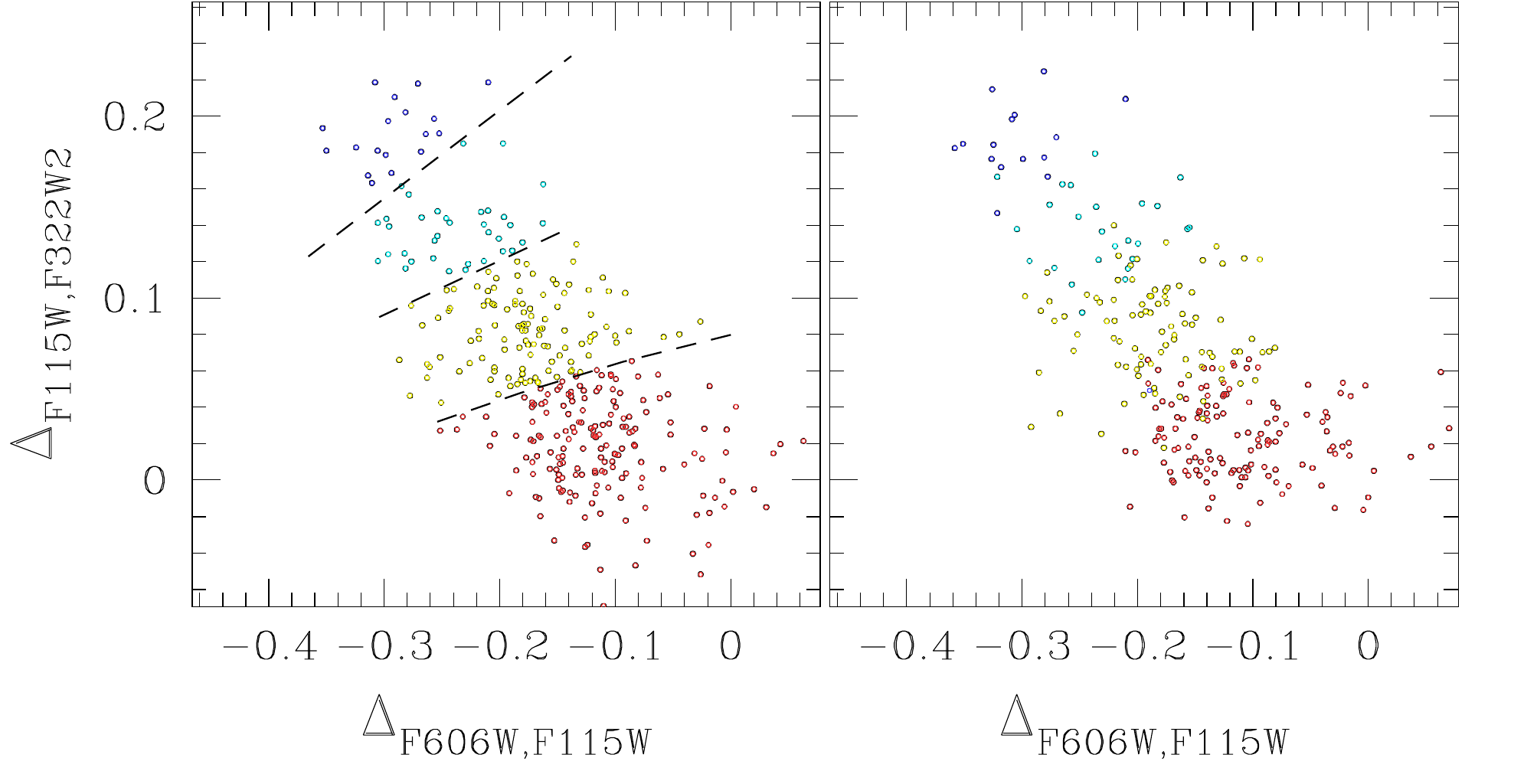} 
    %/home/milone/
 \caption{  $\Delta_{\rm F110W,F160W}$ vs.\,$\Delta_{\rm F606W,F814W}$ (top) and $\Delta_{\rm F115W,F322W2}$ vs.\,$\Delta_{\rm F606W,F115W}$ ChMs (bottom), derived from two distinct subsample of data. The 1P stars, 2P$_{\rm A}$, 2P$_{\rm B}$, and 2P$_{\rm C}$ stars selected from the left-panel ChMs are colored red, yellow, cyan, and blue respectively. See the text for details.}
 \label{fig:j9} 
\end{figure} 
\end{centering}

 To further demonstrate that the ChM plotted in Figure\,\ref{fig:CMDsHST} is consistent with four stellar populations, we used the procedure introduced by \citet{anderson2009a}. 
In a nutshell, we divided the F606W, F814W, F110W, and F160W images into two distinct groups and derived stellar photometry from the images in each group separately. The top panels of Figure\,\ref{fig:j9} show the resulting $\Delta_{\rm F110W,F160W}$ vs.\,$\Delta_{\rm F606W,F814W}$ ChMs.
We used the left-panel ChM to select four groups of stars, including a sample of bonafide 1P stars, and three sub-groups of 2P stars, namely P2$_{\rm A}$, P2$_{\rm B}$, and P2$_{\rm C}$. These stars are colored red, yellow, cyan, and blue, respectively, in the top panels of Figure\,\ref{fig:j9}.

Similarly, we derived the $\Delta_{\rm F115W,F322W2}$ vs.\,$\Delta_{\rm F606W,F115W}$ ChMs plotted in the bottom panels of Figure\,\ref{fig:j9} by using two distinct groups of field-B images in the F115W, F322W2 and F606W filters.
The evidence that the four groups of stars, selected from the left-panel ChMs, show different average colors in the right-panel ChMs demonstrates that the multiple populations of 47\,Tucanae are not artifacts due to observational errors \citep{anderson2009a}. 
%%%%%%%%%%%%%%%%%%%%%%%%%%%%%%%%%%
%%%%%%%%%%%%%%%%%%%%%%%%%%%%%%%%%%

To further confirm the presence of four stellar populations among the M-dwarfs of 47\,Tucanae, we analyze the distribution of the 1P stars, 2P$_{\rm A}$, 2P$_{\rm B}$, and 2P$_{\rm C}$ stars in CMDs constructed with filters that are not used to derive the ChM \citep{milone2010a}. 
%%%%%%%%%%%%%%%%%%%%%%%%%%%%%%%%%%
\begin{centering} 
\begin{figure*} 
    \includegraphics[height=12.0cm,trim={0.5cm 5.0cm 0.0cm 4.0cm},clip]{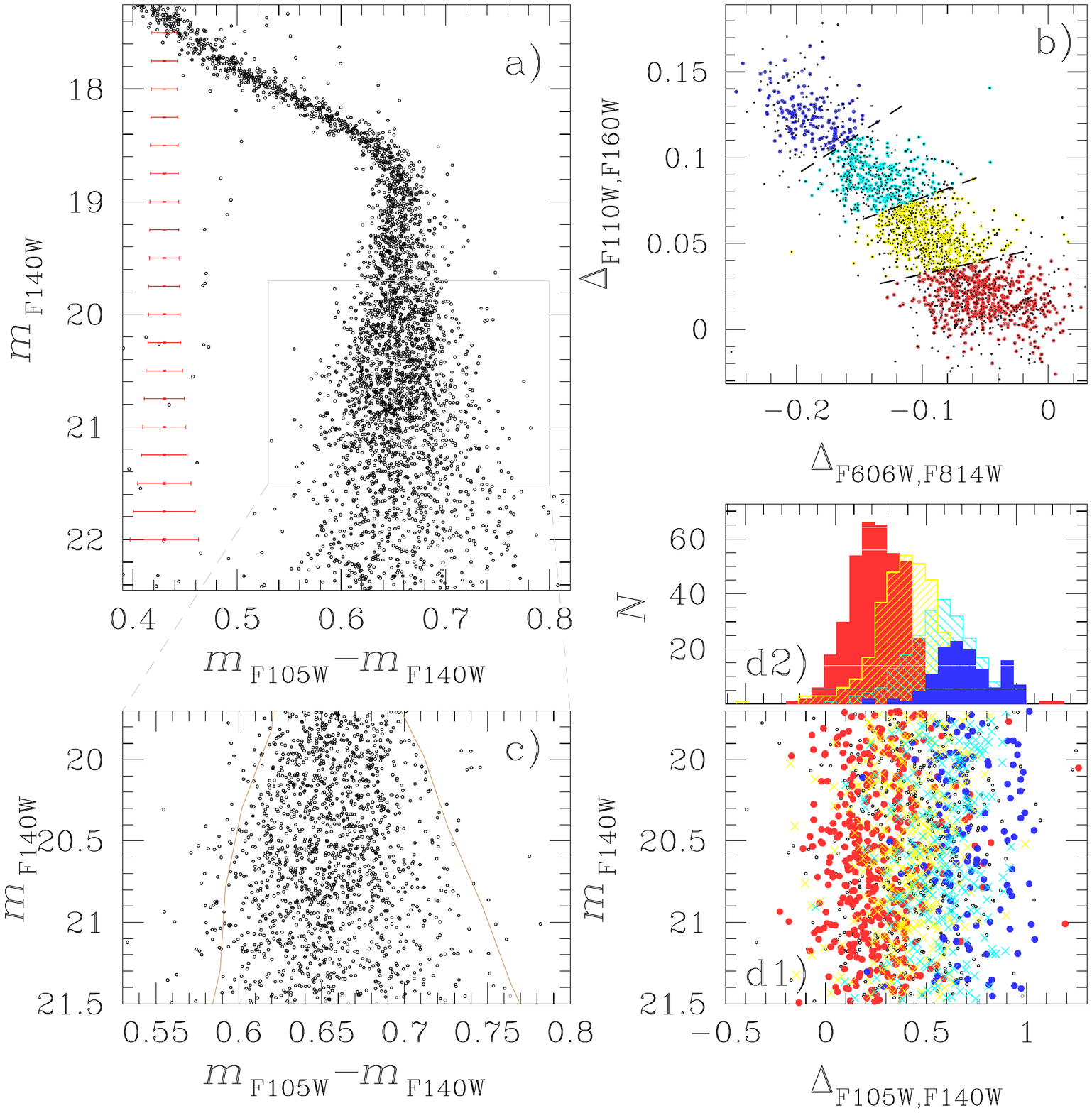}
    %/home/milone/
 \caption{ Panel a shows the $m_{\rm F140W}$ vs.\,$m_{\rm F105W}-m_{\rm F140W}$ CMD of stars in the field A, whereas panel b reproduces the $\Delta_{\rm F110W,F160W}$ vs.\,$\Delta_{\rm F606W,F814W}$ ChM of Figure\,\ref{fig:CMDsHST}. The panel c is a zoom of panel-a CMD, in the region populated by the stars shown in panel b. The brown lines represent the color boundaries of the MS. The verticalized $m_{\rm F140W}$ vs.\,$\Delta_{\rm F105W,F140W}$ diagram is plotted in panel d1, where the 1P, 2P$_{\rm A}$, 2P$_{\rm B}$, and 2P$_{\rm C}$ stars selected from the ChM are colored red, yellow, cyan, and blue respectively. The corresponding $\Delta_{\rm F105W,F140W}$ histogram distributions are shown in panel d2.}
 \label{fig:m10hst} 
\end{figure*} 
\end{centering} 
%%%%%%%%%%%%%%%%%%%%%%%%%%%%%%%%%%
 Figure\,\ref{fig:m10hst} illustrates the procedure  %that we used to further demonstrate that the 
 for the populations identified along the $\Delta_{\rm F110W,F160W}$ vs.\,$\Delta_{\rm F606W,F814W}$ ChM. % hosts four main stellar populations. 
  The $m_{\rm F140W}$ vs.\,$m_{\rm F105W}-m_{\rm F140W}$ CMD is plotted in panel a, whereas the ChM of field-A stars is shown in panel b. 
We derived the red and blue boundaries of the MS (brown lines in panel c) and obtained the verticalized  $m_{\rm F140W}$ vs.\,$\Delta_{\rm F105W,F140W}$ diagram shown in panel d1 \citep[see][for details]{milone2015a}.
We find that the groups of 1P, 2P$_{\rm A}$, 2P$_{\rm B}$, and 2P$_{\rm C}$ stars selected from the ChM show different $\Delta_{\rm F105W,F140W}$ values, as demonstrated by the $\Delta_{\rm F105W,F140W}$ histogram distributions plotted in panel d2.
This fact corroborates the conclusion that the M-dwarfs of field-A stars host four stellar populations.

%%%%%%%%%%%%%%%%%%%%%%%%%%%%%%%%%%
\begin{centering} 
\begin{figure*} 
    \includegraphics[height=12.0cm,trim={0.5cm 5.0cm 0.0cm 4.0cm},clip]{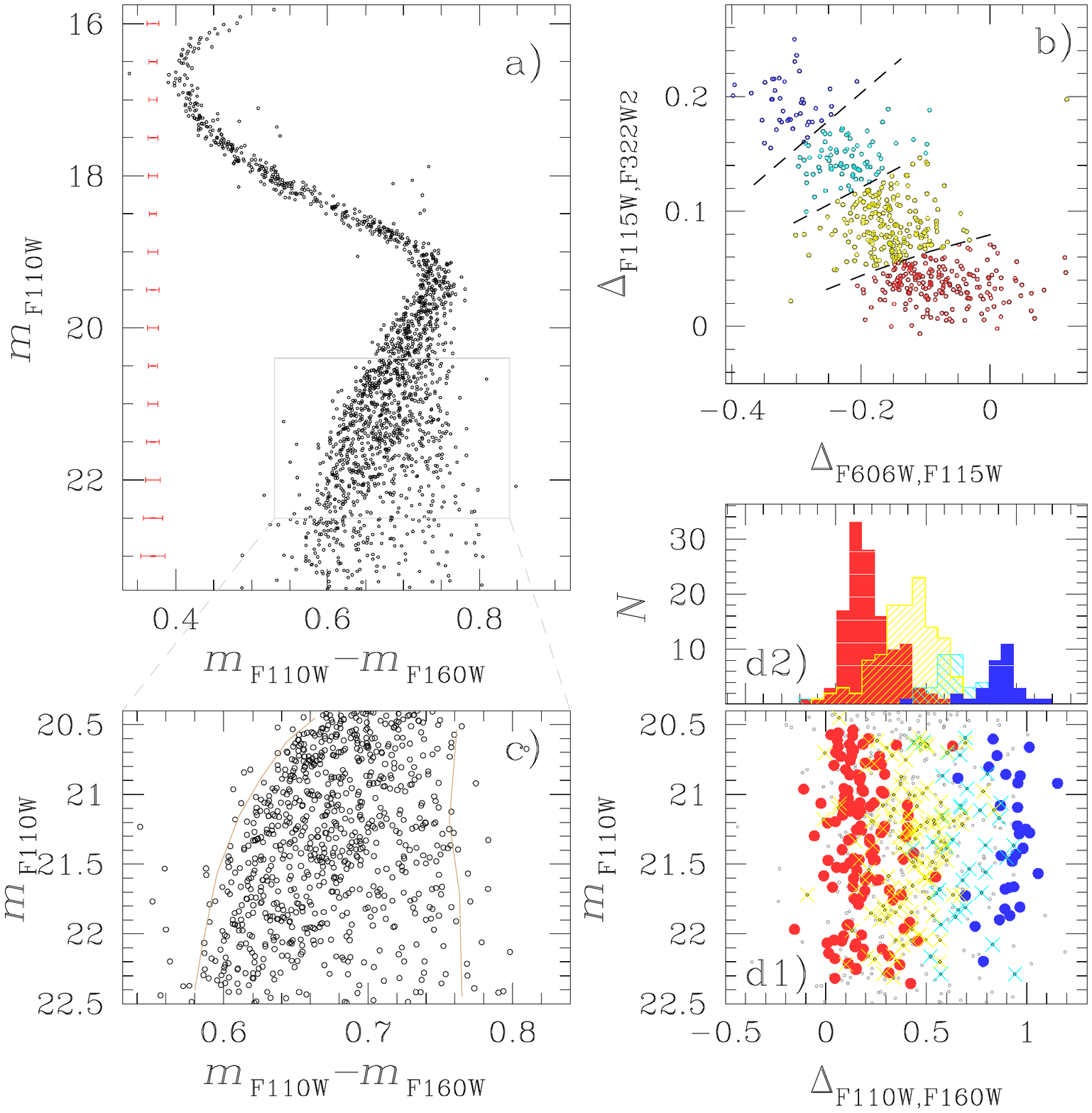}
    %/home/milone/
 \caption{ This figure is similar to Figure\,\ref{fig:m10hst} but for field-B stars. Here, we present the $m_{\rm F110W}$ vs.\,$m_{\rm F110W}-m_{\rm F160W}$ CMD (panels a and c), the $\Delta_{\rm F115W,F322W2}$ vs.\,$\Delta_{\rm F606W,F115W}$ ChM (panel b), the $m_{\rm F110W}$ vs.\,$\Delta_{\rm F110W,F160W}$ diagram (panel d1) and the $\Delta_{\rm F110W,F160W}$ histogram distribution for stars in the four stellar populations of 47\,Tucanae.}
 \label{fig:m10jwst} 
\end{figure*} 
\end{centering} 
%%%%%%%%%%%%%%%%%%%%%%%%%%%%%%%%%%
 As illustrated in Figure\,\ref{fig:m10jwst}, we  derived similar conclusions for field-B stars. In this case, we select the four stellar populations from the $\Delta_{\rm F115W,F322W2}$ vs.\,$\Delta_{\rm F606W,F115W}$ ChM, and investigate their distribution in the $m_{\rm F110W}$ vs.\,$m_{\rm F110W}-m_{\rm F160W}$ CMD. Clearly, the four selected groups of stars exhibit different average $\Delta_{\rm F110W,F160W}$ values.

\end{document}